\documentclass[aps,pra,twocolumn,superscriptaddress,longbibliography]{revtex4-1}
\usepackage{amsmath,amsfonts,amssymb,amscd,mathtools}
 \usepackage[normalem]{ulem}
\usepackage{dsfont}
\usepackage{amsthm}
\usepackage{graphicx}
\usepackage{dcolumn}
\usepackage{bm}
\usepackage{booktabs}
\usepackage{hyperref}
\usepackage[dvipsnames]{xcolor}
\hypersetup{
    bookmarksnumbered=true, 
    unicode=true, 
    pdfstartview={FitH}, 
    pdftitle={}, 
    pdfauthor={}, 
    pdfsubject={}, 
    pdfcreator={}, 
    pdfproducer={}, 
    pdfkeywords={}, 
    pdfnewwindow=true, 
    colorlinks=true, 
    linkcolor=NavyBlue, 
    citecolor=NavyBlue, 
    filecolor=NavyBlue, 
    urlcolor=NavyBlue 
}
\newcounter{one}
\newcommand{\ket}[1]{|{#1}\rangle}
\newcommand{\bra}[1]{\langle{#1}|}

\newcommand{\Tr}[0]{ \mathrm{Tr}}

\newcommand{\eq}[1]{\begin{align} #1 \end{align}}
\newtheorem{theorem}{Theorem}

\theoremstyle{definition}
\newtheorem{definition}[theorem]{Definition}

\newcommand\calD{{\cal D}}
\newcommand\calE{{\cal E}}

\newcommand\calL{{\cal L}}

\newcommand{\beq}{\begin{equation}}
\newcommand{\eeq}{\end{equation}}

\newcommand{\bal}{\begin{equation}\begin{aligned}}
\newcommand{\eal}{\end{aligned}\end{equation}}

\AtBeginDocument{\renewcommand{\selectlanguage}[1]{}}
\makeatletter
\def\@bibdataout@aps{%
 \immediate\write\@bibdataout{@CONTROL{apsrev41Control,author="00",editor="1",pages="0",title="0",year="0"}}%
 \if@filesw\immediate\write\@auxout{\string\citation{apsrev41Control}}\fi
}
\makeatother

\usepackage[lmargin=.7in,rmargin=.7in,tmargin=.7in,bmargin=1in]{geometry}

\begin{document}

\title{Open Quantum Time Crystals without External Thermodynamic Resources: \\
Go and No-Go Theorems}

\author{Yuri Ogami}
\email{ogami.yuri.197@s.kyushu-u.ac.jp}
\affiliation{Joint Graduate School of Mathematics for Innovation, Kyushu University, 744 Motooka, Nishi-ku, Fukuoka, 819-0395, Japan}
\author{Hiroyasu Tajima}
\email{hiroyasu.tajima@inf.kyushu-u.ac.jp}
\affiliation{Department of Informatics, Faculty of Information Science and Electrical Engineering, Kyushu University, 744 Motooka, Nishi-ku, Fukuoka, 819-0395, Japan}
\affiliation{JST, FOREST, 4-1-8 Honcho, Kawaguchi, Saitama, 332-0012, Japan}

\begin{abstract}
Time crystals are phases of matter that spontaneously break time-translation symmetry.
They have been ruled out in the ground and Gibbs states of isolated systems with only short-range interactions, yet they have been proposed and observed experimentally in nonequilibrium settings such as Floquet and open systems.
However, existing proposals for time crystals in open quantum systems require an external supply of thermodynamic resources, such as work or energetic coherence.
Whether coupling to an environment, without any such resource supply, can activate a time crystal absent in the isolated system remains open.
We answer this question systematically through a Go theorem and a no-go theorem.
The Go theorem answers it affirmatively: we construct an explicit model in which coupling a system to a zero-temperature Gibbs environment via a time-independent, energy-conserving interaction, with only short-range Hamiltonians, activates a time crystal in a stationary state that shows none in isolation.
This coupling reproduces a Lindblad master equation satisfying three conditions---time-independent Hamiltonian and jump operators, time-translation covariance, and detailed balance---that hold in standard derivations unless thermodynamic resources are supplied externally.
The no-go theorem shows that, for every Lindblad master equation satisfying these conditions and every stationary state, dissipation can never contribute to activation; only an environment-induced Hamiltonian term can.
Indeed, the periodic two-time correlation function of any such time crystal coincides with that of the isolated system with the open-system Hamiltonian, the sum of the bare isolated Hamiltonian and the environment-induced Hamiltonian, in the thermodynamic limit.
These results reveal both the possibility and the origin of time crystals in open systems without external resource supply.
\end{abstract}

\maketitle

\textit{\textbf{Introduction.---}}
Can time-translation symmetry be broken spontaneously?
In other words, can a many-body system sustain autonomous oscillations in the thermodynamic limit?
In 2012, Wilczek proposed time crystals---phases that spontaneously break this symmetry~\cite{Wilczek2012quantum,Shapere2012classical}.
The proposal immediately sparked an active debate on their realizability~\cite{Bruno2013comment,Wilczek2013reply,Bruno2013impossibility,Nozieres2013time,Khemani2019brief}, and Watanabe and Oshikawa~\cite{Watanabe2015absence,Watanabe2020proof} ruled them out in the ground and Gibbs states of isolated systems obeying a Lieb--Robinson bound~\cite{Lieb1972finite}, including almost all short-range interacting systems.
Beyond isolated equilibrium systems, various time-crystal models have been proposed in Floquet and open systems~\cite{Sacha2015modeling,Khemani2016phase,Else2016floquet,Yao2017discrete,Gong2018discrete,Iemini2018boundary,Buca2019nonstationary,Zaletel2023colloquium}, and several have been observed experimentally~\cite{Zhang2017observation,Choi2017observation,Mi2022time,Kessler2021observation,Kongkhambut2022observation,Wu2024dissipative}.
\begin{figure}[t]
  \centering
  \includegraphics[width=\columnwidth]{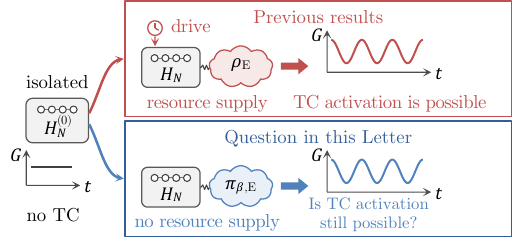}
\caption{Schematic of time-crystal (TC) activation by coupling to an environment.
An isolated system with Hamiltonian $H_N^{(0)}$ (left) exhibits no TC.
Top (red): existing TCs use external driving or a nonequilibrium environment $\rho_{\mathrm E}$, which require a supply of thermodynamic resources.
Bottom (blue): our question---can coupling to a Gibbs environment $\pi_{\beta,\mathrm E}$ at inverse temperature $\beta$ activate a TC without external resource supply?
$H_N$ is the open-system Hamiltonian, and $G$ is the two-time correlation function of a macroscopic order parameter in the thermodynamic limit.}
  \label{fig:concept}
\end{figure}

The open-system proposals are of great interest, but one point has been debated: they all require an external supply of thermodynamic resources.
In Floquet systems~\cite{Else2016floquet,Khemani2016phase,Yao2017discrete,Gong2018discrete,Gambetta2019discrete,Lazarides2020time,RieraCampeny2020time,Kessler2021observation}, the externally driven, time-dependent Hamiltonian involves work input~\cite{DAlessio2014long,Lazarides2014equilibrium,Zhu2019dicke,Cenedese2025thermodynamics} or an external clock~\cite{Goold2016role,Woods2019autonomous}, whereas in time-independent open systems, existing proposals violate detailed balance, time-translation covariance, or both~\cite{Iemini2018boundary,Buca2019nonstationary,Tucker2018shattered,Carollo2024quantum,Yang2025emergent}.
Such violations rely on thermodynamic resources such as temperature differences, nonequilibrium environments, or energetic coherence~\cite{Brandao2013resource,Lostaglio2015description,Lostaglio2015quantum}.
These resources cannot be obtained from an equilibrium environment, and their supply and maintenance require an external system with nonequilibrium free energy.

These observations suggest a conjecture: the activation of time crystals in open systems---their emergence through coupling to an external environment---requires an external supply of thermodynamic resources (Fig.~\ref{fig:concept}).
If so, such resource supply is essential to environment-induced time crystals.
Conversely, if not, environment-induced time crystals can arise without such resource supply.

In this Letter, we prove a Go theorem and a no-go theorem that settle this conjecture for open quantum systems described by master equations.
First, we show that, even without such resource supply, coupling to an environment can activate a time crystal in a stationary state common to the isolated and open systems (Go theorem).
We construct an explicit model in which a system is coupled to a zero-temperature Gibbs environment via a time-independent, energy-conserving interaction, with only short-range Hamiltonians.
The time crystal in this model is robust against changes in each model parameter~\cite{Iemini2018boundary} and against perturbations preserving the dynamical symmetry~\cite{Buca2019nonstationary,Alaeian2022exact}, both discussed for open-system time crystals.

We give several implementations of this model, one of which exactly reproduces the time-crystalline periodicity of the system directly from the microscopic model.
Each implementation reproduces a Lindblad master equation satisfying (i) time independence of the Hamiltonian and jump operators, (ii) time-translation covariance, and (iii) detailed balance, conditions that can be violated in standard derivations of the master equation~\cite{Davies1974markovian,Breuer2007theory,Ciccarello2022quantum} only with an external supply of thermodynamic resources~\cite{Goold2016role,Brandao2013resource,Lostaglio2015quantum}.
Environment-induced time crystals can thus arise without such resource supply.

Next, we prove a general no-go theorem: dissipation can never contribute to such activation; only an environment-induced Hamiltonian term can.
Specifically, under conditions (i)--(iii), any periodic two-time correlation function of the open system coincides with that of the isolated system with the open-system Hamiltonian, the sum of the bare isolated Hamiltonian and the environment-induced Hamiltonian, in the thermodynamic limit.
The theorem is model independent and holds for every stationary state.
Our proof uses techniques from the resource theory of asymmetry, a quantum-information framework for analyzing symmetry breaking~\cite{Bartlett2007reference,Gour2008resource,Marvian2014extending,Marvian2012symmetry,Marvian2013theory,Marvian2022operational,YT,Shitara_Tajima,Yamaguchi2026quantum,Yamaguchi2026quantifying,Kusuki2026resource}: using modes of asymmetry~\cite{Marvian2012symmetry,Marvian2014modes,Lostaglio2015quantum}, we decompose the two-time correlation function into frequency components.

Our results not only show that time crystals can be activated without external resource supply, but also identify the physical requirements on the system and environment for this activation through our universal no-go theorem.
This guides the design of time crystals that rely on neither external driving nor nonequilibrium environments.

\textit{\textbf{Formulation.---}}
\textit{Isolated and Open Dynamics.---}
For each system size $N$, we consider a quantum system on a finite-dimensional Hilbert space $\mathfrak H_N$.
We denote the Hamiltonian of the isolated system by $H_N^{(0)}$.

In the Heisenberg picture, an observable $A_N$ of the isolated system evolves as
\begin{align}
 \frac{\partial}{\partial t}A_N(t)
 &=\mathcal{H}_N^{(0)\dagger}[A_N(t)],\qquad
 \mathcal H_N^{(0)\dagger}[A]:=i[H_N^{(0)},A].
\end{align}

Coupling the system to an environment generally gives rise not only to dissipation but also to an environment-induced contribution to the Hamiltonian.
Denoting this contribution by $K_N$, we write the open-system Hamiltonian as $H_N:=H_N^{(0)}+K_N$.

For the open system, we consider the Lindblad master equation~\cite{Gorini1976completely,Lindblad1976generators,Breuer2007theory}, written in the Heisenberg picture as
\begin{align}
 \frac{\partial}{\partial t}A_N(t)
 &=\mathcal{L}_N^\dagger[A_N(t)],\qquad
 \mathcal L_N^\dagger:=\mathcal H_N^\dagger+\mathcal D_N^\dagger,\\
 \mathcal D_N^\dagger[A]
 &:=\sum_k\gamma_{k,N}\left(
 L_{k,N}^\dagger A L_{k,N}
 -\frac{1}{2}\{L_{k,N}^\dagger L_{k,N},A\}
 \right),
\end{align}
where $\mathcal H_N^\dagger[A]:=i[H_N,A]$,
and $L_{k,N}$ are jump operators labeled by the channel index $k$, with rates $\gamma_{k,N}\geq0$.
We denote the Schrödinger-picture Lindbladian by
$\mathcal L_N=\mathcal H_N+\mathcal D_N$.

The spectral decomposition of $H_N$ is $H_N=\sum_{E\in\calE_N}EP_{E,N}$, where $\calE_N$ is the spectrum of $H_N$ and $P_{E,N}$ is the corresponding spectral projector.
The set of energy differences is $\mathcal{W}_N:=\{E'-E\mid E,E'\in\calE_N\}$.
For inverse temperature $0\leq\beta\leq\infty$, we write the Gibbs state as $\pi_N:=e^{-\beta H_N}/\Tr[e^{-\beta H_N}]$.

To formalize the absence of external resource supply, we impose the following three conditions at inverse temperature $\beta$, which we call the no-resource-supply conditions (NRSC):
\begin{enumerate}
    \item[(i)] Time independence: $H_N$ and $\{\sqrt{\gamma_{k,N}}L_{k,N}\}_k$ are time independent.
    \item[(ii)] Time-translation covariance: for each $k$, there exists $\omega_{k,N}\in\mathcal{W}_N$ such that $[L_{k,N},H_N]=\omega_{k,N}L_{k,N}$.
    \item[(iii)] Detailed balance: for each $k$, there exists a reverse channel $-k$ such that $\gamma_{-k,N}=e^{-\beta\omega_{k,N}}\gamma_{k,N}$ and $L_{k,N}=L_{-k,N}^\dagger$, with $\omega_{k,N}$ defined in (ii).
\end{enumerate}
For $\beta=0,\infty$, the Gibbs state, the detailed-balance relation, and the Lindbladians derived below are understood as the $\beta\to 0,\infty$ limits of their finite-temperature forms.
Standard derivations of the master equation, such as the Born--Markov--secular approximation~\cite{Davies1974markovian,Kossakowski1977quantum,Breuer2007theory} and collision models~\cite{Ciccarello2022quantum}, yield Lindbladians satisfying NRSC when the environment is in equilibrium and the interaction is time independent (see Sec.~VII of the Supplemental Material~\cite{SM}).
Using collision models, we now relate each condition to the external supply of thermodynamic resources.
For the Born--Markov--secular approximation, Sec.~VII A gives the corresponding relation.

First, for a system with Hamiltonian $H$ at finite temperature $0<\beta<\infty$, we quantify the thermodynamic resource of a state $\rho$ by its nonequilibrium free-energy difference relative to the Gibbs state $\pi_\beta:=e^{-\beta H}/\Tr[e^{-\beta H}]$~\cite{Esposito2011second,Brandao2013resource},
\begin{align}
 F_\beta(\rho)-F_\beta(\pi_\beta)
 =\beta^{-1}D(\rho\|\pi_\beta)\geq0,
\end{align}
where $F_\beta(\rho):=\Tr[H\rho]-\beta^{-1}S(\rho)$, and $S(\rho)$ and $D(\rho\|\sigma)$ are the von Neumann entropy and the quantum relative entropy, respectively.
At $\beta=0,\infty$, we instead use $D(\rho\|\pi_\beta)$.
This deviation from the Gibbs state, called athermality, is a standard thermodynamic resource~\cite{Brandao2013resource}.

In a collision model, the system interacts briefly with a sequence of identically prepared ancillas, each uncorrelated with the system, and an appropriate continuous-time limit yields a Lindblad master equation~\cite{Scarani2002thermalizing,Ciccarello2022quantum}.
Let $H_{\mathrm E}$ and $\rho_{\mathrm E}$ be the Hamiltonian and state of each ancilla, $V$ the system--ancilla interaction, and $H_0:=H_N^{(0)}\otimes I_{\mathrm E}+I_S\otimes H_{\mathrm E}$.
In this derivation, NRSC follows from four conditions (Sec.~VII B of the Supplemental Material~\cite{SM}): (C1) $H_N^{(0)}$, $H_{\mathrm E}$, $V$, and $\rho_{\mathrm E}$ are time independent; (C2) $V$ is energy conserving, $[V,H_0]=0$; (C3) $\rho_{\mathrm E}=\pi_{\beta,\mathrm E}$, the Gibbs state of $H_{\mathrm E}$; and (C4) replacing each used ancilla with a fresh one costs zero average power for every input state of the system.

Violating any condition of NRSC requires violating at least one of (C1)--(C4), and each such violation requires an external resource supply.
The time dependence that violates (C1) requires an external clock~\cite{Goold2016role,Woods2019autonomous} asymmetric under time translations~\cite{Marvian2019nobroadcasting,Kwon2018clock}.
If (C2) is violated, the unitary $U:=e^{-i\delta t(H_0+V)}$ of a collision of duration $\delta t$ generally satisfies $[U,H_0]\neq0$, and implementing it for arbitrary states under total energy conservation requires an external system asymmetric under time translations~\cite{TSS,TSS2,TTK}.
Violating (C3) requires a nonequilibrium environment, $\rho_{\mathrm E}\neq\pi_{\beta,\mathrm E}$.
When (C4) is violated, switching the interaction from the used ancilla to a fresh one costs nonzero average power for at least one input state, which requires an external work source~\cite{Ciccarello2022quantum}.
The clock, the external system, and the nonequilibrium environment are all out of their Gibbs states, so at finite temperature preparing them requires the positive nonequilibrium free-energy difference defined above~\cite{Brandao2013resource,Horodecki2013fundamental}.
Therefore, Lindbladians obtained without external resource supply satisfy NRSC.
Our model for the Go theorem satisfies NRSC, and we further verify the absence of external resource supply directly in an explicit implementation including the environment.

\textit{Time Crystals in Isolated Systems.---}
Following Watanabe and Oshikawa~\cite{Watanabe2015absence},
we characterize time crystals through two-time correlation functions of macroscopic order parameters.
We take an order parameter to be
\begin{align}
 A_N:=\sum_x a_{x,N},
 \qquad
 \|a_{x,N}\|_\infty\leq C,
\end{align}
where each $a_{x,N}$ is a Hermitian operator supported on a finite region around site $x$, $\|\cdot\|_\infty$ is the operator norm, and both the region size and $C<\infty$ are independent of $N$ and $x$.
We denote by $\mathfrak A_{\mathrm{loc}}$ the set of such families $\{A_N\}_N$.

For each $N$, let $\rho_{\mathrm{ss},N}$ be a stationary state of the isolated system,
$\mathcal H_N^{(0)}[\rho_{\mathrm{ss},N}]=0$,
and define its two-time correlation function by
\begin{align}
 G_{N,\rho_{\mathrm{ss},N}}^{\mathcal H^{(0)\dagger}}(t;A_N)
 &:=
 \frac{1}{N^2}\Tr\!\left[
 \rho_{\mathrm{ss},N}e^{t\mathcal H_N^{(0)\dagger}}[A_N]A_N
 \right].
\end{align}
Time crystals in isolated systems are defined as follows.
\begin{definition}\label{def:closed_TC}
An isolated system exhibits a time crystal in a family of stationary states
$\{\rho_{\mathrm{ss},N}\}_N$
if there exist
$\{A_N\}_N\in\mathfrak A_{\mathrm{loc}}$,
a nonconstant function $g_{\mathrm{iso}}(t)$,
and a period $\tau>0$ such that
\begin{align}
 \lim_{N\to\infty}
 G_{N,\rho_{\mathrm{ss},N}}^{\mathcal H^{(0)\dagger}}(t;A_N)
 &=g_{\mathrm{iso}}(t)
 \qquad(\forall t\geq0),\\
 g_{\mathrm{iso}}(t+\tau)
 &=g_{\mathrm{iso}}(t)
 \qquad(\forall t\geq0).
\end{align}
\end{definition}
Note that a time crystal is defined relative to a family of stationary states.
Watanabe and Oshikawa~\cite{Watanabe2015absence,Watanabe2020proof} ruled out such time crystals in the ground and Gibbs states of isolated systems with short-range interactions, and related no-go results have been established for other classes of stationary states and Hamiltonians under certain conditions~\cite{Huang2019absence}.
By contrast, time crystals have been proposed in other stationary states and in systems with long-range many-body interactions~\cite{Syrwid2017time,Kozin2019quantum,Schecter2019weak,Iadecola2019quantum}.
All these works define time crystals via two-time correlation functions in stationary states, and we adopt this standard definition.

\textit{Time Crystals in Open Systems.---}
We use the same criterion for open systems.
For each $N$, let $\rho_{\mathrm{ss},N}$ be a stationary state of the open system, $\mathcal L_N[\rho_{\mathrm{ss},N}]=0$, and define its two-time correlation function by
\begin{align}
 G_{N,\rho_{\mathrm{ss},N}}^{\mathcal L^\dagger}(t;A_N)
 :=\frac{1}{N^2}\Tr\!\left[\rho_{\mathrm{ss},N}e^{t\mathcal L_N^\dagger}[A_N]A_N\right].
\end{align}
Stationary states need not be unique, but under NRSC the Gibbs state $\pi_N$ is known to be one of them~\cite{Alicki1976detailed,Kossakowski1977quantum}.
Time crystals in open systems are defined in the same way.
\begin{definition}\label{def:open_TC}
An open system exhibits a time crystal in a family of stationary states $\{\rho_{\mathrm{ss},N}\}_N$ if there exist $\{A_N\}_N\in\mathfrak A_{\mathrm{loc}}$, a nonconstant function $g_{\mathrm{op}}(t)$, and a period $\tau>0$ such that
\begin{align}
 \lim_{N\to\infty}
 G_{N,\rho_{\mathrm{ss},N}}^{\mathcal L^\dagger}(t;A_N)
 &=g_{\mathrm{op}}(t)\qquad(\forall t\geq0),\\
 g_{\mathrm{op}}(t+\tau)
 &=g_{\mathrm{op}}(t)\qquad(\forall t\geq0).
\end{align}
\end{definition}
Because the time evolution in this correlation function is generated by $\mathcal L_N^\dagger$, it captures the dissipative, nonequilibrium character of the dynamics.
This criterion is widely used to define open-system time crystals~\cite{Tucker2018shattered,Wang2025boundary}.

In open systems that violate NRSC, coupling to an environment can induce time crystals~\cite{Gong2018discrete,Iemini2018boundary,Buca2019dissipation,Kessler2021observation,Kongkhambut2022observation,Wu2024dissipative}.
However, because these systems violate NRSC, the effect of external thermodynamic resources cannot be distinguished from that of the coupling itself.
To distinguish them, we compare the isolated and open systems in the same family of stationary states and ask whether activation is possible even under NRSC.

\textit{\textbf{Go theorem.---}}
We now show that coupling to an environment can activate time crystals even under NRSC.
To this end, we construct an explicit model in which only the open system exhibits a time crystal in a stationary state common to both systems.
Further details on the properties of the model and its microscopic implementations are given in Secs.~I--II of the Supplemental Material~\cite{SM}.

\textit{Model.---}
We consider a periodic chain of $N\ge3$ three-level systems on the lattice $\Lambda_N=\{1,\ldots,N\}$.
Using the orthonormal basis $\{\ket{0}_i,\ket{\uparrow}_i,\ket{\downarrow}_i\}$ of site $i$, we define
$n_i=\ket{\uparrow}_i\bra{\uparrow}+\ket{\downarrow}_i\bra{\downarrow}$, $s_i^x=(\ket{\uparrow}_i\bra{\downarrow}+\ket{\downarrow}_i\bra{\uparrow})/2$, $s_i^z=(\ket{\uparrow}_i\bra{\uparrow}-\ket{\downarrow}_i\bra{\downarrow})/2$,
and $S_N^\alpha=\sum_{i=1}^{N}s_i^\alpha$ ($\alpha=x,z$).

The isolated-system and environment-induced Hamiltonians are
\begin{align}
H_N^{(0)}
&=u\sum_{i=1}^{N}(I_i-n_i)
+J\sum_{i=1}^{N}(I_iI_{i+1}-n_in_{i+1}),\notag\\
K_N&=-\Omega S_N^z,\qquad
H_N=H_N^{(0)}+K_N,
\end{align}
where $I_i$ is the identity on site $i$, $n_{N+1}=n_1$, and $u,J,\Omega>0$ are independent of $N$.
Since $H_N^{(0)}$ consists of on-site and nearest-neighbor terms and $K_N$ of on-site terms, $H_N$ is also short-range.
The Lindbladian is
\begin{align}
\mathcal L_N[\rho]
=-i[H_N,\rho]
+\gamma\sum_{i=1}^{N}(W_i\rho W_i-\rho),
\end{align}
with jump operators $W_i=2n_i-I_i$ and a rate $\gamma>0$ independent of $N$.
The Hamiltonian and jump operators are time independent, with $W_i^\dagger=W_i$ and $[W_i,H_N]=0$.
Each $W_i$ is thus a self-adjoint zero-frequency jump operator serving as its own reverse channel with the same rate, so time-translation covariance and detailed balance hold and $\mathcal L_N$ satisfies NRSC.

The Lindbladian also possesses a dynamical symmetry~\cite{Buca2019nonstationary,Alaeian2022exact}: the operators $S_N^+:=\sum_{i=1}^{N}\ket{\uparrow}_i\bra{\downarrow}$ and $S_N^-:=(S_N^+)^\dagger$ satisfy
\begin{align}
[H_N,S_N^\pm]=\mp\Omega S_N^\pm,\qquad
[W_i,S_N^\pm]=0.
\label{eq:go_dynamical_symmetry}
\end{align}
\begin{figure}[t]
  \centering
  \includegraphics[width=\columnwidth]{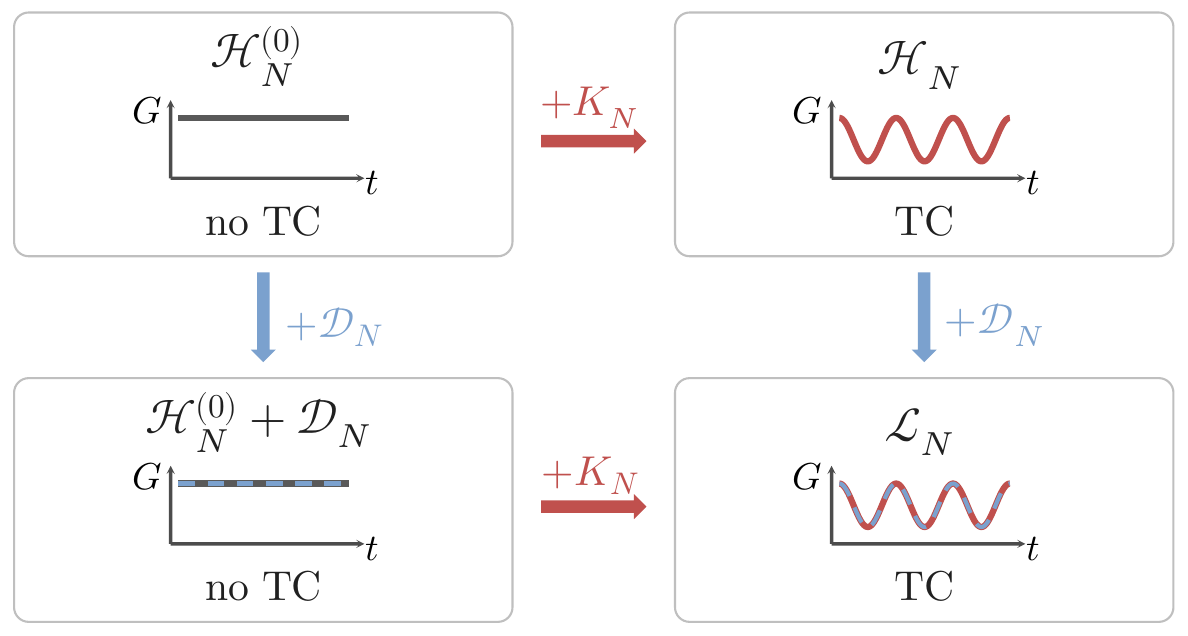}
\caption{Time-crystal activation in our model comes from $K_N$, not from $\mathcal D_N$.
Each panel sketches the thermodynamic limit of the two-time correlation function $G$ of $S_N^x$ in the common stationary state $\rho_{\mathrm{ss},N}$.
The top (bottom) row shows isolated (open) systems, the latter satisfying NRSC.
Red (blue) arrows add $K_N$ ($\mathcal D_N$); dashed blue curves show $G$ with $\mathcal D_N$.
Adding $K_N$ turns $G=1/12$ into $\cos(\Omega t)/12$ [Eqs.~\eqref{eq:go_open_isolated_correlations} and \eqref{eq:go_isolated_correlation}], whereas adding $\mathcal D_N$ leaves $G$ unchanged; by Theorem~\ref{thm:nogo_stationary}, no TC can appear under $\mathcal H_N^{(0)}+\mathcal D_N$.}
  \label{fig:go_model}
\end{figure}

\textit{Activation.---}
As a stationary state common to the isolated and open systems, we take
\begin{align}
\rho_{\mathrm{ss},N}=\frac{1}{N+1}\sum_{q=0}^{N}\ket{D_N^q}\bra{D_N^q},
\label{eq:go_state}
\end{align}
where $\ket{D_N^q}$ is the normalized Dicke state, the uniform superposition of all configurations with $q$ spins up and $N-q$ spins down.
Each Dicke state satisfies $H_N^{(0)}\ket{D_N^q}=0$ and $W_i\ket{D_N^q}=\ket{D_N^q}$ and is an eigenstate of $K_N$, so $\rho_{\mathrm{ss},N}$ is stationary under both dynamics.

In this state, the two-time correlation functions of the order parameter $A_N=S_N^x=(S_N^++S_N^-)/2$ are
\begin{align}
G_{N,\rho_{\mathrm{ss},N}}^{\mathcal L^\dagger}
(t;S_N^x)
&=\frac{N+2}{12N}\cos(\Omega t),
\label{eq:go_open_isolated_correlations}\\
G_{N,\rho_{\mathrm{ss},N}}^{\mathcal H^{(0)\dagger}}
(t;S_N^x)
&=\frac{N+2}{12N}.
\label{eq:go_isolated_correlation}
\end{align}
In the thermodynamic limit, the open-system one thus converges to the nonconstant periodic function $\cos(\Omega t)/12$.
This periodicity stems from $S_N^\pm(t)=e^{\mp i\Omega t}S_N^\pm$, implied by the dynamical symmetry~\eqref{eq:go_dynamical_symmetry}.
Symmetry-induced time-crystalline order of this kind is also discussed for dynamical-symmetry-based~\cite{Buca2019nonstationary,Alaeian2022exact} and symmetry-protected~\cite{Khemani2019brief} time crystals.
By contrast, since $[H_N^{(0)},S_N^x]=0$, the isolated-system correlation function is time independent.
Moreover, as $\rho_{\mathrm{ss},N}$ is a mixture of ground states of the short-range $H_N^{(0)}$, the correlation function of every $\{B_N\}_N\in\mathfrak A_{\mathrm{loc}}$ has a time-independent thermodynamic limit whenever the limit exists.
This confirms time-crystal activation within the same family $\{\rho_{\mathrm{ss},N}\}_N$.
Note that the oscillation in Eq.~\eqref{eq:go_open_isolated_correlations} does not decay even at finite $N$.
Such time crystals have also been discussed in settings that assume external thermodynamic resources~\cite{Else2016floquet,Surace2019floquet,Lyu2020eternal,Buca2019nonstationary,Chinzei2020time}.

In contrast, without $K_N$ the correlation function becomes time independent, even though the dissipative part is kept (Fig.~\ref{fig:go_model}).
Conversely, without the dissipative part, the evolution under $H_N$ alone yields the same periodic correlation function in the same stationary state.
Hence, this activation is caused by the environment-induced Hamiltonian term.

In addition, the time crystal is robust against the perturbations considered for open-system time crystals.
First, it is robust against changes in each model parameter, as discussed in Ref.~\cite{Iemini2018boundary}: it persists when $u,J,\gamma$ vary from site to site below finite bounds independent of $N$ and $i$, and $\Omega$ takes any nonzero value independent of $N$.
Second, it survives perturbations preserving the dynamical symmetry, as discussed in Refs.~\cite{Buca2019nonstationary,Alaeian2022exact}.
If Hermitian perturbations $\delta H_N$ of $H_N$ and perturbations $\delta W_{i,N}$ of $W_i$ satisfy $[\delta H_N,S_N^\pm]=0$ and $[\delta W_{i,N},S_N^\pm]=[\delta W_{i,N}^\dagger,S_N^\pm]=0$, then $\rho_{\mathrm{ss},N}$ stays stationary and the open-system correlation function in Eq.~\eqref{eq:go_open_isolated_correlations} is unchanged.

What remains is to implement this Lindblad dynamics by coupling the system to an equilibrium environment.
As shown in the End Matter and Sec.~II of the Supplemental Material~\cite{SM}, this is achieved with a zero-temperature Gibbs environment, a time-independent, energy-conserving interaction, and only short-range Hamiltonians.
In particular, the implementation in the End Matter exactly reproduces the open-system correlation function in Eq.~\eqref{eq:go_open_isolated_correlations} directly from the microscopic model.

\begin{figure}[t]
  \centering
  \includegraphics[width=0.91\columnwidth]{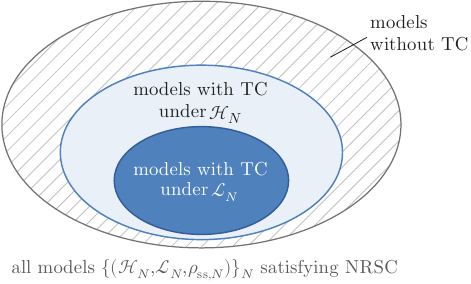}
\caption{Schematic of the no-go theorem.
Each point is a model $\{(\mathcal H_N,\mathcal L_N,\rho_{\mathrm{ss},N})\}_N$: an isolated system $\mathcal H_N$, an open system $\mathcal L_N=\mathcal H_N+\mathcal D_N$ satisfying NRSC, and a stationary state common to them.
Light blue and dark blue indicate models exhibiting a TC under $\mathcal H_N$ and $\mathcal L_N$, respectively, in the common stationary state in the thermodynamic limit.
The hatched region indicates models exhibiting no TC under either dynamics.
Theorem~\ref{thm:nogo_stationary} gives the inclusion relation $(\text{with TC under }\mathcal L_N)\subseteq(\text{with TC under }\mathcal H_N)$.
Consequently, $(\text{without TC})\cup(\text{with TC under }\mathcal H_N)=(\text{all models})$.}
  \label{fig:nogo_venn}
\end{figure}

\textit{\textbf{No-go theorem.---}}
We now show that, under NRSC, dissipation plays no role in time-crystal activation.
To this end, we compare the open-system dynamics with the unitary evolution under the same Hamiltonian $H_N$.

For Gibbs states, we first obtain the following theorem.
\begin{theorem}[No-go theorem for Gibbs states]
\label{thm:nogo_gibbs}
Let $\{\mathcal L_{N}\}_N$ be a family of open systems satisfying NRSC at inverse temperature $0\leq\beta\leq\infty$, with Hamiltonians $\{H_N\}_N$.
If the isolated system with the open-system Hamiltonians $\{H_N\}_N$ does not exhibit a time crystal in the Gibbs states $\{\pi_N\}_N$, neither does the open system in these states.
\end{theorem}

Beyond Gibbs states, the no-go theorem extends to any stationary state of the open system.
Under NRSC, $\mathcal L_{N}[\rho_{\mathrm{ss},N}]=0$ implies $\mathcal H_N[\rho_{\mathrm{ss},N}]=0$~\cite{SM}, so the two systems can be compared in the same stationary state.
\begin{theorem}[No-go theorem for arbitrary stationary states]
\label{thm:nogo_stationary}
Let $\{\mathcal L_{N}\}_N$ be a family of open systems satisfying NRSC at inverse temperature $0\leq\beta\leq\infty$, with Hamiltonians $\{H_N\}_N$.
Let $\{\rho_{\mathrm{ss},N}\}_N$ be any family of stationary states of these open systems.
If the isolated system with the open-system Hamiltonians $\{H_N\}_N$ does not exhibit a time crystal in $\{\rho_{\mathrm{ss},N}\}_N$, neither does the open system in these states.
\end{theorem}

These theorems have two implications (Fig.~\ref{fig:nogo_venn}).
First, under NRSC, activation occurs only through a change in the Hamiltonian; no activation requires the dissipator $\calD_N$.
Indeed, any time crystal under NRSC also appears in the isolated system with $\{H_N\}_N$ in the same stationary state, so $\calD_N$ plays no role in activation.
Second, as a special case of the first, activation does not occur under NRSC when coupling to the environment leaves the Hamiltonian unchanged.
Indeed, then $K_N=0$ and $H_N=H^{(0)}_N$, so $\{\calL_N\}_N$ exhibits no time crystal unless the isolated system with $H^{(0)}_N$ does in the same state.
For example, when the Watanabe--Oshikawa theorem rules out time crystals in the Gibbs states of the isolated system, Theorem~\ref{thm:nogo_gibbs} extends this absence to the same Gibbs states of the corresponding open system satisfying NRSC and $K_N=0$.
Theorem~\ref{thm:nogo_stationary} further applies to any stationary state of the open system, not only to the Gibbs and ground states covered by the Watanabe--Oshikawa theorem.

We outline the proof for nonzero-temperature Gibbs states in the End Matter.
The full proof, including the zero-temperature and arbitrary-stationary-state cases, is in Secs.~IV--VI of the Supplemental Material~\cite{SM}.

\textbf{\textit{Discussion and summary.---}}
In this Letter, we formalized the absence of external resource supply as NRSC and constructed a Lindblad model satisfying NRSC in which the isolated system exhibits no time crystal in a stationary state but the open system does.
We then implemented this dynamics with a zero-temperature Gibbs environment.
This implementation provides a concrete example in which time crystals are activated with energy conservation, only short-range interactions, and no external thermodynamic resources.

In contrast, we proved a general no-go theorem: under NRSC, such activation can arise only from an environment-induced Hamiltonian term, not from dissipation.
It holds for every stationary state, including Gibbs states, and every inverse temperature $0\leq\beta\leq\infty$, without assuming a local Hamiltonian or a unique stationary state.

Our no-go theorem extends the Watanabe--Oshikawa absence of time crystals in ground states~\cite{Watanabe2015absence} to open systems satisfying NRSC with the same Hamiltonian.
The time crystal in our model does not contradict this extended result:
$\rho_{\mathrm{ss},N}$ in Eq.~\eqref{eq:go_state} is a long-range correlated mixture of ground states of $H_N^{(0)}$, so the Watanabe--Oshikawa theorem rules out a time crystal in the isolated system.
However, $K_N$ lifts the degeneracy of the Dicke states, so $\rho_{\mathrm{ss},N}$ remains stationary but is no longer a ground state of $H_N=H_N^{(0)}+K_N$, and the extended result does not apply.

As our model also illustrates, the no-go theorem does not forbid spontaneous time-translation symmetry breaking itself.
Existing Floquet and boundary time crystals, for instance, use external thermodynamic resources and violate NRSC, and are therefore consistent with the theorem~\cite{Else2016floquet,Khemani2016phase,Iemini2018boundary,Carollo2024quantum}.
Our model is also consistent with the theorem: in the same stationary state, the original Hamiltonian $H_N^{(0)}$ yields no time crystal, whereas the unitary evolution under $H_N=H_N^{(0)}+K_N$ alone reproduces the periodic two-time correlation of the open system.

Both of our theorems leave room for further development.
The no-go theorem applies to Lindblad master equations satisfying NRSC; whether it extends to more general dynamics, such as non-Markovian ones~\cite{Breuer2016colloquium}, remains open.
The Go theorem, by contrast, is not restricted to Lindblad dynamics.
Our microscopic implementation conserves energy, involves only short-range interactions, and uses a zero-temperature Gibbs environment, so it shows directly that time crystals can be activated without external thermodynamic resources.
However, our model is not a direct experimental proposal; finding a laboratory realization is another direction for future work.


\begin{acknowledgments}
YO was supported by the WISE Program (MEXT) at Kyushu University.
HT was supported by JSPS Grants-in-Aid for Scientific Research No. JP25K00924, and MEXT
KAKENHI Grant-in-Aid for Transformative Research
Areas B “Quantum Energy Innovation” Grant Numbers
24H00830 and 24H00831, JST MOONSHOT No. JPMJMS256E, JST FOREST No. JPMJFR2365, and Royal Society International Collaboration Awards 2025 Flexigrant number ICA/R2/252240.
\end{acknowledgments}

\bibliography{GO_and_NOGO_TC/go_and_nogo_ref}


\clearpage
\section*{\NoCaseChange{End Matter}}
\suppressfloats[t]

\begin{figure}[t]
 \centering
 \includegraphics[width=\columnwidth]{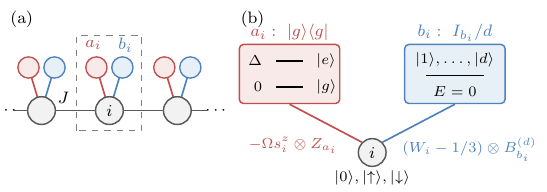}
\caption{Microscopic implementation of the Lindbladian by coupling to a zero-temperature Gibbs environment.
(a) Each system site $i$ (gray) on a one-dimensional lattice is coupled to a two-level system $a_i$ (red) and a $d$-level system $b_i$ (blue).
Lines between system sites represent nearest-neighbor interactions of strength $J$.
(b) System site $i$, its environments $a_i$ and $b_i$, and their interactions.
The two-level system $a_i$ has energies $0$ and $\Delta$ and is in the state $|g\rangle_{a_i}\langle g|$.
The $d$-level system $b_i$ has a zero Hamiltonian and is in the state $I_{b_i}/d$.
}
 \label{fig:go_microscopic}
\end{figure}

\textit{\textbf{Go theorem: Microscopic realization.---}}
We construct a microscopic model in which the Lindbladian $\mathcal L_N$ of the main text arises from coupling the system to a zero-temperature Gibbs environment.
Each system site $i$ is coupled to an environment consisting of a two-level system $a_i$ and a $d$-level system $b_i$ with $d\geq2$ (Fig.~\ref{fig:go_microscopic}).
Using the orthonormal bases $\{|g\rangle_{a_i},|e\rangle_{a_i}\}$ of $a_i$ and $\{|j\rangle_{b_i}\}_{j=1}^{d}$ of $b_i$, we define the environment Hamiltonian and state as
\begin{align}
H_{\mathrm E,N,d}
&=\Delta\sum_{i=1}^{N}|e\rangle_{a_i}\langle e|,\\
\eta_{\mathrm E,N,d}
&=\bigotimes_{i=1}^{N}
\left(|g\rangle_{a_i}\langle g|\otimes\frac{I_{b_i}}{d}\right),
\end{align}
where $\Delta>0$ is a constant independent of $N$ and $d$, and $b_i$ has a zero Hamiltonian.
Hence, $\eta_{\mathrm E,N,d}$ is the zero-temperature Gibbs state of $H_{\mathrm E,N,d}$.
The system--environment interaction is
\begin{align}
V_{N,d}
&=
-\Omega\sum_{i=1}^{N}s_i^z\otimes Z_{a_i}
+
\sum_{i=1}^{N}
\left(W_i-\frac13 I_i\right)\otimes B_{b_i}^{(d)},
\end{align}
where $Z_{a_i}=|g\rangle_{a_i}\langle g|-|e\rangle_{a_i}\langle e|$ and
\begin{align}
B_{b_i}^{(d)}
&=\sum_{j=1}^{d}b_j^{(d)}|j\rangle_{b_i}\langle j|,\\
b_j^{(d)}
&=\gamma\tan\!\left[
\frac{\pi(2j-d-1)}{2d}
\right].
\end{align}
Since each $a_i$ is prepared in $|g\rangle_{a_i}$, an eigenstate of $Z_{a_i}$ with eigenvalue $1$, the first term in $V_{N,d}$ reduces to the system Hamiltonian term $-\Omega\sum_i s_i^z=K_N$.
As shown below, the second term reproduces the dissipative part of $\mathcal L_N$ for $d\to\infty$.
On the lattice in which $a_i$ and $b_i$ are attached to system site $i$ (Fig.~\ref{fig:go_microscopic}(a)), each term of $H_{\mathrm E,N,d}$ and $V_{N,d}$ acts on a single site or two adjacent sites, so both are short-range.
Moreover, all Hamiltonians are time independent, and the interaction is energy conserving since $[V_{N,d},H_N^{(0)}\otimes I_{\mathrm E}+I_S\otimes H_{\mathrm E,N,d}]=0$.

The total Hamiltonian is $H_{\mathrm{tot},N,d}:=H_N^{(0)}\otimes I_{\mathrm E}+I_S\otimes H_{\mathrm E,N,d}+V_{N,d}$, and we write $U_{N,d}(t):=e^{-itH_{\mathrm{tot},N,d}}$.
With the total system initially in $\rho\otimes\eta_{\mathrm E,N,d}$, the reduced state of the system at time $t$ is
\begin{align}
\Phi_{N,d,t}[\rho]
:=\Tr_{\mathrm E}\!\left[
U_{N,d}(t)(\rho\otimes\eta_{\mathrm E,N,d})
U_{N,d}^{\dagger}(t)
\right].
\end{align}
For any initial state $\rho$, fixed system size $N$, and finite $T>0$, we have
\begin{align}
\sup_{0\leq t\leq T}
\left\|\Phi_{N,d,t}[\rho]-e^{t\mathcal L_N}[\rho]\right\|_1
\longrightarrow0
\qquad(d\to\infty).
\end{align}
Thus, the system--environment coupling above realizes the Lindblad dynamics, including both its Hamiltonian and dissipative parts.

Moreover, $\rho_{\mathrm{ss},N}$ is also stationary under the reduced map $\Phi_{N,d,t}$.
In this state, the two-time correlation function of $S_N^x$ under $\Phi_{N,d,t}$ equals the open-system one in Eq.~\eqref{eq:go_open_isolated_correlations} exactly, for all finite $N$, $d$ and all $t\geq0$.
Hence, the thermodynamic limit at fixed finite $d$ yields the same periodic two-time correlation.

\textit{\textbf{No-go: Methods of Proof.---}}
We now outline our proof methods.

The properties introduced below are proved in
Secs.~III--IV of the Supplemental Material~\cite{SM}.
In the following proof outline, we consider nonzero temperatures, $0\leq\beta<\infty$.

\begin{figure}[t]
  \centering
  \includegraphics[width=\columnwidth]{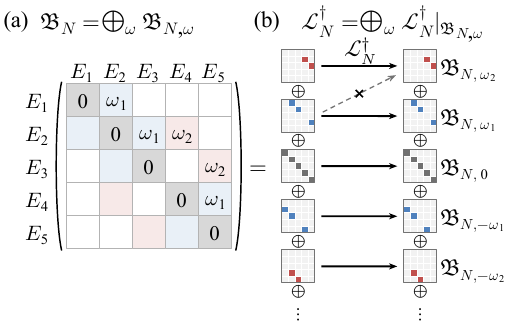}
  \caption{Decomposition of the operator space and the Lindbladian into modes of asymmetry.
(a) Operator space $\mathfrak B_N$ represented in the energy eigenbasis, decomposed as $\mathfrak B_N=\bigoplus_\omega\mathfrak B_{N,\omega}$, where $\mathfrak B_{N,\omega}$ consists of operators whose nonzero components have energy difference $E'-E=\omega$.
Gray indicates the mode $0$; blue and red indicate $\omega_1$ and $\omega_2$ above the diagonal and $-\omega_1$ and $-\omega_2$ below it.
Other frequencies are uncolored and omitted from the direct sum on the right.
(b) By time-translation covariance, a Lindbladian $\mathcal L_N^\dagger$ satisfying NRSC maps each $\mathfrak B_{N,\omega}$ to itself (solid arrows) and has no components between different modes (dashed arrow with a cross), so $\mathcal L_N^\dagger=\bigoplus_\omega\mathcal L_N^\dagger|_{\mathfrak B_{N,\omega}}$.}
  \label{fig:modes}
\end{figure}

\textit{Modes of Asymmetry.---}
Let $\mathfrak B_N$ be the space of operators on $\mathfrak H_N$.
For $X_N\in\mathfrak B_N$, we define its $\omega$-mode (Fig.~\ref{fig:modes}) as
\eq{
X_{N,\omega}:=\sum_EP_{E,N}X_NP_{E+\omega,N},
}
with $P_{E,N}:=0$ if $E$ is not an eigenvalue of $H_N$.
Correspondingly, the subspace of operators with energy difference $\omega\in\mathcal{W}_N$ is
\begin{align}
 \mathfrak B_{N,\omega}
 :=\{X_N\in\mathfrak B_N\mid
 \mathcal H_N^\dagger[X_N]=-i\omega X_N\}.
\end{align}
By definition, $e^{t\mathcal H_N^\dagger}[X_{N,\omega}]=e^{-i\omega t}X_{N,\omega}$, so $X_{N,\omega}$ is the component of $X_N$ that oscillates at frequency $\omega$ under the unitary dynamics generated by $H_N$.
In this sense, each $\omega$ labels a mode of asymmetry with respect to time-translation symmetry~\cite{Marvian2014modes,Marvian2012symmetry,Lostaglio2015quantum}.

From the definition, the mode decomposition has two basic properties.
\begin{enumerate}
 \item[(M1)] The whole operator space is the direct sum of the subspaces for frequencies $\omega\in \mathcal{W}_N$:
\begin{align}
  \mathfrak B_N=
  \bigoplus_{\omega\in\mathcal{W}_N}\mathfrak B_{N,\omega}.
 \end{align}

 \item[(M2)] The following two relations hold:
\begin{align}
  X_{N,\omega}&\in\mathfrak B_{N,\omega},\quad
  Y_{N,\omega'}\in\mathfrak B_{N,\omega'}\notag\\
  &\Longrightarrow\quad
  X_{N,\omega}Y_{N,\omega'}\in\mathfrak B_{N,\omega+\omega'},\\
  X_{N,\omega}&\in\mathfrak B_{N,\omega}
  \quad\Longrightarrow\quad
  X_{N,\omega}^\dagger\in\mathfrak B_{N,-\omega}.
 \end{align}
\end{enumerate}
By (M1), any $X_N\in\mathfrak B_N$ is uniquely decomposed as $X_N=\sum_{\omega\in\mathcal{W}_N}X_{N,\omega}$.
Moreover, (M2) gives the transformation rules of mode labels under operator products and adjoints.

\textit{Spectral Decomposition in Each Mode.---}
The mode decomposition simplifies the eigenvalue problem of the Lindbladian: since $\mathcal H_N^\dagger$ acts on each $\mathfrak B_{N,\omega}$ as multiplication by $-i\omega$, it suffices to diagonalize the dissipative part $\mathcal D_N^\dagger$ within each mode.

For $0\leq\beta<\infty$, the Gibbs state $\pi_N$ has full rank, so
\begin{align}
 \langle X_N,Y_N\rangle_{\pi_N}
 :=\Tr\!\left[\pi_N X_N^\dagger Y_N\right]
\end{align}
defines an inner product on $\mathfrak B_N$, under which different modes are orthogonal.
Moreover, for a Lindbladian satisfying NRSC, $\mathcal D_N^\dagger$ is self-adjoint on each mode with respect to this inner product~\cite{Alicki1976detailed,Carlen2017gradient}.
Therefore, we can choose an orthonormal basis $\{R_{N,\omega,a}\}_{\omega,a}$ of $\mathfrak B_N$ consisting of eigenoperators of $\mathcal D_N^\dagger$, with $R_{N,\omega,a}\in\mathfrak B_{N,\omega}$, such that
\begin{align}
 \mathcal L_N^\dagger[R_{N,\omega,a}]
 =(-i\omega-\lambda_{N,\omega,a})R_{N,\omega,a},
 \label{eq:lindblad_spectral}
\end{align}
where $-\lambda_{N,\omega,a}\leq0$ is the corresponding eigenvalue of $\mathcal D_N^\dagger$ and $a$ labels the basis elements within each mode.

\textit{Proof of the no-go theorem.---}
We now sketch the proof of Theorem~\ref{thm:nogo_gibbs} for $0\leq\beta<\infty$.
We compare a Lindbladian $\mathcal L_N$ satisfying NRSC with the unitary evolution under the same Hamiltonian $H_N$ in the Gibbs state $\pi_N$.
For $\{A_N\}_N\in\mathfrak A_{\mathrm{loc}}$, expanding $A_N$ in the eigenoperator basis of Eq.~(\ref{eq:lindblad_spectral}) yields
\begin{align}
 G_{N,\pi_N}^{\mathcal L^\dagger}(t;A_N)
 &=\sum_{\omega,a}v_{N,\omega,a}
 e^{(i\omega-\lambda_{N,\omega,a})t},
 \label{eq:gibbs_open_spectral}\\
 G_{N,\pi_N}^{\mathcal H^\dagger}(t;A_N)
 &=\sum_{\omega,a}v_{N,\omega,a}e^{i\omega t},
 \label{eq:gibbs_closed_spectral}
\end{align}
where $v_{N,\omega,a}:=N^{-2}|\langle R_{N,\omega,a},A_N\rangle_{\pi_N}|^2\geq0$.
From these expressions, it follows that if the thermodynamic limit of the open-system correlation function exists for all $t\geq0$ and is periodic, then for any $t\geq0$
\begin{align}
 \lim_{N\to\infty}
 \left|
 G_{N,\pi_N}^{\mathcal L^\dagger}(t;A_N)
 -G_{N,\pi_N}^{\mathcal H^\dagger}(t;A_N)
 \right|=0.
 \label{eq:gibbs_correlation_reduction}
\end{align}

Using this, we prove Theorem~\ref{thm:nogo_gibbs} by contradiction.
Suppose that the isolated system with Hamiltonian $H_N$ does not exhibit a time crystal in $\{\pi_N\}_N$ but the open system satisfying NRSC does.

By Definition~\ref{def:open_TC}, for some $\{A_N\}_N\in\mathfrak A_{\mathrm{loc}}$, the open-system correlation function converges to a nonconstant periodic function in the thermodynamic limit.
By Eq.~\eqref{eq:gibbs_correlation_reduction}, the isolated-system correlation function for the same $\{\pi_N\}_N$ and $\{A_N\}_N$ converges to the same function.
Hence, by Definition~\ref{def:closed_TC}, this isolated system also exhibits a time crystal in $\{\pi_N\}_N$, contradicting the assumption.
This proves Theorem~\ref{thm:nogo_gibbs} for $0\leq\beta<\infty$.

Proof details, including the absolute-zero and arbitrary-stationary-state cases,
are given in
Secs.~IV--VI
of the Supplemental Material~\cite{SM}.

\end{document}



\setcounter{theorem}{0}
\renewcommand{\thetheorem}{S.\arabic{theorem}}
\setcounter{figure}{0}
\renewcommand{\thefigure}{S.\arabic{figure}}
\renewcommand{\theequation}{S.\arabic{equation}}

\title{Supplemental Material for \\
``Open Quantum Time Crystals without External Thermodynamic Resources: \\
Go and No-Go Theorems''}
\author{Yuri Ogami}
\email{ogami.yuri.197@s.kyushu-u.ac.jp}
\affiliation{Joint Graduate School of Mathematics for Innovation, Kyushu University, 744 Motooka, Nishi-ku, Fukuoka, 819-0395, Japan}
\author{Hiroyasu Tajima}
\email{hiroyasu.tajima@inf.kyushu-u.ac.jp}
\affiliation{Department of Informatics, Faculty of Information Science and Electrical Engineering, Kyushu University, 744 Motooka, Nishi-ku, Fukuoka, 819-0395, Japan}
\affiliation{JST, FOREST, 4-1-8 Honcho, Kawaguchi, Saitama, 332-0012, Japan}

\maketitle

Part I of this Supplemental Material presents the models for the Go theorem and their microscopic derivations, and Part II proves the no-go theorem.
In Part I, Section~\ref{sec:go_model} presents a Lindblad model of time-crystal activation and examines its two-time correlation function and stability against perturbations.
Section~\ref{sec:go_derivations} derives the Lindblad dynamics of Sec.~\ref{sec:go_model} from continuous coupling to an equilibrium environment and from a collision model.
In these models, the system and environment Hamiltonians and the interaction between them all consist of short-range interactions, and the interaction is time independent and conserves energy.

Section~\ref{sec:resource_theory_of_asymmetry} reviews the resource theory of asymmetry and presents the basic properties of modes of asymmetry and their application to Lindbladians satisfying NRSC.
Section~\ref{sec:asymmetry_spectral_details} establishes basic properties of Lindbladians and proves the no-go theorem for Gibbs states at nonzero temperatures.
Section~\ref{sec:v4:stationary_structure} derives the properties of stationary states at nonzero temperatures and at absolute zero and introduces the operator spaces and induced superoperators used to prove the no-go theorem.
Section~\ref{sec:v4:spectral_nogo} uses these results to prove the no-go theorems for arbitrary stationary states and for Gibbs states at absolute zero.
Section~\ref{sec:nrsc_lindbladian} derives Lindbladians satisfying NRSC from the Born--Markov--secular approximation and from a collision model.
\tableofcontents
\vspace{\baselineskip}

\newpage
\part{Go theorem}

In this part, we show that coupling to an environment can activate a time crystal absent in the isolated system, even without an external supply of thermodynamic resources.
We prove this by explicit construction.
In Sec.~\ref{sec:go_model}, we first construct a Lindbladian that satisfies all three NRSC conditions---time independence, time-translation covariance, and detailed balance---and indeed activates a time crystal.
In this model, the time crystal is activated by the environment-induced Hamiltonian term $K_N$.
In Sec.~\ref{sec:go_derivations}, we then show that this Lindbladian arises from microscopic models with no external supply of thermodynamic resources.
Namely, the environment is in a zero-temperature Gibbs state, the system--environment interaction is time independent and conserves energy, and all Hamiltonians consist of short-range interactions.

\section{Time-crystal activation in a Lindblad model}\label{sec:go_model}
In this section, we verify that time-crystal activation indeed occurs under a Lindblad equation that simultaneously satisfies time independence of the Hamiltonian and jump operators, time-translation covariance, and detailed balance.
We first define the model, verify NRSC, and construct a state that is stationary for both the isolated and open systems.
We then compute the two-time correlation function of the order parameter $S_N^x$ in this stationary state and show that, in the open system, it keeps oscillating periodically even in the thermodynamic limit, whereas in the isolated system it is time independent.
We further show that the isolated system exhibits no time crystal for any choice of order parameter.
Finally, we show that this time crystal is robust against variations of the model parameters and against perturbations preserving the dynamical symmetry.

\subsection{Model and stationary states}\label{subsec:go_setup}
We first recall the model used in the main text.
We consider a chain of $N\geq3$ three-level systems, one on each site of the one-dimensional lattice $\Lambda_N=\{1,\ldots,N\}$.
Each site $i$ has the orthonormal basis $\{\ket{0}_i,\ket{\uparrow}_i,\ket{\downarrow}_i\}$, and the Hilbert space of the system is $\mathfrak H_N=(\mathbb C^3)^{\otimes N}$.
We define the on-site operators
\begin{align}
n_i&=\ket{\uparrow}_i\bra{\uparrow}
+\ket{\downarrow}_i\bra{\downarrow},\notag\\
s_i^z&=\frac12\left(
\ket{\uparrow}_i\bra{\uparrow}
-\ket{\downarrow}_i\bra{\downarrow}
\right),\notag\\
s_i^+&=\ket{\uparrow}_i\bra{\downarrow},
\qquad
s_i^-=\ket{\downarrow}_i\bra{\uparrow},\notag\\
s_i^x&=\frac12(s_i^++s_i^-),
\qquad
s_i^y=\frac1{2i}(s_i^+-s_i^-),
\end{align}
and the collective operators
\begin{align}
S_N^\alpha=\sum_{i=1}^Ns_i^\alpha
\quad(\alpha=x,y,z),\qquad
S_N^\pm=\sum_{i=1}^Ns_i^\pm.
\end{align}

The isolated-system Hamiltonian is
\begin{align}
H_N^{(0)}
=u\sum_{i=1}^N(I_i-n_i)
+J\sum_{i=1}^N(I_iI_{i+1}-n_in_{i+1}),
\end{align}
where $n_{N+1}=n_1$ (periodic boundary conditions) and $u,J>0$ are independent of $N$.
Here, $I_i$ denotes the identity on site $i$.
By definition, $H_N^{(0)}$ consists of on-site terms and nearest-neighbor two-site interactions.
The environment-induced Hamiltonian term and the open-system Hamiltonian are
\begin{align}
K_N&=-\Omega S_N^z,\qquad \Omega>0,\notag\\
H_N&=H_N^{(0)}+K_N,
\end{align}
with $\Omega$ also independent of $N$.
Since $K_N$ is a sum of on-site terms, $H_N$ also consists of short-range interactions.
As shown in Sec.~\ref{sec:go_derivations}, this Lindblad equation, including $K_N$, can indeed be realized by short-range interactions without external resources.

The jump operators and the Lindbladian are
\begin{align}
W_i&=2n_i-I_i,\notag\\
\mathcal L_N[\rho]
&=-i[H_N,\rho]
+\gamma\sum_{i=1}^N(W_i\rho W_i-\rho),
\label{eq:go_lindbladian}
\end{align}
where $\gamma>0$ is independent of $N$. We denote the dissipative part by $\mathcal D_N[\rho]$.
Since $n_i^2=n_i$, we have $W_i^\dagger=W_i$ and $W_i^2=I_i$.
Moreover, $[W_i,H_N]=0$ because $n_i$ commutes with all $n_j$ and $s_j^\alpha$.

We now check NRSC for Eq.~\eqref{eq:go_lindbladian}.
For each site $i$, we take two channels labeled by $\sigma=\pm$, with jump operators $L_{\sigma i,N}=W_i$ and rates $\gamma_{\sigma i,N}=\gamma/2$.
The dissipative part with these channels is
\begin{align}
\sum_{i=1}^N\sum_{\sigma=\pm}\gamma_{\sigma i,N}
\left(L_{\sigma i,N}\rho L_{\sigma i,N}^\dagger-\frac12\{L_{\sigma i,N}^\dagger L_{\sigma i,N},\rho\}\right)
&=\sum_{i=1}^N\sum_{\sigma=\pm}\frac{\gamma}{2}
\left(W_i\rho W_i^\dagger-\frac12\{W_i^\dagger W_i,\rho\}\right)\notag\\
&=\gamma\sum_{i=1}^N\left(W_i\rho W_i-\frac12\{W_i^2,\rho\}\right)\notag\\
&=\gamma\sum_{i=1}^N(W_i\rho W_i-\rho),
\end{align}
which is the dissipative part of Eq.~\eqref{eq:go_lindbladian}.
The first equality follows from $L_{\sigma i,N}=W_i$ and $\gamma_{\sigma i,N}=\gamma/2$, the second from adding the two identical terms with $W_i^\dagger=W_i$, and the last from $W_i^2=I_i$.
Since $[W_i,H_N]=0$, both channels have zero frequency, $\omega_{\sigma i,N}=0$, giving condition (ii), time-translation covariance.
Since $W_i$ is Hermitian and the two channels have equal rates at zero frequency, condition (iii), detailed balance, also holds.
The Hamiltonian and jump operators are time independent, which gives condition (i) as well.

The Lindbladian also possesses a dynamical symmetry~\cite{Buca2019nonstationary,Alaeian2022exact}: the operators $S_N^\pm$ satisfy
\begin{align}
[H_N,S_N^\pm]=\mp\Omega S_N^\pm,\qquad [W_i,S_N^\pm]=0.
\label{eq:go_dynamical_symmetry}
\end{align}
Indeed, the on-site relations $[n_i,s_i^\pm]=0$ and $[s_i^z,s_i^\pm]=\pm s_i^\pm$ imply $[H_N^{(0)},S_N^\pm]=0$, $[W_i,S_N^\pm]=0$, and $[S_N^z,S_N^\pm]=\pm S_N^\pm$, and $K_N=-\Omega S_N^z$ then gives Eq.~\eqref{eq:go_dynamical_symmetry}.

To compare the two-time correlation functions of the isolated and open systems in the same state, we construct a state that is stationary under both dynamics.
For $X\subset\Lambda_N$, let $\sigma_i(X)=\uparrow$ if $i\in X$ and $\sigma_i(X)=\downarrow$ otherwise, and define, for $q=0,\ldots,N$, the Dicke states
\begin{align}
\ket{D_N^q}
=\binom Nq^{-1/2}
\sum_{\substack{X\subset\Lambda_N\\|X|=q}}
\bigotimes_{i=1}^N\ket{\sigma_i(X)}_i.
\end{align}
Terms with different $X$ differ at some site and are thus orthogonal.
Since each $\ket{D_N^q}$ has $\binom Nq$ terms and different $q$ have different numbers of up spins, we have $\braket{D_N^q}{D_N^{q'}}=\delta_{q,q'}$.
With these states, we define
\begin{align}
\rho_{\mathrm{ss},N}&=\frac{P_N}{N+1},\label{eq:go_stationary_state}\\
P_N&=\sum_{q=0}^N
\ket{D_N^q}\bra{D_N^q}.\notag
\end{align}

Since no Dicke state contains $\ket{0}_i$, we have $n_i\ket{D_N^q}=\ket{D_N^q}$ for all $i$ and $q$, so that
\begin{align}
H_N^{(0)}\ket{D_N^q}&=0,\notag\\
W_i\ket{D_N^q}&=\ket{D_N^q}.
\label{eq:go_dicke_H0_W}
\end{align}
Since each configuration has $q$ up spins and $N-q$ down spins, we also have
\begin{align}
S_N^z\ket{D_N^q}
=\left(\frac q2-\frac{N-q}{2}\right)\ket{D_N^q}
=\left(q-\frac N2\right)\ket{D_N^q}.
\label{eq:go_dicke_Sz}
\end{align}
These relations give
\begin{align}
[H_N^{(0)},\rho_{\mathrm{ss},N}]
=[H_N,\rho_{\mathrm{ss},N}]=0,
\qquad
W_i\rho_{\mathrm{ss},N}W_i=\rho_{\mathrm{ss},N},
\end{align}
so $\mathcal L_N[\rho_{\mathrm{ss},N}]=0$, and $\rho_{\mathrm{ss},N}$ is stationary for both the isolated and open systems.

This state differs from the zero-temperature Gibbs state of $H_N^{(0)}$.
Indeed, $H_N^{(0)}\geq0$ because $n_i$ and $n_in_{i+1}$ are projectors, so Eq.~\eqref{eq:go_dicke_H0_W} shows that every Dicke state is a ground state of $H_N^{(0)}$.
Since $u>0$, the ground space is spanned by all configurations of $\ket{\uparrow}_i$ and $\ket{\downarrow}_i$, and the zero-temperature Gibbs state is its uniform mixture,
\begin{align}
\pi_{\infty,N}^{(0)}
=\frac{1}{2^N}\prod_{i=1}^N n_i,
\end{align}
whereas $\rho_{\mathrm{ss},N}$ in Eq.~\eqref{eq:go_stationary_state} is the uniform mixture of the Dicke states.

\subsection{Two-time correlation functions}\label{subsec:go_correlations}
We compute the two-time correlation functions of the order parameter $S_N^x$ in the isolated and open systems.
The Heisenberg-picture Lindbladian is
\begin{align}
\mathcal L_N^\dagger[A]
=i[H_N,A]
+\gamma\sum_i(W_iAW_i-A).
\end{align}
By Eq.~\eqref{eq:go_dynamical_symmetry},
\begin{align}
\mathcal D_N^\dagger[S_N^\pm]&=0,
\qquad i[H_N,S_N^\pm]=\mp i\Omega S_N^\pm,\notag\\
\mathcal L_N^\dagger[S_N^\pm]
&=\mp i\Omega S_N^\pm.
\label{eq:go_spin_eigenoperators}
\end{align}
This gives
$
e^{t\mathcal L_N^\dagger}[S_N^\pm]
=e^{\mp i\Omega t}S_N^\pm
$.
With $S_N^x=(S_N^++S_N^-)/2$, we obtain
\begin{align}
e^{t\mathcal L_N^\dagger}[S_N^x]
=\cos(\Omega t)S_N^x+\sin(\Omega t)S_N^y.
\label{eq:go_spin_x_evolution}
\end{align}

Since $s_i^x$ is Hermitian, supported on one site, and satisfies $\|s_i^x\|_\infty=1/2$, the family $\{S_N^x\}_N$ meets the definition of order parameters in the main text.
Using this evolution, its two-time correlation function is
\begin{align}
G_{N,\rho_{\mathrm{ss},N}}^{\mathcal L^\dagger}
(t;S_N^x)
&=\frac1{N^2}
\Tr\!\left[
\rho_{\mathrm{ss},N}
e^{t\mathcal L_N^\dagger}[S_N^x]S_N^x
\right]\notag\\
&=\frac{\cos(\Omega t)}{N^2}
\Tr[\rho_{\mathrm{ss},N}(S_N^x)^2]+
\frac{\sin(\Omega t)}{N^2}
\Tr[\rho_{\mathrm{ss},N}S_N^yS_N^x].
\end{align}

We evaluate the expectation values using the Dicke-state ladder relations
\begin{align}
S_N^+\ket{D_N^q}
&=\sqrt{(N-q)(q+1)}\ket{D_N^{q+1}},
\label{eq:go_dicke_ladder_plus}\\
S_N^-\ket{D_N^q}
&=\sqrt{q(N-q+1)}\ket{D_N^{q-1}},
\label{eq:go_dicke_ladder}
\end{align}
with $S_N^+\ket{D_N^N}=S_N^-\ket{D_N^0}=0$.
Equation~\eqref{eq:go_dicke_ladder_plus} follows from the definition of the Dicke states.
Indeed, in the term of $\ket{D_N^q}$ labeled by $X$, the operator $s_j^+$ gives a nonzero contribution only for $j\notin X$, and for each resulting set $Y=X\cup\{j\}$, the site $j$ can be any of the $q+1$ sites in $Y$.
Thus,
\begin{align}
S_N^+\ket{D_N^q}
&=(q+1)\binom Nq^{-1/2}
\sum_{\substack{Y\subset\Lambda_N\\|Y|=q+1}}
\bigotimes_i\ket{\sigma_i(Y)}_i\notag\\
&=(q+1)
\sqrt{\frac{\binom N{q+1}}{\binom Nq}}
\ket{D_N^{q+1}}\notag\\
&=\sqrt{(N-q)(q+1)}
\ket{D_N^{q+1}}.
\end{align}
Similarly, for $S_N^-$, each resulting set has $q-1$ sites and arises from $N-q+1$ pairs of an original configuration and an acting site, giving Eq.~\eqref{eq:go_dicke_ladder}.

Applying $S_N^\pm$ twice gives either zero or a multiple of $\ket{D_N^{q\pm2}}$.
Orthogonality between Dicke states with different $q$ then gives
\begin{align}
\bra{D_N^q}(S_N^\pm)^2\ket{D_N^q}&=0,
\end{align}
and hence
\begin{align}
\bra{D_N^q}(S_N^x)^2\ket{D_N^q}
&=\frac14
\bra{D_N^q}
S_N^+S_N^-+S_N^-S_N^+
\ket{D_N^q}\notag\\
&=\frac14\{q(N-q+1)+(N-q)(q+1)\},\notag\\
\bra{D_N^q}S_N^yS_N^x\ket{D_N^q}
&=\frac1{4i}
\bra{D_N^q}
S_N^+S_N^--S_N^-S_N^+
\ket{D_N^q}\notag\\
&=\frac{2q-N}{4i}.
\end{align}
Using Eq.~\eqref{eq:go_spin_x_evolution} and the definition of $\rho_{\mathrm{ss},N}$, we obtain
\begin{align}
G_{N,\rho_{\mathrm{ss},N}}^{\mathcal L^\dagger}(t;S_N^x)
&=\frac1{N^2}\Tr\!\left[
\rho_{\mathrm{ss},N}
e^{t\mathcal L_N^\dagger}[S_N^x]S_N^x
\right]\notag\\
&=\frac1{N^2(N+1)}\sum_{q=0}^N
\bra{D_N^q}\bigl[
\cos(\Omega t)(S_N^x)^2
+\sin(\Omega t)S_N^yS_N^x
\bigr]\ket{D_N^q}\notag\\
&=\frac1{4N^2(N+1)}\sum_{q=0}^N
\left[
\{2q(N-q)+N\}\cos(\Omega t)
+\frac{2q-N}{i}\sin(\Omega t)
\right]\notag\\
&=\frac{N+2}{12N}\cos(\Omega t),
\label{eq:go_open_correlation}
\end{align}
where the last step uses $\sum_{q=0}^Nq(N-q)=N(N+1)(N-1)/6$ and $\sum_{q=0}^N(2q-N)=0$.
Hence,
\begin{align}
\lim_{N\to\infty}
G_{N,\rho_{\mathrm{ss},N}}^{\mathcal L^\dagger}
(t;S_N^x)
=\frac1{12}\cos(\Omega t).
\end{align}
The limit is nonconstant and periodic with period $2\pi/\Omega$, so the open system exhibits a time crystal in this family of states.

For the isolated system, we define $\mathcal H_N^{(0)\dagger}[A]:=i[H_N^{(0)},A]$. Since $[H_N^{(0)},S_N^x]=0$, we have $e^{t\mathcal H_N^{(0)\dagger}}[S_N^x]=S_N^x$, and the same order parameter has the time-independent correlation
\begin{align}
G_{N,\rho_{\mathrm{ss},N}}^{\mathcal H^{(0)\dagger}}
(t;S_N^x)
=\frac{N+2}{12N}.
\end{align}

\subsection{Absence of time crystals in the isolated system}\label{subsec:go_isolated}
As shown in Sec.~\ref{subsec:go_correlations}, the two-time correlation function of $S_N^x$ is time independent in the isolated system.
In this subsection, we prove that this holds for any order parameter, namely, that the isolated system exhibits no time crystal in the family of stationary states $\{\rho_{\mathrm{ss},N}\}_N$.

Let $\{B_N\}_N\in\mathfrak A_{\mathrm{loc}}$ be a general family of order parameters.
Since $H_N^{(0)}\geq0$ and $H_N^{(0)}\ket{D_N^q}=0$ by Eq.~\eqref{eq:go_dicke_H0_W}, each Dicke state is a pure ground state of the short-range Hamiltonian $H_N^{(0)}$. The Watanabe--Oshikawa result~\cite{Watanabe2015absence} therefore applies.
We define the isolated-system correlation in each Dicke state by
\begin{align}
G_{N,q}(t):=
\frac1{N^2}
\bra{D_N^q}
e^{t\mathcal H_N^{(0)\dagger}}[B_N]B_N
\ket{D_N^q}.
\end{align}
Then the definition of $\rho_{\mathrm{ss},N}$ gives
\begin{align}
&G_{N,\rho_{\mathrm{ss},N}}^{\mathcal H^{(0)\dagger}}(t;B_N)
-G_{N,\rho_{\mathrm{ss},N}}^{\mathcal H^{(0)\dagger}}(0;B_N)\notag\\
&\quad=\frac1{N+1}\sum_{q=0}^{N}
[G_{N,q}(t)-G_{N,q}(0)]
\xrightarrow[N\to\infty]{}0.
\label{eq:go_isolated_correlation_bound}
\end{align}
For each fixed $t$, the Watanabe--Oshikawa bound is uniform over ground states, giving the convergence above.
Whenever the thermodynamic limit exists, the correlation is time independent. Thus, the isolated system has no time crystal in this family of states, whereas the open system does, as shown by Eq.~\eqref{eq:go_open_correlation}.

\subsection{Robustness of the time crystal}\label{subsec:go_robustness}
In this subsection, we show that the time crystal activated by coupling to the environment is robust against perturbations previously studied for time crystals in open systems.
Specifically, we consider variations of the model parameters and perturbations preserving the dynamical symmetry.

\subsubsection{Parameter variations}
First, as a perturbation, we vary the model coefficients from site to site and show that the time crystal is preserved.
We replace the parameters $u,J,\gamma$ of the Lindblad equation by site-dependent values $u_i,J_i,\gamma_i$.
We assume $0<u_i\leq u_+$, $0<J_i\leq J_+$, and $0<\gamma_i\leq\gamma_+$.
The upper bounds $u_+,J_+,\gamma_+<\infty$ are independent of $N$ and $i$.
We also replace $\Omega$ by a nonzero value $\widetilde\Omega=\Omega+\delta\Omega$ independent of $N$.
We write the perturbed Hamiltonians and Lindbladian as
\begin{align}
\widetilde H_N^{(0)}
&=\sum_i u_i(I_i-n_i)
+\sum_i J_i(I_iI_{i+1}-n_in_{i+1}),\notag\\
\widetilde H_N
&=\widetilde H_N^{(0)}
-\widetilde\Omega S_N^z,\notag\\
\mathcal L_N^\prime[\rho]
&=-i[\widetilde H_N,\rho]
+\sum_i\gamma_i(W_i\rho W_i-\rho).
\end{align}
As in Eqs.~\eqref{eq:go_dicke_H0_W} and \eqref{eq:go_dicke_Sz}, $\widetilde H_N^{(0)}P_N=0$, $W_iP_N=P_N$, and $[S_N^z,P_N]=0$, so
\begin{align}
\mathcal L_N^\prime[\rho_{\mathrm{ss},N}]=0,
\qquad
[\widetilde H_N^{(0)},\rho_{\mathrm{ss},N}]=0.
\end{align}
Since $[\widetilde H_N^{(0)},S_N^\pm]=[W_i,S_N^\pm]=0$, we obtain
\begin{align}
\mathcal L_N^{\prime\dagger}[S_N^\pm]
=\mp i\widetilde\Omega S_N^\pm.
\end{align}
Thus, the perturbation replaces $\Omega$ by $\widetilde\Omega$ in the action of the Lindbladian on $S_N^\pm$, so the two-time correlation function is given by Eq.~\eqref{eq:go_open_correlation} with $\Omega$ replaced by $\widetilde\Omega$.
Its amplitude is unchanged, while its period becomes $2\pi/|\widetilde\Omega|$.
The periodic oscillations persist in the thermodynamic limit, so the time crystal is robust against these parameter variations.

\subsubsection{Perturbations preserving dynamical symmetry}

Next, as a perturbation, we vary the Hamiltonian and jump operators while preserving the dynamical symmetry~\cite{Buca2019nonstationary,Alaeian2022exact} and show that the time crystal is also preserved.
As shown in Eq.~\eqref{eq:go_dynamical_symmetry}, the unperturbed Lindbladian has the dynamical symmetry.
Specifically, we consider the perturbation
\begin{align}
H_N^{(0)}&\to \widetilde H_N^{(0)}:=H_N^{(0)}+\delta H_N,
\qquad
W_i\to\widetilde W_{i,N}:=W_i+\delta W_{i,N},\notag\\
\Omega&\to\widetilde\Omega,
\qquad
\gamma\to\widetilde\gamma_{i,N}.
\end{align}
Here, $\delta H_N$ is Hermitian, and $\delta H_N$ and $\delta W_{i,N}$ satisfy
\begin{align}
[\delta H_N,S_N^\pm]=[\delta W_{i,N},S_N^\pm]=[\delta W_{i,N}^\dagger,S_N^\pm]=0.
\end{align}
Also, $\widetilde\Omega\neq0$ is independent of $N$, and $\widetilde\gamma_{i,N}\geq0$ is the rate of $\widetilde W_{i,N}$.
Namely, we add to $H_N^{(0)}$ and $W_i$ only terms commuting with $S_N^\pm$, and may change $\Omega$ and the rates.
The perturbed Hamiltonian and Lindbladian are
\begin{align}
\widetilde H_N
&=\widetilde H_N^{(0)}-\widetilde\Omega S_N^z,\notag\\
\mathcal L_N^\prime[\rho]
&=-i[\widetilde H_N,\rho]+\sum_i\widetilde\gamma_{i,N}
\left(
\widetilde W_{i,N}\rho\widetilde W_{i,N}^\dagger
-\frac12\{
\widetilde W_{i,N}^\dagger\widetilde W_{i,N},\rho
\}
\right).
\end{align}
These conditions imply $[\widetilde H_N,S_N^\pm]=\mp\widetilde\Omega S_N^\pm$ and $[\widetilde W_{i,N},S_N^\pm]=[\widetilde W_{i,N}^\dagger,S_N^\pm]=0$.
Namely, the perturbed Lindbladian also has the dynamical symmetry, with frequency $\widetilde\Omega$.
This perturbation leaves $\rho_{\mathrm{ss},N}$ stationary, as shown at the end of this subsection.
Since $S_N^\pm$ commute with $\widetilde W_{i,N}$ and $\widetilde W_{i,N}^\dagger$, we have $\widetilde W_{i,N}^\dagger S_N^\pm\widetilde W_{i,N}=\frac12\{\widetilde W_{i,N}^\dagger\widetilde W_{i,N},S_N^\pm\}$, so the Heisenberg-picture dissipative part vanishes on $S_N^\pm$.
Together with $[\widetilde H_N,S_N^\pm]=\mp\widetilde\Omega S_N^\pm$, this gives
\begin{align}
\mathcal L_N^{\prime\dagger}[S_N^\pm]
=\mp i\widetilde\Omega S_N^\pm.
\label{eq:go_perturbed_eigenoperators}
\end{align}
Thus, Eq.~\eqref{eq:go_perturbed_eigenoperators} coincides with Eq.~\eqref{eq:go_spin_eigenoperators} up to $\Omega\to\widetilde\Omega$, and the stationary state $\rho_{\mathrm{ss},N}$ is unchanged, so the two-time correlation function is given by Eq.~\eqref{eq:go_open_correlation} with $\Omega$ replaced by $\widetilde\Omega$: $G_{N,\rho_{\mathrm{ss},N}}^{\mathcal L^{\prime\dagger}}(t;S_N^x)=\frac{N+2}{12N}\cos(\widetilde\Omega t)$.
Its amplitude is unchanged, while its period becomes $2\pi/|\widetilde\Omega|$.
The periodic oscillations persist in the thermodynamic limit, so the time crystal is also robust against perturbations preserving the dynamical symmetry.

Finally, we show that $\rho_{\mathrm{ss},N}$ remains stationary under the perturbation.
Let $X_N$ denote $\widetilde H_N^{(0)}$ or $\widetilde W_{i,N}$.
The conditions on the perturbation and $[H_N^{(0)},S_N^\pm]=[W_i,S_N^\pm]=0$ give $[X_N,S_N^\pm]=0$.
Furthermore, $S_N^z=\frac12[S_N^+,S_N^-]$ and the Jacobi identity imply $[X_N,S_N^z]=0$.\footnote{By the Jacobi identity, $[X_N,S_N^z]=\frac12[X_N,[S_N^+,S_N^-]]=\frac12\bigl([[X_N,S_N^+],S_N^-]+[S_N^+,[X_N,S_N^-]]\bigr)=0$.}

The relation $[X_N,S_N^z]=0$ implies that $X_N$ preserves the eigenspace of $S_N^z$ with eigenvalue $N/2$.
Since $S_N^z$ reaches its maximum $N/2$ only when all sites are in $\ket{\uparrow}$, this eigenspace is one dimensional, spanned by $\ket{D_N^N}=\ket{\uparrow}^{\otimes N}$, so we can write
\begin{align}
X_N\ket{D_N^N}=a_{X,N}\ket{D_N^N},\notag
\end{align}
for some coefficient $a_{X,N}$.
Then $[X_N,S_N^-]=0$ gives
\begin{align}
X_N(S_N^-)^{N-q}\ket{D_N^N}
&=(S_N^-)^{N-q}X_N\ket{D_N^N}\notag\\
&=a_{X,N}(S_N^-)^{N-q}\ket{D_N^N}.
\end{align}
By Eq.~\eqref{eq:go_dicke_ladder}, $(S_N^-)^{N-q}\ket{D_N^N}$ is a nonzero multiple of $\ket{D_N^q}$, so
\begin{align}
X_N\ket{D_N^q}
=a_{X,N}\ket{D_N^q}
\qquad(q=0,\ldots,N).
\end{align}
The operator $X_N^\dagger$ satisfies the same commutation relations, $[X_N^\dagger,S_N^\pm]=-[X_N,S_N^\mp]^\dagger=0$, so the same argument gives $X_N^\dagger P_N=a_{X,N}^*P_N$, where $a_{X,N}^*=\bra{D_N^N}X_N^\dagger\ket{D_N^N}$.
Taking the adjoint gives $P_NX_N=a_{X,N}P_N$, so that
\begin{align}
X_NP_N=P_NX_N=a_{X,N}P_N.
\label{eq:go_scalar_action}
\end{align}
Setting $X_N=\widetilde W_{i,N}$ for each jump operator and using this relation and its adjoint, we find
\begin{align}
&X_N\rho_{\mathrm{ss},N}X_N^\dagger
-\frac12\{X_N^\dagger X_N,\rho_{\mathrm{ss},N}\}\notag\\
&\qquad=
|a_{X,N}|^2\rho_{\mathrm{ss},N}
-|a_{X,N}|^2\rho_{\mathrm{ss},N}=0.
\end{align}
The Hamiltonian term also vanishes, because Eq.~\eqref{eq:go_scalar_action} with $X_N=\widetilde H_N^{(0)}$ and $[S_N^z,P_N]=0$ give $[\widetilde H_N,\rho_{\mathrm{ss},N}]=0$. Therefore, $\mathcal L_N^\prime[\rho_{\mathrm{ss},N}]=0$.

\section{Derivation of the Lindbladian from coupling to an equilibrium environment}\label{sec:go_derivations}
In this section, we show that the Lindbladian activating the time crystal in Sec.~\ref{sec:go_model} can be realized without an external supply of thermodynamic resources during the open-system dynamics.
Namely, we derive it from an environment in a zero-temperature Gibbs state and a time-independent, energy-conserving, short-range interaction Hamiltonian.
We give three derivations.
First, we construct a model in which the system remains coupled to the same environment under these conditions (Sec.~\ref{subsec:go_microscopic}), and show that it reproduces the two-time correlation function of the time crystal exactly and that the reduced dynamics converges to the Lindblad evolution as the number of environment levels increases.
We then derive the same Lindbladian more simply from a collision model (Sec.~\ref{subsec:go_collision}).
Finally, we construct an implementation in which the initial coupling also requires no average work (Sec.~\ref{subsec:go_microscopic2}).

\subsection{Implementation with continuous coupling to the environment}\label{subsec:go_microscopic}
We construct a microscopic model that exactly reproduces the two-time correlation function in Eq.~\eqref{eq:go_open_correlation} through an energy-conserving coupling to a zero-temperature Gibbs environment.
We also show that the reduced dynamics converges to the Lindblad evolution as the number of environment levels increases.

\subsubsection{Microscopic model and initial environment state}
Each system site $i$ has an associated two-level system $a_i$ and a $d$-level system $b_i$, where $d\geq2$ is an integer. Their orthonormal bases are $\{\ket{g}_{a_i},\ket{e}_{a_i}\}$ and $\{\ket{j}_{b_i}\}_{j=1}^d$.
The environment Hilbert space and Hamiltonian are
\begin{align}
\mathfrak H_{\mathrm E,N,d}
&=\bigotimes_{i=1}^N
\left(\mathbb C^2_{a_i}\otimes\mathbb C^d_{b_i}\right),\notag\\
H_{\mathrm E,N,d}
&=\Delta\sum_{i=1}^N
\ket{e}_{a_i}\bra{e}.
\end{align}
Here, $H_{\mathrm E,N,d}$ contains only on-site terms, $\Delta>0$ is independent of $N,d$, and $b_i$ has zero Hamiltonian.

We initialize the environment in the zero-temperature Gibbs state of $H_{\mathrm E,N,d}$,
\begin{align}
\eta_{\mathrm E,N,d}
=\bigotimes_{i=1}^N
\left(
\ket{g}_{a_i}\bra{g}
\otimes\frac{I_{b_i}}d
\right).
\end{align}

We take the interaction between the system and environment to be
\begin{align}
V_{N,d}
=-\Omega\sum_{i=1}^Ns_i^z\otimes Z_{a_i}
+\sum_{i=1}^N(W_i-I_i/3)\otimes B_{b_i}^{(d)},
\end{align}
where the environment operators and their eigenvalues are
\begin{align}
Z_{a_i}
&=\ket{g}_{a_i}\bra{g}
-\ket{e}_{a_i}\bra{e},\notag\\
B_{b_i}^{(d)}
&=\sum_{j=1}^db_j^{(d)}
\ket{j}_{b_i}\bra{j},\notag\\
b_j^{(d)}
&=\gamma\tan\!\left[
\frac{\pi(2j-d-1)}{2d}
\right].
\end{align}
We place the environment sites $a_i$ and $b_i$ next to site $i$ of the one-dimensional system chain. Then each term in $H_N^{(0)}$, $H_{\mathrm E,N,d}$, and $V_{N,d}$ acts on a single site or two neighboring sites. Thus, all three consist of short-range interactions.

The system operators $s_i^z$ and $W_i-I_i/3$ commute with $H_N^{(0)}$, and the environment operators $Z_{a_i}$ and $B_{b_i}^{(d)}$ commute with $H_{\mathrm E,N,d}$, giving
\begin{align}
[V_{N,d},
H_N^{(0)}\otimes I_{\mathrm E}
+I_S\otimes H_{\mathrm E,N,d}]=0.
\end{align}
Thus, the interaction conserves the sum of the system and environment Hamiltonians.

By the definition of the eigenvalues, $b_{d+1-j}^{(d)}=-b_j^{(d)}$ and $\Tr B_{b_i}^{(d)}=0$. Since $\Tr[Z_{a_i}\ket{g}_{a_i}\bra{g}]=1$ and $\Tr[B_{b_i}^{(d)}I_{b_i}/d]=0$ for the environment state, we obtain
\begin{align}
\Tr_{\mathrm E}\!\left[
V_{N,d}(I_S\otimes\eta_{\mathrm E,N,d})
\right]
=-\Omega\sum_i s_i^z
=K_N.
\end{align}

Finally, we define the total Hamiltonian and evolution operator by
\begin{align}
H_{\mathrm{tot},N,d}
&=H_N^{(0)}\otimes I_{\mathrm E}
+I_S\otimes H_{\mathrm E,N,d}+V_{N,d},\notag\\
U_{N,d}(t)&=e^{-itH_{\mathrm{tot},N,d}}.
\end{align}
\subsubsection{System dynamics and two-time correlation functions}

We derive the reduced system map from the evolution under the total Hamiltonian.
For the initial environment state $\eta_{\mathrm E,N,d}$, the reduced map and its adjoint acting on system observables are defined by
\begin{align}
\Phi_{N,d,t}[\rho]
&:=\Tr_{\mathrm E}\!\left[U_{N,d}(t)(\rho\otimes\eta_{\mathrm E,N,d})U_{N,d}^\dagger(t)\right],\notag\\
\Phi_{N,d,t}^\dagger[A]
&:=\Tr_{\mathrm E}\!\left[
(I_S\otimes\eta_{\mathrm E,N,d})U_{N,d}^\dagger(t)(A\otimes I_{\mathrm E})U_{N,d}(t)
\right].
\label{eq:go_exact_reduced_maps}
\end{align}
To evaluate these maps, we write $\ket{G}_a:=\bigotimes_{i=1}^N\ket{g}_{a_i}$ and $\ket{\boldsymbol j}_b:=\bigotimes_{i=1}^N\ket{j_i}_{b_i}$, with $\boldsymbol j=(j_1,\ldots,j_N)$, and expand the environment state as
\begin{align}
\eta_{\mathrm E,N,d}
=\frac1{d^N}\sum_{\boldsymbol j}
\ket{G}_a\bra{G}\otimes\ket{\boldsymbol j}_b\bra{\boldsymbol j}.
\label{eq:go_environment_state_components}
\end{align}
Each state in this expansion satisfies
\begin{align}
H_{\mathrm E,N,d}(\ket{G}_a\otimes\ket{\boldsymbol j}_b)&=0,\notag\\
Z_{a_i}\ket{G}_a&=\ket{G}_a,\notag\\
B_{b_i}^{(d)}\ket{\boldsymbol j}_b&=b_{j_i}^{(d)}\ket{\boldsymbol j}_b.
\label{eq:go_environment_eigenvalues}
\end{align}
Substituting Eq.~\eqref{eq:go_environment_state_components} into Eq.~\eqref{eq:go_exact_reduced_maps}, we evaluate the partial trace using Eq.~\eqref{eq:go_environment_eigenvalues} and $K_N=-\Omega\sum_i s_i^z$ to obtain
\begin{align}
\Phi_{N,d,t}[\rho]
&=\frac1{d^N}\sum_{\boldsymbol j}
e^{-itH_{\boldsymbol j}}\rho e^{itH_{\boldsymbol j}},
\qquad
H_{\boldsymbol j}:=H_N+\sum_i b_{j_i}^{(d)}(W_i-I_i/3).
\label{eq:go_exact_reduced_map_sum}
\end{align}
In particular, Eqs.~\eqref{eq:go_dicke_H0_W} and \eqref{eq:go_dicke_Sz} give $[H_N^{(0)},\rho_{\mathrm{ss},N}]=[K_N,\rho_{\mathrm{ss},N}]=[W_i,\rho_{\mathrm{ss},N}]=0$, and hence $[H_{\boldsymbol j},\rho_{\mathrm{ss},N}]=0$ for every $\boldsymbol j$.
Therefore, Eq.~\eqref{eq:go_exact_reduced_map_sum} and this relation give
\begin{align}
\Phi_{N,d,t}[\rho_{\mathrm{ss},N}]
&=\frac1{d^N}\sum_{\boldsymbol j}
e^{-itH_{\boldsymbol j}}\rho_{\mathrm{ss},N}e^{itH_{\boldsymbol j}}\notag\\
&=\frac1{d^N}\sum_{\boldsymbol j}\rho_{\mathrm{ss},N}
=\rho_{\mathrm{ss},N}.
\label{eq:go_reduced_stationarity}
\end{align}
Thus, coupling the isolated system to the environment preserves $\rho_{\mathrm{ss},N}$ as a stationary state.

We first calculate the evolution of $S_N^x$ to evaluate the two-time correlation in this stationary state.
Both $S_N^x$ and $S_N^y$ commute with $H_N^{(0)}$ and all $W_i-I_i/3$. The adjoint of Eq.~\eqref{eq:go_exact_reduced_map_sum}, together with Eq.~\eqref{eq:go_spin_x_evolution}, therefore gives
\begin{align}
\Phi_{N,d,t}^\dagger[S_N^x]
&=\frac1{d^N}\sum_{\boldsymbol j}e^{itH_{\boldsymbol j}}S_N^xe^{-itH_{\boldsymbol j}}
=e^{itK_N}S_N^xe^{-itK_N}\notag\\
&=\cos(\Omega t)S_N^x+\sin(\Omega t)S_N^y
=e^{t\mathcal L_N^\dagger}[S_N^x].
\label{eq:go_exact_spin_x_evolution}
\end{align}
We define the system's two-time correlation function in $\rho_{\mathrm{ss},N}$ as
\begin{align}
C_{N,d}(t)
&:=\frac1{N^2}\Tr_S\!\left[
\rho_{\mathrm{ss},N}\Phi_{N,d,t}^\dagger[S_N^x]S_N^x
\right].
\label{eq:go_exact_system_correlation}
\end{align}
Substituting Eq.~\eqref{eq:go_exact_spin_x_evolution} and using the same expectation values as in Eq.~\eqref{eq:go_open_correlation}, we obtain
\begin{align}
C_{N,d}(t)
&=\frac{\cos(\Omega t)}{N^2}
\Tr[
\rho_{\mathrm{ss},N}(S_N^x)^2]+
\frac{\sin(\Omega t)}{N^2}
\Tr[
\rho_{\mathrm{ss},N}S_N^yS_N^x]\notag\\
&=\frac{N+2}{12N}\cos(\Omega t)
=G_{N,\rho_{\mathrm{ss},N}}^{\mathcal L^\dagger}(t;S_N^x).
\label{eq:go_total_correlation}
\end{align}
The result coincides exactly with the open-system correlation function in Eq.~\eqref{eq:go_open_correlation} for any finite $N,d$ and all times.
Moreover,
\begin{align}
\lim_{N\to\infty}C_{N,d}(t)
=\frac1{12}\cos(\Omega t),
\end{align}
so the periodicity of the two-time correlation function follows from the total Hamiltonian dynamics without approximation.

\subsubsection{Convergence to the Lindblad evolution}\label{subsec:go_limit}

We prove that, for fixed $N$, the reduced map converges to the Lindblad evolution as $d\to\infty$, uniformly on every finite interval $0\leq t\leq T$. The relations $[H_N,W_i]=0$ and $[W_i,W_k]=0$ allow us to factorize the unitary evolution as
\begin{align}
e^{-itH_{\boldsymbol j}}
&=e^{-itH_N}\prod_{i=1}^{N}
e^{-itb_{j_i}^{(d)}(W_i-I_i/3)}.
\end{align}
We introduce the maps
\begin{align}
\mathcal U_{N,t}[X]
&:=e^{-itH_N}Xe^{itH_N},\notag\\
\mathcal Q_{i,d,t}[X]
&:=\frac1d\sum_{j=1}^d
 e^{-itb_j^{(d)}(W_i-I_i/3)}X e^{itb_j^{(d)}(W_i-I_i/3)}.
\end{align}
Writing $\mathcal C_i[X]:=W_iXW_i$ and $\id[X]:=X$, we have $\mathcal D_N=\gamma\sum_i(\mathcal C_i-\id)$. The corresponding single-site Lindblad map is $\mathcal Q_{i,t}^{\mathrm L}:=e^{\gamma t(\mathcal C_i-\id)}$. Averaging over each $j_i$ independently in Eq.~\eqref{eq:go_exact_reduced_map_sum}, and using $[H_N,W_i]=0$ for the Lindblad evolution, we can write the reduced map and the Lindblad evolution, respectively, as
\begin{align}
\Phi_{N,d,t}
&=\mathcal U_{N,t}\circ\prod_{i=1}^N\mathcal Q_{i,d,t},
\label{eq:go_reduced_map_factorization}\\
e^{t\mathcal L_N}
&=\mathcal U_{N,t}\circ\prod_{i=1}^N\mathcal Q_{i,t}^{\mathrm L},
\label{eq:go_target_map_factorization}
\end{align}
where
\begin{align}
\prod_{i=1}^{N}\mathcal Q_{i,d,t}
&:=\mathcal Q_{1,d,t}\circ\cdots\circ\mathcal Q_{N,d,t},
\end{align}
and the same composition convention applies to $\mathcal Q_{i,t}^{\mathrm L}$.

To estimate the difference between the two evolutions, we compare $\mathcal Q_{i,d,t}$ and $\mathcal Q_{i,t}^{\mathrm L}$.
We first compute $\mathcal Q_{i,d,t}$ explicitly.
Using $W_i^2=I_i$, we have, for any real $b$,
\begin{align}
e^{-itb(W_i-I_i/3)}Xe^{itb(W_i-I_i/3)}
&=e^{itb/3}e^{-itbW_i}Xe^{itbW_i}e^{-itb/3}\notag\\
&=e^{-itbW_i}Xe^{itbW_i}\notag\\
&=\bigl(\cos(tb)I_i-i\sin(tb)W_i\bigr)X
\bigl(\cos(tb)I_i+i\sin(tb)W_i\bigr)\notag\\
&=\cos^2(tb)X+\sin^2(tb)W_iXW_i
-i\sin(tb)\cos(tb)[W_i,X].
\end{align}
Since $b_{d+1-j}^{(d)}=-b_j^{(d)}$, we have $\sum_{j=1}^d\sin(tb_j^{(d)})\cos(tb_j^{(d)})=0$.
With
\begin{align}
\phi_d(s):=\frac1d\sum_{j=1}^d\cos(sb_j^{(d)}),
\end{align}
the finite-environment map becomes
\begin{align}
\mathcal Q_{i,d,t}
&=\left[\frac1d\sum_{j=1}^d\cos^2(tb_j^{(d)})\right]\id
+\left[\frac1d\sum_{j=1}^d\sin^2(tb_j^{(d)})\right]\mathcal C_i\notag\\
&=\frac1{2d}\sum_{j=1}^d[1+\cos(2tb_j^{(d)})]\id
+\frac1{2d}\sum_{j=1}^d[1-\cos(2tb_j^{(d)})]\mathcal C_i\notag\\
&=\frac{1+\phi_d(2t)}2\id
+\frac{1-\phi_d(2t)}2\mathcal C_i.
\label{eq:go_finite_environment_channel}
\end{align}

The identity $\mathcal C_i^2=\id$ similarly gives the Lindblad map as
\begin{align}
\mathcal Q_{i,t}^{\mathrm L}
&=e^{-\gamma t}e^{\gamma t\mathcal C_i}\notag\\
&=e^{-\gamma t}\left[
\sum_{n=0}^\infty\frac{(\gamma t)^{2n}}{(2n)!}\id
+\sum_{n=0}^\infty\frac{(\gamma t)^{2n+1}}{(2n+1)!}\mathcal C_i
\right]\notag\\
&=e^{-\gamma t}
\left[
\cosh(\gamma t)\id
+\sinh(\gamma t)\mathcal C_i
\right]\notag\\
&=\frac{1+e^{-2\gamma t}}2\id
+\frac{1-e^{-2\gamma t}}2\mathcal C_i.
\label{eq:go_lindblad_channel}
\end{align}
Thus, the maps have the same form and differ only in the coefficients of $\id$ and $\mathcal C_i$. Their comparison reduces to showing $\phi_d(2t)\to e^{-2\gamma t}$.

First, we write $e^{-2\gamma t}$ as an integral.
The Cauchy distribution\footnote{Averaging unitary evolution over the Cauchy distribution is known to yield a dephasing Lindblad equation~\cite{Kropf2016effective}.} with location parameter $0$ and scale parameter $\gamma$ has the probability density
\begin{align}
p_\gamma(y):=\frac{\gamma}{\pi(y^2+\gamma^2)},
\end{align}
which satisfies, for $t\geq0$,
\begin{align}
e^{-2\gamma t}
=\int_{-\infty}^{\infty}p_\gamma(y)\cos(2ty)\,dy.
\end{align}
To relate this integral to $\phi_d(2t)$, we change variables to $y=\gamma\tan[\pi(x-\tfrac12)]$.
Since $p_\gamma(y)\,dy=dx$,
\begin{align}
\int_0^1
\cos\!\left[2\gamma t\tan\pi\left(x-\tfrac12\right)\right]dx\label{eq:go_cauchy_integral_evaluation}
=e^{-2\gamma t}.
\end{align}
Next, we express $\phi_d(2t)$ using the integrand $f_t(x):=\cos[2\gamma t\tan\pi(x-\tfrac12)]$ of Eq.~\eqref{eq:go_cauchy_integral_evaluation}.
We divide $[0,1]$ into intervals $I_j:=[(j-1)/d,j/d]$ of width $1/d$, with midpoints $x_j:=(j-\tfrac12)/d$.
Since $b_j^{(d)}=\gamma\tan[\pi(x_j-\tfrac12)]$, the definition of $\phi_d$ gives
\begin{align}
\phi_d(2t)
&=\frac1d\sum_{j=1}^{d}\cos(2tb_j^{(d)})
=\frac1d\sum_{j=1}^{d}f_t(x_j)
=\sum_{j=1}^{d}\int_{I_j}f_t(x_j)\,dx.
\label{eq:go_midpoint_sum_integral}
\end{align}

We now estimate the difference between this sum and $\int_0^1f_t(x)\,dx$.
Fixing a finite $T>0$, we have, for $0\leq t\leq T$,
\begin{align}
\begin{aligned}
|\phi_d(2t)-e^{-2\gamma t}|
&=\left|\sum_{j=1}^d\int_{I_j}
\bigl[f_t(x_j)-f_t(x)\bigr]\,dx\right|\\
&\leq\sum_{j=1}^d\int_{I_j}
|f_t(x_j)-f_t(x)|\,dx.
\end{aligned}
\end{align}
We bound each $|f_t(x_j)-f_t(x)|$ by the mean value theorem.
Since $|f_t'|$ is not uniformly bounded near the endpoints, we fix $0<\delta<1/2$ and treat separately the intervals $I_j$ not wholly contained in $[\delta,1-\delta]$.
Using $|f_t(x_j)-f_t(x)|\leq2$, we bound the contribution from these intervals as
\begin{align}
\begin{aligned}
\sum_{I_j\not\subset[\delta,1-\delta]}
\int_{I_j}|f_t(x_j)-f_t(x)|\,dx
&\leq
\sum_{I_j\not\subset[\delta,1-\delta]}
\int_{I_j}2\,dx\\
&=
2\sum_{I_j\not\subset[\delta,1-\delta]}\frac1d\\
&\leq
2\left(2\delta+\frac2d\right)
=4\delta+\frac4d.
\end{aligned}
\end{align}
The intervals near the left and right endpoints are contained in $[0,\delta+1/d]$ and $[1-\delta-1/d,1]$, respectively.
Their total length is at most $2(\delta+1/d)$, which gives the last inequality.

For the remaining intervals, we have
\begin{align}
|f_t'(x)|
\leq\frac{2\pi\gamma T}{\sin^2(\pi x)}
\leq\frac{2\pi\gamma T}{\sin^2(\pi\delta)},
\end{align}
uniformly over $0\leq t\leq T$.
Applying the mean value theorem, we bound the contribution from each interval as
\begin{align}
\begin{aligned}
\int_{I_j}|f_t(x_j)-f_t(x)|\,dx
&\leq
\frac{2\pi\gamma T}{\sin^2(\pi\delta)}
\int_{I_j}|x-x_j|\,dx\\
&=
\frac{2\pi\gamma T}{\sin^2(\pi\delta)}
\,2\int_0^{1/(2d)}u\,du\\
&=\frac{\pi\gamma T}{2d^2\sin^2(\pi\delta)}.
\end{aligned}
\end{align}
The first equality follows because $x_j$ is the midpoint of an interval of width $1/d$.
Since there are at most $d$ such intervals, their contributions sum to at most $\pi\gamma T/[2d\sin^2(\pi\delta)]$.

Combining these estimates and taking the supremum over $0\leq t\leq T$, we obtain
\begin{align}
\epsilon_d(T)
\label{eq:go_dephasing_uniform_limit}:=\sup_{0\leq t\leq T}|\phi_d(2t)-e^{-2\gamma t}|
\leq4\delta+\frac4d+
\frac{\pi\gamma T}{2d\sin^2(\pi\delta)}.
\end{align}
For example, choosing $\delta=d^{-1/3}$ for sufficiently large $d$ and using $\sin(\pi\delta)\geq2\delta$, we obtain
\begin{align}
\epsilon_d(T)
&\leq
\left(4+\frac{\pi\gamma T}{8}\right)d^{-1/3}
+\frac4d
\longrightarrow0
\qquad(d\to\infty).
\end{align}

For $0\leq t\leq T$, we use the triangle inequality and the unitarity of $W_i$ to obtain
\begin{align}
\|(\mathcal Q_{i,d,t}-\mathcal Q_{i,t}^{\mathrm L})[X]\|_1
&=\frac{|\phi_d(2t)-e^{-2\gamma t}|}{2}\|X-W_iXW_i\|_1\notag\\
&\leq\frac{\epsilon_d(T)}2\bigl(\|X\|_1+\|W_iXW_i\|_1\bigr)\notag\\
&=\epsilon_d(T)\|X\|_1.
\label{eq:go_single_site_channel_bound}
\end{align}
Moreover, neither map increases the trace norm:\footnote{For $0\leq p\leq1$, the triangle inequality and the unitarity of $W_i$ give
\begin{align*}
\|pX+(1-p)W_iXW_i\|_1
&\leq p\|X\|_1+(1-p)\|W_iXW_i\|_1\\
&=p\|X\|_1+(1-p)\|X\|_1\\
&=\|X\|_1.
\end{align*}
Substituting $p=(1+\phi_d(2t))/2$ and $p=(1+e^{-2\gamma t})/2$ gives Eq.~\eqref{eq:go_channel_contractions} for the two maps, respectively.} for any operator $X$,
\begin{align}
\|\mathcal Q_{i,d,t}[X]\|_1
&\leq\|X\|_1,\qquad
\|\mathcal Q_{i,t}^{\mathrm L}[X]\|_1\leq\|X\|_1.
\label{eq:go_channel_contractions}
\end{align}

Replacing one site map at a time, we expand the difference of the product maps as
\begin{align}
\left(\prod_{i=1}^N\mathcal Q_{i,d,t}
-\prod_{i=1}^N\mathcal Q_{i,t}^{\mathrm L}\right)[X]
&=\sum_{j=1}^N\left(\prod_{i<j}\mathcal Q_{i,d,t}\right)\circ
(\mathcal Q_{j,d,t}-\mathcal Q_{j,t}^{\mathrm L})\circ
\left(\prod_{i>j}\mathcal Q_{i,t}^{\mathrm L}\right)[X].
\label{eq:go_product_difference_expansion}
\end{align}
Equation~\eqref{eq:go_product_difference_expansion} and the triangle inequality give
\begin{align}
\left\|\left(\prod_{i=1}^N\mathcal Q_{i,d,t}
-\prod_{i=1}^N\mathcal Q_{i,t}^{\mathrm L}\right)[X]\right\|_1
&\leq\sum_{j=1}^N\left\|
(\mathcal Q_{j,d,t}-\mathcal Q_{j,t}^{\mathrm L})
\left[\left(\prod_{i>j}\mathcal Q_{i,t}^{\mathrm L}\right)[X]\right]
\right\|_1\notag\displaybreak[1]\\
&\leq\epsilon_d(T)\sum_{j=1}^N
\left\|\left(\prod_{i>j}\mathcal Q_{i,t}^{\mathrm L}\right)[X]\right\|_1\notag\\
&\leq N\epsilon_d(T)\|X\|_1.
\label{eq:go_product_channel_bound}
\end{align}
Equation~\eqref{eq:go_channel_contractions} gives the first and third inequalities, and Eq.~\eqref{eq:go_single_site_channel_bound} gives the second.\footnote{For the first inequality, we set
$Y_j:=(\mathcal Q_{j,d,t}-\mathcal Q_{j,t}^{\mathrm L})[(\prod_{i>j}\mathcal Q_{i,t}^{\mathrm L})[X]]$
and use
\begin{align*}
\left\|\sum_{j=1}^N\left(\prod_{i<j}\mathcal Q_{i,d,t}\right)[Y_j]\right\|_1
&\leq\sum_{j=1}^N\left\|\left(\prod_{i<j}\mathcal Q_{i,d,t}\right)[Y_j]\right\|_1
\leq\sum_{j=1}^N\|Y_j\|_1,
\end{align*}
which follows from the triangle inequality and Eq.~\eqref{eq:go_channel_contractions}.}

We can now compare the full reduced map with the Lindblad evolution.
Equations~\eqref{eq:go_reduced_map_factorization} and \eqref{eq:go_target_map_factorization} imply, for every initial state $\rho$,
\begin{align}
\sup_{0\leq t\leq T}\|\Phi_{N,d,t}[\rho]-e^{t\mathcal L_N}[\rho]\|_1
&=\sup_{0\leq t\leq T}\left\|
\left(\prod_i\mathcal Q_{i,d,t}-\prod_i\mathcal Q_{i,t}^{\mathrm L}\right)[\rho]
\right\|_1\notag\\
&\leq N\epsilon_d(T)\|\rho\|_1
=N\epsilon_d(T)\longrightarrow0.
\label{eq:go_reduced_map_limit}
\end{align}
The equality follows from $\|UXU^\dagger\|_1=\|X\|_1$ for any unitary $U$, and the inequality from Eq.~\eqref{eq:go_product_channel_bound}; the last step uses $\|\rho\|_1=1$.
For fixed $N$, $\epsilon_d(T)\to0$ as $d\to\infty$. Thus, the reduced map converges to the Lindblad evolution uniformly in trace norm on every finite time interval.

\subsection{Derivation of the Lindbladian using a collision model}\label{subsec:go_collision}
We now derive the Lindbladian in Eq.~\eqref{eq:go_lindbladian} more simply from a collision model~\cite{Scarani2002thermalizing,Ciccarello2022quantum}.
Each interaction lasts $\delta t>0$, and we use a fresh environment in the same state for each round.
The environment associated with each system site $i$ consists of two two-level systems $a_i,b_i$, with orthonormal bases $\{\ket{g}_{a_i},\ket{e}_{a_i}\}$ and $\{\ket{0}_{b_i},\ket{1}_{b_i}\}$. Its Hamiltonian is
\begin{align}
H_{\mathrm E,N}^{\mathrm{col}}
=\Delta\sum_{i=1}^N\ket{e}_{a_i}\bra{e},
\label{eq:go_col_environment_H}
\end{align}
with $\Delta>0$ independent of $N$ and zero Hamiltonian for $b_i$.
We initialize the environment in the zero-temperature Gibbs state of this Hamiltonian,
\begin{align}
\eta_{\mathrm E,N}^{\mathrm{col}}
=\bigotimes_{i=1}^N\left(
\ket{g}_{a_i}\bra{g}\otimes\frac{I_{b_i}}2
\right).
\label{eq:go_col_environment_state}
\end{align}
The system may start in any state $\rho$. Before each interaction, it is in a product state with the environment used in that round.

We take the interaction to be
\begin{align}
V_{N,\delta t}^{\mathrm{col}}
=-\Omega\sum_{i=1}^Ns_i^z\otimes Z_{a_i}
+\sqrt{\frac{\gamma}{\delta t}}
\sum_{i=1}^N(W_i-I_i/3)\otimes Z_{b_i},
\label{eq:go_col_interaction}
\end{align}
where $Z_{b_i}=\ket{0}_{b_i}\bra{0}-\ket{1}_{b_i}\bra{1}$ and $Z_{a_i}=\ket{g}_{a_i}\bra{g}-\ket{e}_{a_i}\bra{e}$. We use the same lattice arrangement as in Sec.~\ref{subsec:go_microscopic}.
Each term in $H_N^{(0)}$, $H_{\mathrm E,N}^{\mathrm{col}}$, and $V_{N,\delta t}^{\mathrm{col}}$ then acts on a single site or two neighboring sites. Thus, all three consist of short-range interactions. The Hamiltonian is time independent throughout each interaction, and every round uses the same operators and coefficients.

We have $[W_i,H_N^{(0)}]=[s_i^z,H_N^{(0)}]=0$ and $[Z_{a_i},H_{\mathrm E,N}^{\mathrm{col}}]=[Z_{b_i},H_{\mathrm E,N}^{\mathrm{col}}]=0$.
Therefore, writing $H_{\mathrm f,N}^{\mathrm{col}}=H_N^{(0)}\otimes I_{\mathrm E}+I_S\otimes H_{\mathrm E,N}^{\mathrm{col}}$, we obtain
\begin{align}
[V_{N,\delta t}^{\mathrm{col}},H_{\mathrm f,N}^{\mathrm{col}}]=0,
\label{eq:go_col_energy_conservation}
\end{align}
so the interaction conserves energy.
The resulting reduced map for one interaction is
\begin{align}
\Phi_{N,\delta t}^{\mathrm{col}}[\rho]
&=\Tr_{\mathrm E}\!\left[
U_{N,\delta t}^{\mathrm{col}}
(\rho\otimes\eta_{\mathrm E,N}^{\mathrm{col}})
U_{N,\delta t}^{\mathrm{col}\dagger}
\right],\\
U_{N,\delta t}^{\mathrm{col}}
&=e^{-i\delta t(H_{\mathrm f,N}^{\mathrm{col}}+V_{N,\delta t}^{\mathrm{col}})}.
\label{eq:go_col_one_step}
\end{align}

To evaluate the one-step reduced map to first order in $\delta t$, we write the coefficient of $\delta t^{-1/2}$ in the interaction as
\begin{align}
H_{\mathrm I,N}^{\mathrm{col}}
:=\sqrt\gamma\sum_{i=1}^N(W_i-I_i/3)\otimes Z_{b_i}.
\end{align}
Keeping $N$ fixed, we expand $U_{N,\delta t}^{\mathrm{col}}$ and its adjoint to second order in $\sqrt{\delta t}$ as
\begin{align}
U_{N,\delta t}^{\mathrm{col}}
&=I-i\sqrt{\delta t}\,H_{\mathrm I,N}^{\mathrm{col}}
-i\delta t\left(H_{\mathrm f,N}^{\mathrm{col}}-\Omega\sum_i s_i^z\otimes Z_{a_i}\right)
-\frac{\delta t}{2}(H_{\mathrm I,N}^{\mathrm{col}})^2
+O((\delta t)^{3/2}),\notag\\
U_{N,\delta t}^{\mathrm{col}\dagger}
&=I+i\sqrt{\delta t}\,H_{\mathrm I,N}^{\mathrm{col}}
+i\delta t\left(H_{\mathrm f,N}^{\mathrm{col}}-\Omega\sum_i s_i^z\otimes Z_{a_i}\right)
-\frac{\delta t}{2}(H_{\mathrm I,N}^{\mathrm{col}})^2
+O((\delta t)^{3/2}).
\label{eq:go_col_unitary_expansion}
\end{align}
Applying these expansions to $\rho_{S\mathrm E}=\rho\otimes\eta_{\mathrm E,N}^{\mathrm{col}}$, we find
\begin{align}
U_{N,\delta t}^{\mathrm{col}}\rho_{S\mathrm E}U_{N,\delta t}^{\mathrm{col}\dagger}
&=\rho_{S\mathrm E}
-i\sqrt{\delta t}[H_{\mathrm I,N}^{\mathrm{col}},\rho_{S\mathrm E}]\notag\\
&\quad-i\delta t\left[H_{\mathrm f,N}^{\mathrm{col}}-\Omega\sum_i s_i^z\otimes Z_{a_i},\rho_{S\mathrm E}\right]\notag\\
&\quad+\delta t\left(
H_{\mathrm I,N}^{\mathrm{col}}\rho_{S\mathrm E}H_{\mathrm I,N}^{\mathrm{col}}
-\frac12\{(H_{\mathrm I,N}^{\mathrm{col}})^2,\rho_{S\mathrm E}\}
\right)+O((\delta t)^{3/2}).
\label{eq:go_col_state_expansion}
\end{align}

Next, we take the partial trace over the environment term by term.
The initial environment state in Eq.~\eqref{eq:go_col_environment_state} gives
\begin{align}
\Tr_{\mathrm E}[\eta_{\mathrm E,N}^{\mathrm{col}}Z_{a_i}]
&={}_{a_i}\bra{g}Z_{a_i}\ket{g}_{a_i}=1,\\
\Tr_{\mathrm E}[\eta_{\mathrm E,N}^{\mathrm{col}}Z_{b_i}]
&=\frac12\Tr Z_{b_i}=0,\\
\Tr_{\mathrm E}[\eta_{\mathrm E,N}^{\mathrm{col}}Z_{b_i}Z_{b_j}]
&=\begin{cases}
\frac12\Tr Z_{b_i}^2=1&i=j,\\
\frac12\Tr Z_{b_i}\,\frac12\Tr Z_{b_j}=0&i\neq j
\end{cases}
=\delta_{ij}.
\label{eq:go_col_bath_moments}
\end{align}
For the term proportional to $\sqrt{\delta t}$, the partial trace therefore gives
\begin{align}
\Tr_{\mathrm E}\!\left[[H_{\mathrm I,N}^{\mathrm{col}},\rho_{S\mathrm E}]\right]
&=\sqrt\gamma\sum_i[W_i-I_i/3,\rho]
\Tr_{\mathrm E}[\eta_{\mathrm E,N}^{\mathrm{col}}Z_{b_i}]
=0.
\label{eq:go_col_first_order_vanishes}
\end{align}
For the commutator proportional to $\delta t$, the partial trace gives
\begin{align}
&\Tr_{\mathrm E}\!\left[\left[
H_{\mathrm f,N}^{\mathrm{col}}-\Omega\sum_i s_i^z\otimes Z_{a_i},\rho_{S\mathrm E}
\right]\right]\notag\\
&\quad=[H_N^{(0)},\rho]
+\rho\,\Tr[[H_{\mathrm E,N}^{\mathrm{col}},\eta_{\mathrm E,N}^{\mathrm{col}}]]
-\Omega\sum_i[s_i^z,\rho]
\Tr_{\mathrm E}[\eta_{\mathrm E,N}^{\mathrm{col}}Z_{a_i}]\notag\\
&\quad=\left[H_N^{(0)}-\Omega\sum_i s_i^z,\rho\right]
=[H_N^{(0)}+K_N,\rho].
\label{eq:go_col_hamiltonian_term}
\end{align}
This shows that the coupling to $a_i$ generates $K_N$.

We evaluate the remaining three terms using Eq.~\eqref{eq:go_col_bath_moments}.
The relation $\Tr_{\mathrm E}[Z_{b_i}\eta_{\mathrm E,N}^{\mathrm{col}}Z_{b_j}]=\delta_{ij}$ gives
\begin{align}
\Tr_{\mathrm E}\!\left[
H_{\mathrm I,N}^{\mathrm{col}}\rho_{S\mathrm E}H_{\mathrm I,N}^{\mathrm{col}}
\right]
&=\gamma\sum_{i,j}(W_i-I_i/3)\rho(W_j-I_j/3)
\Tr_{\mathrm E}[Z_{b_i}\eta_{\mathrm E,N}^{\mathrm{col}}Z_{b_j}]\notag\\
&=\gamma\sum_i(W_i-I_i/3)\rho(W_i-I_i/3).
\label{eq:go_col_jump_term}
\end{align}
Similarly, we obtain
\begin{align}
\Tr_{\mathrm E}\!\left[(H_{\mathrm I,N}^{\mathrm{col}})^2\rho_{S\mathrm E}\right]
&=\gamma\sum_{i,j}(W_i-I_i/3)(W_j-I_j/3)\rho
\Tr_{\mathrm E}[Z_{b_i}Z_{b_j}\eta_{\mathrm E,N}^{\mathrm{col}}]\notag\\
&=\gamma\sum_i(W_i-I_i/3)^2\rho,
\label{eq:go_col_anticommutator_term_left}\\
\Tr_{\mathrm E}\!\left[\rho_{S\mathrm E}(H_{\mathrm I,N}^{\mathrm{col}})^2\right]
&=\gamma\sum_{i,j}\rho(W_i-I_i/3)(W_j-I_j/3)
\Tr_{\mathrm E}[\eta_{\mathrm E,N}^{\mathrm{col}}Z_{b_i}Z_{b_j}]\notag\\
&=\gamma\sum_i\rho(W_i-I_i/3)^2.
\label{eq:go_col_anticommutator_term_right}
\end{align}
Adding these three contributions and using $W_i^2=I_i$, we find
\begin{align}
&\Tr_{\mathrm E}\!\left[
H_{\mathrm I,N}^{\mathrm{col}}\rho_{S\mathrm E}H_{\mathrm I,N}^{\mathrm{col}}
-\frac12\{(H_{\mathrm I,N}^{\mathrm{col}})^2,\rho_{S\mathrm E}\}
\right]\notag\\
&\quad=\gamma\sum_i\left[
(W_i-I_i/3)\rho(W_i-I_i/3)
-\frac12\{(W_i-I_i/3)^2,\rho\}
\right]\notag\\
&\quad=\gamma\sum_i\left[
W_i\rho W_i-\frac13\{W_i,\rho\}+\frac19\rho
-\frac12\{W_i^2-\tfrac23W_i+I_i/9,\rho\}
\right]\notag\\
&\quad=\gamma\sum_i\left(W_i\rho W_i-\frac12\{W_i^2,\rho\}\right)
=\gamma\sum_i(W_i\rho W_i-\rho).
\label{eq:go_col_dissipator}
\end{align}
Substituting Eqs.~\eqref{eq:go_col_first_order_vanishes}, \eqref{eq:go_col_hamiltonian_term}, and \eqref{eq:go_col_dissipator} into the partial trace of Eq.~\eqref{eq:go_col_state_expansion} yields
\begin{align}
\Phi_{N,\delta t}^{\mathrm{col}}[\rho]
&=\rho-i\delta t[H_N^{(0)}+K_N,\rho]
+\gamma\delta t\sum_i(W_i\rho W_i-\rho)
+O((\delta t)^{3/2})\notag\\
&=\rho+\delta t\mathcal L_N[\rho]+O((\delta t)^{3/2}).
\label{eq:go_col_lindblad_limit}
\end{align}
Because each round uses a fresh environment, the system state $\rho_S(t)$ at time $t$ satisfies $\rho_S(t+\delta t)=\Phi_{N,\delta t}^{\mathrm{col}}[\rho_S(t)]$.
Subtracting $\rho_S(t)$, dividing by $\delta t$, and taking the limit $\delta t\to0$ gives
\begin{align}
\frac{d\rho_S(t)}{dt}
&=\lim_{\delta t\to0}
\frac{\Phi_{N,\delta t}^{\mathrm{col}}[\rho_S(t)]-\rho_S(t)}{\delta t}
=\mathcal L_N[\rho_S(t)].
\label{eq:go_col_master_equation}
\end{align}
We therefore recover the Lindblad equation with the Lindbladian in Eq.~\eqref{eq:go_lindbladian}.

\subsection{Implementation requiring no average work at the initial coupling}\label{subsec:go_microscopic2}

In the preceding two subsections, we constructed a Lindbladian that activates time crystals using an equilibrium environment and a time-independent interaction Hamiltonian that conserves energy.
This Lindbladian satisfies NRSC.
That is, we have constructed models that activate time crystals although no external resource is supplied to the system during the open-system dynamics.\footnote{In the continuous-coupling model, the total Hamiltonian is time independent after the coupling is introduced, so no work supply is required during the subsequent evolution.
For the collision model, let $\rho_n$ be the system state immediately before the $n$th interaction, $U$ the evolution operator for one interaction, and $\eta_{\mathrm E}$ the initial environment state.
Since $K_N$ commutes with $H_N^{(0)}$, $s_i^z$, and $W_i$, it commutes with the total Hamiltonian, so $U^\dagger(K_N\otimes I_{\mathrm E})U=K_N\otimes I_{\mathrm E}$. Using the definition of the partial trace and cyclicity of the trace, we obtain
\begin{align*}
\Tr[\rho_{n+1}K_N]
&=\Tr_{S\mathrm E}\!\left[
(\rho_n\otimes\eta_{\mathrm E})
U^\dagger(K_N\otimes I_{\mathrm E})U
\right]
=\Tr[\rho_nK_N].
\end{align*}
The mean interaction energy is conserved during each interaction, and at the start of round $n$ it equals $\Tr[(\rho_n\otimes\eta_{\mathrm E})V_{N,\delta t}^{\mathrm{col}}]=\Tr[\rho_nK_N]$.
Since detaching the environment removes the interaction term from the Hamiltonian, the average work of detachment at the end of round $n$ is $-\Tr[\rho_nK_N]$.
The net average work of this detachment and the coupling in round $n+1$ is therefore
\begin{align*}
\overline W_{\mathrm{off}}^{(n)}
+\overline W_{\mathrm{on}}^{(n+1)}
&=-\Tr[\rho_nK_N]
+\Tr[\rho_{n+1}K_N]
=0.
\end{align*}} Thus, the two models already meet the goal of realizing time crystals without an external resource supply during the dynamics.
Here we consider a stricter condition by also examining the work needed to couple the system and environment before the dynamics begins.
We define work as the change in the mean energy of the total system due to an external change in the Hamiltonian.
For instantaneous coupling, the work is the expectation value of the difference between the total Hamiltonians before and after coupling. For the models above, it is
\begin{align}
\overline W_{\mathrm{on}} &=\Tr[\rho\otimes\eta_{\mathrm E}(H_{\mathrm{tot}}-H_N^{(0)}\otimes I_{\mathrm E}-I_S\otimes H_{\mathrm E})]\notag\\
&=\Tr[\rho\otimes\eta_{\mathrm E}V]\notag\\
&=\Tr[\rho K_N].
\end{align}
We now construct a model in which the average work of the initial coupling is also zero for every initial system state.

\subsubsection{Microscopic model and initial environment state}

\begin{figure}[tbp]
  \centering
  \includegraphics[width=0.65\linewidth]{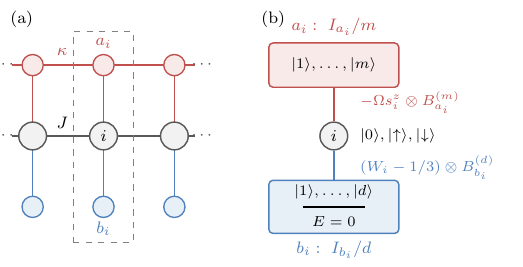} 
\caption{Continuous-coupling implementation requiring no average work at the initial coupling.
(a) Each three-level system site $i$ (gray) is placed between an $m$-level system $a_i$ (red) above and a $d$-level system $b_i$ (blue) below.
Gray lines between system sites represent interactions with coefficient $J$, and red lines between neighboring $a_i$ sites represent $\kappa\bigl(I_{a_ia_{i+1}}-\sum_{q=1}^{m}\ket{q,q}\bra{q,q}\bigr)$.
The zero-temperature Gibbs state of the $a_i$ chain is $\eta_{\mathrm A,N,m}=m^{-1}\sum_{q=1}^{m}\ket{Q_q}\bra{Q_q}$, with $\ket{Q_q}=\ket{q}^{\otimes N}$.
(b) Interaction between each system site and the environment.
The state $I_{a_i}/m$ is the reduced state of $a_i$, whereas the full $a_i$ chain is in the correlated state given above.
The operator $B_{a_i}^{(m)}$ has eigenvalue $1$ on $\ket{q}$ for $q<m$ and eigenvalue $-(m-1)$ on $\ket{m}$.
Each $b_i$ has zero Hamiltonian and state $I_{b_i}/d$.
}
\label{fig:go_microscopic_centered}
\end{figure}

As shown in Fig.~\ref{fig:go_microscopic_centered}, we couple each system site $i$ to an $m$-level system $a_i$ and a $d$-level system $b_i$.
For integers $m,d\geq2$, we use the orthonormal bases $\{\ket{q}_{a_i}\}_{q=1}^m$ and $\{\ket{j}_{b_i}\}_{j=1}^d$, respectively.
The environment Hilbert space is $\mathfrak H_{\mathrm E,N,m,d}=\bigotimes_{i=1}^N(\mathbb C^m_{a_i}\otimes\mathbb C^d_{b_i})$.
We define its Hamiltonian by
\begin{align}
H_{\mathrm E,N,m,d}
=\kappa\sum_{i=1}^{N-1}
\left(I_{a_ia_{i+1}}-\sum_{q=1}^m
\ket{q,q}_{a_i a_{i+1}}\bra{q,q}\right).
\label{eq:go_environment_Hamiltonian}
\end{align}
Here, $\kappa>0$ is independent of $N,m,d$, and each $b_i$ has zero Hamiltonian.
Let $H_{\mathrm A,N,m}$ denote the Hamiltonian on the $a_i$ sites in Eq.~\eqref{eq:go_environment_Hamiltonian}.
Each term contributes an energy $\kappa$ when the basis indices of neighboring sites $a_i,a_{i+1}$ differ.
Thus, the ground-state subspace of $H_{\mathrm A,N,m}$ is spanned by $\ket{Q_q}_a:=\bigotimes_{i=1}^N\ket{q}_{a_i}$, $q=1,\ldots,m$, and the zero-temperature Gibbs state of the full environment is
\begin{align}
\eta_{\mathrm E,N,m,d}
&=\frac1m\sum_{q=1}^m
\ket{Q_q}_a\bra{Q_q}
\otimes\frac{I_b}{d^N},
\label{eq:go_environment_Gibbs_state}
\end{align}
where $I_b=\bigotimes_{i=1}^N I_{b_i}$.

We define the system--environment interaction as
\begin{align}
V_{N,m,d}
=-\Omega\sum_{i=1}^Ns_i^z\otimes B_{a_i}^{(m)}
+\sum_{i=1}^N(W_i-I_i/3)\otimes B_{b_i}^{(d)}.
\label{eq:go_centered_interaction_definition}
\end{align}
Here, $B_{b_i}^{(d)}$ and its eigenvalues $b_j^{(d)}$ are the same as in Sec.~\ref{subsec:go_microscopic}.
The Hermitian operators $B_{a_i}^{(m)}$ and $B_{b_i}^{(d)}$ are defined by
\begin{align}
B_{a_i}^{(m)}
&=\sum_{q=1}^{m-1}\ket{q}_{a_i}\bra{q}
-(m-1)\ket{m}_{a_i}\bra{m},
\label{eq:go_shift_bath_operator}\\
B_{b_i}^{(d)}
&=\sum_{j=1}^db_j^{(d)}\ket{j}_{b_i}\bra{j},
\qquad
b_j^{(d)}=\gamma\tan\!\left[
\frac{\pi(2j-d-1)}{2d}\right].
\label{eq:go_dephasing_bath_operator}
\end{align}
The time-independent total Hamiltonian is
\begin{align}
H_{\mathrm{tot},N,m,d}
=H_N^{(0)}\otimes I_{\mathrm E}
+I_S\otimes H_{\mathrm E,N,m,d}
+V_{N,m,d},
\label{eq:go_total_Hamiltonian}
\end{align}
and we write $U_{N,m,d}(t):=e^{-itH_{\mathrm{tot},N,m,d}}$.

We place $a_i$ and $b_i$ next to site $i$ of the system chain, and $a_i$ next to $a_{i+1}$.
Then each term in $H_N^{(0)}$, $H_{\mathrm E,N,m,d}$, and $V_{N,m,d}$ acts on a single site or two neighboring sites.
Thus, all three consist of short-range interactions.
Moreover, $s_i^z$ and $W_i-I_i/3$ commute with $H_N^{(0)}$, while $B_{a_i}^{(m)}$ and $B_{b_i}^{(d)}$ commute with $H_{\mathrm E,N,m,d}$, giving
\begin{align}
[V_{N,m,d},
H_N^{(0)}\otimes I_{\mathrm E}]
=[V_{N,m,d},
I_S\otimes H_{\mathrm E,N,m,d}]=0.
\label{eq:go_bare_energy_conservation}
\end{align}
The interaction therefore conserves the sum of the system and environment Hamiltonians.

We compute the average work required to couple the environment.
Using Eqs.~\eqref{eq:go_environment_Gibbs_state}, \eqref{eq:go_shift_bath_operator}, and \eqref{eq:go_dephasing_bath_operator}, we obtain
\begin{align}
&\Tr_{\mathrm E}\!\left[
V_{N,m,d}(I_S\otimes\eta_{\mathrm E,N,m,d})
\right]\notag\\
&=-\Omega\sum_{i=1}^Ns_i^z
\frac1m\sum_{q=1}^m
{}_a\!\bra{Q_q}B_{a_i}^{(m)}\ket{Q_q}_a
+\sum_{i=1}^N(W_i-I_i/3)\frac1d\sum_{j=1}^db_j^{(d)}\notag\\
&=-\Omega\sum_{i=1}^Ns_i^z
\frac{(m-1)-(m-1)}m
+\sum_{i=1}^N(W_i-I_i/3)\frac1d\sum_{j=1}^db_j^{(d)}
=0.
\label{eq:go_centered_interaction}
\end{align}
The last equality follows from $b_{d+1-j}^{(d)}=-b_j^{(d)}$.
Thus, for any initial system state $\rho$, the average work required to introduce the interaction is
\begin{align}
\overline W_{\mathrm{on}}
&=\Tr_{S\mathrm E}\!\left[
V_{N,m,d}(\rho\otimes\eta_{\mathrm E,N,m,d})
\right]\notag\\
&=\Tr_S\!\left[
\rho\,\Tr_{\mathrm E}\!\left[
V_{N,m,d}(I_S\otimes\eta_{\mathrm E,N,m,d})
\right]\right]=0.
\label{eq:go_mean_switching_work}
\end{align}

\subsubsection{System dynamics and two-time correlation functions}

We derive the reduced system map from the evolution under the total Hamiltonian.
For the initial environment state $\eta_{\mathrm E,N,m,d}$, the reduced map and its adjoint acting on system observables are defined by
\begin{align}
\Phi_{N,m,d,t}[\rho]
&:=\Tr_{\mathrm E}\!\left[U_{N,m,d}(t)(\rho\otimes\eta_{\mathrm E,N,m,d})U_{N,m,d}^\dagger(t)\right],\notag\\
\Phi_{N,m,d,t}^\dagger[A]
&:=\Tr_{\mathrm E}\!\left[
(I_S\otimes\eta_{\mathrm E,N,m,d})U_{N,m,d}^\dagger(t)(A\otimes I_{\mathrm E})U_{N,m,d}(t)
\right].
\label{eq:go_exact_reduced_maps2}
\end{align}
To evaluate these maps, we write $\ket{\boldsymbol j}_b:=\bigotimes_{i=1}^N\ket{j_i}_{b_i}$, with $\boldsymbol j=(j_1,\ldots,j_N)$, and expand the environment state as
\begin{align}
\eta_{\mathrm E,N,m,d}
=\frac1{md^N}\sum_{q=1}^m\sum_{\boldsymbol j}
\ket{Q_q}_a\bra{Q_q}
\otimes\ket{\boldsymbol j}_b\bra{\boldsymbol j}.
\label{eq:go_environment_state_components2}
\end{align}
Each state in this expansion satisfies
\begin{align}
H_{\mathrm E,N,m,d}
\bigl(\ket{Q_q}_a\otimes\ket{\boldsymbol j}_b\bigr)
&=0,\notag\\
B_{a_i}^{(m)}\ket{Q_q}_a
&=c_q^{(m)}\ket{Q_q}_a,
\qquad
c_q^{(m)}
=
\begin{cases}
1,&1\le q<m,\\
-(m-1),&q=m,
\end{cases}\notag\\
B_{b_i}^{(d)}\ket{\boldsymbol j}_b
&=b_{j_i}^{(d)}\ket{\boldsymbol j}_b.
\end{align}
Substituting Eq.~\eqref{eq:go_environment_state_components2} into Eq.~\eqref{eq:go_exact_reduced_maps2}, we evaluate the partial trace using these relations and $K_N=-\Omega\sum_i s_i^z$ to obtain
\begin{align}
\Phi_{N,m,d,t}[\rho]
&=\Tr_{\mathrm E}\!\left[
U_{N,m,d}(t)(\rho\otimes\eta_{\mathrm E,N,m,d})U_{N,m,d}^\dagger(t)
\right]\notag\\
&=\frac1{md^N}\sum_{q=1}^m\sum_{\boldsymbol j}
e^{-it[H_N^{(0)}+c_q^{(m)}K_N+\sum_i b_{j_i}^{(d)}(W_i-I_i/3)]}
\rho\,
e^{it[H_N^{(0)}+c_q^{(m)}K_N+\sum_i b_{j_i}^{(d)}(W_i-I_i/3)]}.
\label{eq:go_exact_reduced_map_sum2}
\end{align}
Here, we used $-\Omega\sum_{i=1}^Nc_q^{(m)}s_i^z=c_q^{(m)}K_N$.
In particular, Eqs.~\eqref{eq:go_dicke_H0_W} and \eqref{eq:go_dicke_Sz} give $[H_N^{(0)},\rho_{\mathrm{ss},N}]=[K_N,\rho_{\mathrm{ss},N}]=[W_i,\rho_{\mathrm{ss},N}]=0$, and hence
\begin{align}
\Phi_{N,m,d,t}[\rho_{\mathrm{ss},N}]
&=\frac1{md^N}\sum_{q=1}^m\sum_{\boldsymbol j}
\rho_{\mathrm{ss},N}
=\rho_{\mathrm{ss},N}.
\label{eq:go_total_stationarity2}
\end{align}
Thus, $\rho_{\mathrm{ss},N}$ is also a stationary state in this model.

We now calculate the two-time correlation function of $S_N^x$ in this stationary state,
\begin{align}
C_{N,m,d}(t)
&:=\frac1{N^2}\Tr_S\!\left[
\rho_{\mathrm{ss},N}\Phi_{N,m,d,t}^\dagger[S_N^x]S_N^x
\right].
\label{eq:go_exact_system_correlation2}
\end{align}

Both $S_N^x$ and $S_N^y$ commute with $H_N^{(0)}$ and all $W_i-I_i/3$.
Using $i[K_N,S_N^x]=\Omega S_N^y$ and $i[K_N,S_N^y]=-\Omega S_N^x$, we obtain
\begin{align}
e^{itc_q^{(m)}K_N}S_N^xe^{-itc_q^{(m)}K_N}
=\cos(c_q^{(m)}\Omega t)S_N^x
+\sin(c_q^{(m)}\Omega t)S_N^y.
\end{align}
Since $[H_N^{(0)},S_N^x]=[W_i-I_i/3,S_N^x]=0$, the adjoint of Eq.~\eqref{eq:go_exact_reduced_map_sum2} gives
\begin{align}
\Phi_{N,m,d,t}^\dagger[S_N^x]
&=\frac1m\sum_{q=1}^m
e^{itc_q^{(m)}K_N}S_N^xe^{-itc_q^{(m)}K_N}\notag\\
&=\frac1m\sum_{q=1}^m\left[
\cos(c_q^{(m)}\Omega t)S_N^x
+\sin(c_q^{(m)}\Omega t)S_N^y
\right].
\label{eq:go_exact_spin_x_evolution2}
\end{align}
Using $\Tr[\rho_{\mathrm{ss},N}(S_N^x)^2]=N(N+2)/12$ and $\Tr[\rho_{\mathrm{ss},N}S_N^yS_N^x]=0$, we substitute Eq.~\eqref{eq:go_exact_spin_x_evolution2} into Eq.~\eqref{eq:go_exact_system_correlation2} to find
\begin{align}
C_{N,m,d}(t)
&=\frac1{mN^2}\sum_{q=1}^m
\cos(c_q^{(m)}\Omega t)
\Tr[\rho_{\mathrm{ss},N}(S_N^x)^2]\notag\\
&=\frac{N+2}{12N}
\left[
\left(1-\frac1m\right)\cos(\Omega t)
+\frac1m\cos((m-1)\Omega t)
\right].
\label{eq:go_total_correlation2}
\end{align}
This expression is exact for all finite $N,m,d$ and all times, and is independent of $d$.
Taking $N\to\infty$ at fixed finite $m,d$, we obtain
\begin{align}
\lim_{N\to\infty}C_{N,m,d}(t)
=\frac1{12}
\left[
\left(1-\frac1m\right)\cos(\Omega t)
+\frac1m\cos((m-1)\Omega t)
\right].
\end{align}
Thus, a nonconstant periodic function with period $2\pi/\Omega$ remains.

\subsubsection{Convergence to the Lindblad evolution}\label{subsec:go_limit2}

We show that the reduced map derived in the preceding subsection converges to the Lindblad evolution introduced above as $m,d\to\infty$ at fixed $N$, uniformly on every finite interval $0\leq t\leq T$.
We do so by verifying that the couplings to $a_i$ and $b_i$ yield, in the limit, the dynamics generated by $K_N$ and the dissipative term, respectively.
As above, since $[n_i,n_j]=[n_i,s_j^z]=0$ for all $i,j$, we have
\begin{align}
[H_N^{(0)},K_N]=[H_N^{(0)},W_i]=[K_N,W_i]=[W_i,W_j]=0
\qquad(i,j=1,\ldots,N).
\label{eq:go_commutation2}
\end{align}
Hence, the exponentials in Eq.~\eqref{eq:go_exact_reduced_map_sum2} factorize as
\begin{align}
e^{-it[H_N^{(0)}+c_q^{(m)}K_N+\sum_i b_{j_i}^{(d)}(W_i-I_i/3)]}
=e^{-itH_N^{(0)}}\,e^{-itc_q^{(m)}K_N}\prod_{i=1}^N e^{-itb_{j_i}^{(d)}(W_i-I_i/3)}.
\label{eq:go_exponential_factorization2}
\end{align}
We write $\mathcal U_{N,t}^{(0)}[X]:=e^{-itH_N^{(0)}}Xe^{itH_N^{(0)}}$ and define the map induced by coupling to $a_i$ by
\begin{align}
\mathcal S_{N,m,t}[X]
&:=\frac{m-1}{m}e^{-itK_N}Xe^{itK_N}
+\frac1m e^{it(m-1)K_N}Xe^{-it(m-1)K_N}.
\label{eq:go_centered_shift_channel}
\end{align}
We also write $\mathcal U_{K_N,t}[X]:=e^{-itK_N}Xe^{itK_N}$ and use the maps $\mathcal Q_{i,d,t}$ and $\mathcal Q_{i,t}^{\mathrm L}$ defined in Sec.~\ref{subsec:go_limit}.
Substituting Eq.~\eqref{eq:go_exponential_factorization2} into Eq.~\eqref{eq:go_exact_reduced_map_sum2} and using Eq.~\eqref{eq:go_commutation2} for the Lindblad evolution, we can write the reduced map for the finite environment and the Lindblad evolution, respectively, as
\begin{align}
\Phi_{N,m,d,t}
&=\mathcal U_{N,t}^{(0)}\circ\mathcal S_{N,m,t}\circ
\prod_{i=1}^N\mathcal Q_{i,d,t},
\label{eq:go_reduced_map_factorization2}\\
e^{t\mathcal L_N}
&=\mathcal U_{N,t}^{(0)}\circ\mathcal U_{K_N,t}\circ
\prod_{i=1}^N\mathcal Q_{i,t}^{\mathrm L}.
\label{eq:go_target_map_factorization2}
\end{align}
Products denote composition, and all maps appearing here commute.

First, we bound the difference between $\mathcal S_{N,m,t}$, induced by coupling to $a_i$, and $\mathcal U_{K_N,t}$, generated by $K_N$.
Using Eq.~\eqref{eq:go_centered_shift_channel} and invariance of the trace norm under unitary conjugation, we obtain, for any $X$ and $t$,
\begin{align}
\|(\mathcal S_{N,m,t}-\mathcal U_{K_N,t})[X]\|_1
&=\frac1m\left\|
e^{it(m-1)K_N}Xe^{-it(m-1)K_N}
-e^{-itK_N}Xe^{itK_N}
\right\|_1\notag\\
&\leq\frac2m\|X\|_1.
\label{eq:go_shift_channel_bound}
\end{align}
Thus, $\mathcal S_{N,m,t}$ converges uniformly to the dynamics generated by $K_N$ as $m\to\infty$.

The maps $\mathcal Q_{i,d,t}$ induced by coupling to $b_i$ and $\mathcal Q_{i,t}^{\mathrm L}$ generated by dissipation are the same as in Sec.~\ref{subsec:go_microscopic}, so the bounds in Eqs.~\eqref{eq:go_channel_contractions} and \eqref{eq:go_product_channel_bound} apply. Combining these with Eq.~\eqref{eq:go_shift_channel_bound}, we obtain, for any initial state $\rho$,
\begin{align}
&\sup_{0\leq t\leq T}
\|\Phi_{N,m,d,t}[\rho]-e^{t\mathcal L_N}[\rho]\|_1\notag\\
&=\sup_{0\leq t\leq T}\left\|
\left(\mathcal S_{N,m,t}\prod_i\mathcal Q_{i,d,t}
-\mathcal U_{K_N,t}\prod_i\mathcal Q_{i,t}^{\mathrm L}\right)[\rho]
\right\|_1\notag\\
&=\sup_{0\leq t\leq T}\left\|
\left[(\mathcal S_{N,m,t}-\mathcal U_{K_N,t})\prod_i\mathcal Q_{i,d,t}
+\mathcal U_{K_N,t}\left(\prod_i\mathcal Q_{i,d,t}
-\prod_i\mathcal Q_{i,t}^{\mathrm L}\right)\right][\rho]
\right\|_1\notag\\
&\leq\sup_{0\leq t\leq T}\left\{
\left\|\left[(\mathcal S_{N,m,t}-\mathcal U_{K_N,t})\prod_i\mathcal Q_{i,d,t}\right][\rho]\right\|_1
+\left\|\left(\prod_i\mathcal Q_{i,d,t}
-\prod_i\mathcal Q_{i,t}^{\mathrm L}\right)[\rho]\right\|_1
\right\}\notag\\
&\leq\frac2m\sup_{0\leq t\leq T}
\left\|\left(\prod_i\mathcal Q_{i,d,t}\right)[\rho]\right\|_1
+N\epsilon_d(T)\|\rho\|_1\notag\\
&\leq\frac2m+N\epsilon_d(T).
\label{eq:go_centered_reduced_map_limit}
\end{align}
The first equality follows from Eqs.~\eqref{eq:go_reduced_map_factorization2} and \eqref{eq:go_target_map_factorization2} and the invariance of the trace norm under $\mathcal U_{N,t}^{(0)}$, and the first inequality from the triangle inequality and the invariance under $\mathcal U_{K_N,t}$.
The second inequality uses Eq.~\eqref{eq:go_shift_channel_bound} for the first term and Eq.~\eqref{eq:go_product_channel_bound} for the second. The last inequality uses Eq.~\eqref{eq:go_channel_contractions} and $\|\rho\|_1=1$.
Since $\epsilon_d(T)\to0$, we have $2/m+N\epsilon_d(T)\to0$ as $m,d\to\infty$ at fixed $N$.
Thus, the reduced map converges uniformly in trace norm to the target Lindblad evolution on every finite time interval.

\newpage
\part{No-go theorem}

In this part, we prove a general no-go theorem: without an external supply of thermodynamic resources, a time crystal can be activated only by an environment-induced Hamiltonian.
This generalizes the finding of Part I that the environment-induced Hamiltonian term $K_N$ activates the time crystal.
In Sec.~\ref{sec:resource_theory_of_asymmetry}, we first review the resource theory of asymmetry and modes of asymmetry, which underlie the proof.
In Sec.~\ref{sec:asymmetry_spectral_details}, we then use modes of asymmetry to prove the no-go theorem for Gibbs stationary states at nonzero temperatures.
The subsequent sections extend the theorem to zero-temperature Gibbs states and to arbitrary stationary states.
In Sec.~\ref{sec:v4:stationary_structure}, we prepare for this by deriving general properties of stationary states of Lindblad equations.
Building on this, we prove these extensions in Sec.~\ref{sec:v4:spectral_nogo}.
Finally, in Sec.~\ref{sec:nrsc_lindbladian}, we show that, without an external supply of thermodynamic resources, the Born--Markov--secular approximation and a collision model yield Lindbladians satisfying NRSC; this justifies using NRSC to represent the absence of such a supply.

\section{Resource theory of asymmetry}
\label{sec:resource_theory_of_asymmetry}
\label{sec:sm:asymmetry_modes}

In this section, we review the basics of the resource theory of asymmetry and modes of asymmetry.
Section~\ref{subsec:rta_free} defines free states and free operations, and Sec.~\ref{subsec:rta_modes} presents the basic properties of modes of asymmetry.
In Sec.~\ref{subsec:mode_decomposition_properties}, we apply these properties to Lindbladians satisfying NRSC.
Throughout this section, we consider only finite-dimensional systems.

In a resource theory, states that can be prepared at no cost under given physical constraints are called free states, and CPTP maps that can be implemented at no cost are called free operations~\cite{Chitambar2019quantum}.
The choice of free states and free operations depends on the physical situation, but free operations are generally required to satisfy at least the following properties:
\begin{enumerate}
 \item[(F1)] If $\mathcal E_1$ and $\mathcal E_2$ are free operations, then so is $\mathcal E_2\circ\mathcal E_1$.
 \item[(F2)] The identity map is a free operation.
 \item[(F3)] If $\mathcal E$ is a free operation and $\sigma$ is a free state, then $\mathcal E[\sigma]$ is also a free state.
\end{enumerate}
By (F1) and (F3), nonfree states cannot be generated from free states by free operations.
Such states are regarded as resources that are valuable under the given constraints.
A resource monotone is a nonnegative function $R$ that measures the amount of resource and does not increase under free operations:
\begin{align}
 R(\mathcal E[\rho])&\leq R(\rho)
 \qquad(\mathcal E:\text{free operation}),\notag\\
 R(\sigma)&=0
 \qquad(\sigma:\text{free state}).
 \label{eq:rta:monotone}
\end{align}
Nonnegativity reflects that $R$ quantifies an amount, and monotonicity that free operations cannot increase the resource.
Resource measures need not be scalar valued, and in some resource theories they take values in positive semidefinite matrices~\cite{Kudo_Tajima,Yamaguchi2026quantum}.
The resource theory of asymmetry, the subject of this section, also provides such an example for a general compact Lie group~\cite{Yamaguchi2026quantum,Yamaguchi2026quantifying}.

In general, there are infinitely many resource monotones, and it is not obvious which one should be used to quantify the resource content of a state.
A standard criterion for choosing among them is asymptotic conversion in the i.i.d. setting, where many copies of one state are converted into many copies of another~\cite{Chitambar2019quantum}.
Given nonfree states $\rho$ and $\rho'$ and a real number $r\geq0$, we say that $\rho$ is asymptotically convertible to $\rho'$ at rate $r$ if there exists a sequence of free operations $\{\mathcal E_n\}_{n=1}^{\infty}$ such that
\begin{align}
 \lim_{n\to\infty}\frac12\bigl\|\mathcal E_n[\rho^{\otimes n}]-\rho'^{\otimes\lfloor rn\rfloor}\bigr\|_1=0,
 \label{eq:rta:iid_conversion}
\end{align}
where $\|\cdot\|_1$ denotes the trace norm.
The supremum over such $r$ is called the optimal conversion rate from $\rho$ to $\rho'$.
For a fixed class of states, such as the set of all pure states, we call a resource monotone or a set of resource monotones an i.i.d.-complete measure if it determines the optimal conversion rate between any two states in the class.
That is, an i.i.d.-complete measure quantifies the resource of a state by how many copies of other states can be produced from it.

A familiar example in physics is quantum thermodynamics, in which an inverse temperature $\beta$ is fixed and the Gibbs state at that temperature is the free state.
The free operations are thermal operations: a Gibbs state at inverse temperature $\beta$, which is a free state, is freely prepared, interacts with the target system through an energy-conserving unitary, and is then traced out~\cite{Janzing2000thermodynamic,Horodecki2013fundamental}.
This setting describes free access to an infinitely large heat bath at a fixed temperature.
In this resource theory, the deviation from equilibrium is the resource, and the nonequilibrium free energy is its i.i.d.-complete measure for states without energetic coherence~\cite{Brandao2013resource}.

Another example is the resource theory of entanglement.
Its free states are separable states, i.e., probabilistic mixtures of product states, and its free operations are local operations and classical communication (LOCC), which consist of local operations, local measurements, and classical communication of the measurement outcomes~\cite{Horodecki2009quantum}.
In this theory, entanglement is the resource, and the entanglement entropy is its i.i.d.-complete measure for pure bipartite states~\cite{Bennett1996concentrating}.

Resource theories are thus well suited to quantifying abstract properties of states, such as the deviation from equilibrium and entanglement.
The resource theory of asymmetry, introduced below, applies this framework to quantify symmetry breaking.

\subsection{Symmetric states and covariant operations}
\label{subsec:rta_free}

The resource theory of asymmetry treats asymmetry as a resource under quantum operations constrained by symmetry~\cite{Bartlett2007reference,Gour2008resource,Marvian2012symmetry}.
This framework has long been developed as a theory of symmetry breaking for weak symmetry~\cite{Marvian2013theory,Marvian2014extending,Marvian2022operational,YT,Shitara_Tajima,Yamaguchi2026quantum,Yamaguchi2026quantifying} and has recently been extended to strong symmetry~\cite{Kusuki2026resource}.
In the weak-symmetry setting, i.i.d.-complete measures of symmetry breaking have been obtained for pure states under arbitrary finite groups~\cite{Shitara_Tajima,Mitsuhashi2026general} and for arbitrary, possibly mixed, states under arbitrary compact Lie groups~\cite{Yamaguchi2026quantum,Yamaguchi2026quantifying}.
For strong symmetry, standard measures of symmetry breaking under arbitrary compact Lie groups have also recently been obtained for arbitrary, possibly mixed, states~\cite{Kusuki2026resource}.
In this paper, however, we restrict attention to weak symmetry for the groups ${\rm U}(1)$ and $\bbR$.
For other groups and the corresponding resource measures, see the references above.

When the symmetry group is ${\rm U}(1)$ or $\bbR$, its action can be written in terms of a single Hermitian operator $H$ as
\begin{align}
 \mathcal U_\theta[\rho]:=e^{-iH\theta}\rho\,e^{iH\theta}.
 \label{eq:rta:symmetry_transformation}
\end{align}
If all eigenvalue differences of $H$ are integer multiples of some positive number, $\mathcal U_\theta$ is periodic in $\theta$, and the symmetry group can be identified with ${\rm U}(1)$.
Otherwise, the symmetry group is $\bbR$.
We call $H$ the generator of the symmetry.
For the time-translation symmetry considered in this paper, the generator is $H_N$.

As free states, we use symmetric states, defined as follows.
\begin{definition}[Symmetric state]
\label{def:rta:symmetric_state}
A state $\rho$ is called symmetric with respect to $H$ if
\begin{align}
 \mathcal U_\theta[\rho]=\rho
 \label{eq:rta:symmetric_state}
\end{align}
for any $\theta\in\bbR$.
\end{definition}
Equation~\eqref{eq:rta:symmetric_state} means that $\rho$ commutes with $e^{-iH\theta}$ for any $\theta$, which is equivalent to $[\rho,H]=0$.
In terms of the spectral decomposition $H=\sum_h h\,\Pi_h$, where $\Pi_h$ is the projection onto the eigenspace of $H$ with eigenvalue $h$, this condition reads
\begin{align}
 \rho=\sum_h\Pi_h\rho\,\Pi_h.
 \label{eq:rta:no_coherence}
\end{align}
That is, symmetric states have no coherence between different eigenspaces of $H$.
For time translation, where $H$ is the Hamiltonian of the system, symmetric states are those invariant under the isolated-system dynamics, such as energy eigenstates and Gibbs states.

As free operations, we use covariant operations, defined as follows.
\begin{definition}[Covariant operation]
\label{def:rta:covariant_operation}
A CPTP map $\mathcal E$ from a system $S$ to a system $S'$ is called a covariant operation if
\begin{align}
 \mathcal E\circ\mathcal U^{S}_\theta=\mathcal U^{S'}_\theta\circ\mathcal E
 \label{eq:rta:covariance}
\end{align}
for any $\theta\in\bbR$.
Here, $\mathcal U^{S}_\theta$ and $\mathcal U^{S'}_\theta$ are the symmetry transformations generated by $H_S$ on the input system and by $H_{S'}$ on the output system, respectively.
\end{definition}
Equation~\eqref{eq:rta:covariance} means that applying the symmetry transformation before or after the operation gives the same result.
Covariant operations satisfy (F1)--(F3).
Properties (F1) and (F2) follow immediately from Eq.~\eqref{eq:rta:covariance}.
For (F3), let $\sigma$ be a symmetric state and $\mathcal E$ a covariant operation.
Then, for any $\theta$,
\begin{align}
 \mathcal U^{S'}_\theta\bigl[\mathcal E[\sigma]\bigr]
 =\mathcal E\bigl[\mathcal U^{S}_\theta[\sigma]\bigr]
 =\mathcal E[\sigma],
 \label{eq:rta:free_to_free}
\end{align}
so $\mathcal E[\sigma]$ is also a symmetric state.
Therefore, covariant operations alone cannot create asymmetric states from symmetric states, and asymmetry is a resource in this framework.

Any covariant operation $\mathcal E$ can be implemented with a unitary satisfying a conservation law and a symmetric state of an ancilla~\cite{Keyl1999optimal,Marvian2012symmetry,Marvian2013theory}.
More precisely, there exist ancillas $E$ and $E'$ with $SE=S'E'$, a unitary $V$ satisfying $V(H_S+H_E)V^\dagger=H_{S'}+H_{E'}$, and an ancilla state $\sigma_E$ satisfying $[\sigma_E,H_E]=0$ such that
\begin{align}
 \mathcal E[\rho_S]
 =\Tr_{E'}\!\left[V(\rho_S\otimes\sigma_E)V^\dagger\right],
 \label{eq:rta:dilation}
\end{align}
where $H_S$, $H_{S'}$, $H_E$, and $H_{E'}$ are the generators of the symmetry transformations~\eqref{eq:rta:symmetry_transformation} on the respective systems.
Conversely, any map of this form is a covariant operation.
The condition on $V$ expresses conservation of the total generator.
For time translation, each generator is the Hamiltonian of the corresponding system, and this condition is energy conservation.
In this case, implementing a noncovariant operation in the form of Eq.~\eqref{eq:rta:dilation} requires either an ancilla state with $[\sigma_E,H_E]\neq0$ or a unitary with $V(H_S+H_E)V^\dagger\neq H_{S'}+H_{E'}$.

A standard resource monotone of asymmetry is the SLD quantum Fisher information of the family $\{\mathcal U_\theta[\rho]\}_{\theta\in\bbR}$~\cite{Yadin2016general,Takagi_skew,Marvian2022operational,YT}:
\begin{align}
 \mathcal F_\rho(H):=2\sum_{i,j:\,p_i+p_j>0}\frac{(p_i-p_j)^2}{p_i+p_j}\bigl|\bra{i}H\ket{j}\bigr|^2,
 \label{eq:rta:qfi}
\end{align}
where $\rho=\sum_ip_i\dm{i}$ is the spectral decomposition of $\rho$.
The quantity $\mathcal F_\rho(H)$ satisfies Eq.~\eqref{eq:rta:monotone} and vanishes if and only if $\rho$ is symmetric.

\subsection{Modes of asymmetry}
\label{subsec:rta_modes}

States and operators can be decomposed into components according to the frequency of the phase they acquire under the symmetry transformation.
The corresponding subspaces are called modes of asymmetry.
Let $\mathfrak B$ be the space of all operators on the system, and let $\mathcal{W}_H$ be the set of eigenvalue differences of $H$.
For each $\omega\in\bbR$, we define the mode space
\begin{align}
 \mathfrak B_\omega:=\{A\in\mathfrak B\mid[H,A]=-\omega A\}.
 \label{eq:rta:mode_spaces}
\end{align}
Then $\mathfrak B_\omega\neq\{0\}$ only if $\omega\in\mathcal{W}_H$.
The space $\mathfrak B_\omega$ is a simultaneous eigenspace of the symmetry transformations:
\begin{align}
 A\in\mathfrak B_\omega
 \quad\Longleftrightarrow\quad
 \mathcal U_\theta[A]=e^{i\omega\theta}A
 \qquad(\forall\theta\in\bbR).
 \label{eq:rta:mode_eigen}
\end{align}
\begin{proposition}[Basic properties of modes of asymmetry~\cite{Marvian2012symmetry,Marvian2014modes,Lostaglio2015quantum}]
\label{prop:rta:modes}
The mode spaces $\{\mathfrak B_\omega\}_{\omega\in\mathcal{W}_H}$ satisfy the following.
\begin{enumerate}
 \item[(a)] Any $A\in\mathfrak B$ is uniquely decomposed as
 \begin{align}
  A=\sum_{\omega\in\mathcal{W}_H}A_\omega,
  \qquad
  A_\omega:=\sum_h\Pi_hA\,\Pi_{h+\omega}\in\mathfrak B_\omega,
  \label{eq:rta:mode_components}
 \end{align}
 where $\Pi_{h+\omega}:=0$ if $h+\omega$ is not an eigenvalue of $H$.
 That is, $\mathfrak B=\bigoplus_{\omega\in\mathcal{W}_H}\mathfrak B_\omega$.
 \item[(b)] If $A\in\mathfrak B_\omega$ and $A'\in\mathfrak B_{\omega'}$, then $AA'\in\mathfrak B_{\omega+\omega'}$ and $A^\dagger\in\mathfrak B_{-\omega}$.
 \item[(c)] A linear map $\mathcal E$ on $\mathfrak B$ satisfies $\mathcal E\circ\mathcal U_\theta=\mathcal U_\theta\circ\mathcal E$ for any $\theta\in\bbR$ if and only if $\mathcal E[\mathfrak B_\omega]\subseteq\mathfrak B_\omega$ for any $\omega\in\mathcal{W}_H$.
\end{enumerate}
\end{proposition}
For details, see Ref.~\cite{Marvian2014modes}.

By Eq.~\eqref{eq:rta:no_coherence} and (a), a state $\rho$ is symmetric if and only if $\rho_\omega=0$ for all $\omega\neq0$.
That is, the asymmetry of a state lies in its mode components with $\omega\neq0$.
Moreover, by (c), a covariant operation from a system to itself maps each mode component into the mode space of the same frequency.
Therefore, covariant operations cannot create mode components at frequencies absent from the input state.

\subsection{Application to Lindbladians satisfying NRSC}
\label{subsec:mode_decomposition_properties}
\label{subsec:covariance_mode_preservation}

Throughout this subsection, we fix $N$ and denote the space of all operators by $\mathfrak B_N$.
The dual of a superoperator is defined by
\begin{align}
 \Tr[\mathcal M_N[\rho_N]X_N]
 =\Tr[\rho_N\mathcal M_N^\dagger[X_N]].
 \label{eq:v4:superoperator_duality}
\end{align}
We apply Proposition~\ref{prop:rta:modes} to $H=H_N$ to derive the time-translation covariance and the mode decomposition of Lindbladians satisfying NRSC.
Let $\calE_N$ be the set of eigenvalues of $H_N$, and write the spectral decomposition of the Hamiltonian and the set of energy differences as
\begin{align}
 H_N&=\sum_{E\in\calE_N} EP_{E,N},\notag\\
 \mathcal{W}_N&:=\{E'-E\mid E,E'\in\calE_N\}.
 \label{eq:sm:hamiltonian_resolution}
\end{align}
Here, $P_{E,N}$ is the projection onto the entire eigenspace with eigenvalue $E$, and
\begin{align}
 P_{E,N}P_{E',N}&=\delta_{E,E'}P_{E,N},\notag\\
 \sum_E P_{E,N}&=I.
 \label{eq:v4:spectral_projection_properties}
\end{align}
For $E$ that is not an eigenvalue, we set $P_{E,N}:=0$.
We also write the time translation as
\begin{align}
 \mathcal U_{N,t}[X_N]:=e^{-iH_Nt}X_Ne^{iH_Nt},
 \qquad t\in\bbR.
 \label{eq:rta:time_translation}
\end{align}
Then $\mathcal U_{N,t}^\dagger=\mathcal U_{N,-t}=e^{t\mathcal H_N^\dagger}$, where $\mathcal H_N^\dagger[X_N]=i[H_N,X_N]$.

We write the mode components and mode spaces with respect to $H_N$ as
\begin{align}
 X_{N,\omega}&:=\sum_E P_{E,N}X_NP_{E+\omega,N},
 \label{eq:sm:mode_component}\\
 \mathfrak B_{N,\omega}&:=\{X_N\in\mathfrak B_N\mid[H_N,X_N]=-\omega X_N\},
 \label{eq:sm:mode_space}
\end{align}
with $\mathfrak B_{N,\nu}=\{0\}$ for $\nu\notin\mathcal{W}_N$.
Applying (a) and (b) of Proposition~\ref{prop:rta:modes} to $H=H_N$, with $h$ and $\Pi_h$ replaced by $E$ and $P_{E,N}$, we obtain properties (M1) and (M2) of modes of asymmetry used in the main text.
In particular, (a) gives
\begin{align}
 \mathfrak B_N
 =\bigoplus_{\omega\in\mathcal{W}_N}\mathfrak B_{N,\omega},
 \label{eq:sm:mode_direct_sum}
\end{align}
so any $X_N\in\mathfrak B_N$ is uniquely decomposed as $X_N=\sum_{\omega\in\mathcal{W}_N}X_{N,\omega}$.

We now consider Lindbladians satisfying NRSC.
Denoting by $\mathcal D_{L_{k,N}}^\dagger$ the dissipative action of each jump operator, we write the Lindbladian as
\begin{align}
 \mathcal L_N^\dagger
 &=\mathcal H_N^\dagger+\mathcal D_N^\dagger,\notag\\
 \mathcal H_N^\dagger[X_N]
 &=i[H_N,X_N],
 \label{eq:sm:gksl_decomposition}\\
 \mathcal D_N^\dagger[X_N]
 &=\sum_k\gamma_{k,N}\mathcal D_{L_{k,N}}^\dagger[X_N],
 \qquad\gamma_{k,N}\geq0,
 \label{eq:sm:dissipator_sum}\\
 \mathcal D_{L_{k,N}}^\dagger[X_N]
 &:=L_{k,N}^\dagger X_NL_{k,N}
 -\frac12\{L_{k,N}^\dagger L_{k,N},X_N\}.
 \label{eq:sm:single_jump_dissipator}
\end{align}
Since channels with zero rate do not contribute to $\mathcal D_N^\dagger$, we omit them without loss of generality and assume $\gamma_{k,N}>0$ in what follows.
By condition (ii) of NRSC, i.e., time-translation covariance, each jump operator satisfies
\begin{align}
 [L_{k,N},H_N]=\omega_{k,N}L_{k,N},
 \qquad\omega_{k,N}\in\mathcal{W}_N.
 \label{eq:sm:jump_covariance_condition}
\end{align}

\begin{lemma}[Time-translation covariance of the dissipative part]
\label{lem:sm:covariance_mode_preservation}
Under condition (ii) of NRSC, for any $t\in\bbR$,
\begin{align}
 \mathcal U_{N,t}\circ\mathcal D_N^\dagger
 =\mathcal D_N^\dagger\circ\mathcal U_{N,t}.
 \label{eq:sm:dissipator_time_translation_covariance}
\end{align}
\end{lemma}

\noindent\textit{Proof of Lemma~\ref{lem:sm:covariance_mode_preservation}.---}
By Eqs.~\eqref{eq:sm:jump_covariance_condition} and \eqref{eq:rta:mode_eigen},
\begin{align}
 \mathcal U_{N,t}[L_{k,N}]
 &=e^{i\omega_{k,N}t}L_{k,N},\\
 \mathcal U_{N,t}[L_{k,N}^\dagger]
 &=e^{-i\omega_{k,N}t}L_{k,N}^\dagger.
\end{align}
Using these relations and $\mathcal U_{N,t}[L_{k,N}^\dagger L_{k,N}]=L_{k,N}^\dagger L_{k,N}$, we obtain, for any $X_N\in\mathfrak B_N$,
\begin{align}
 \mathcal U_{N,t}\!\left[\mathcal D_{L_{k,N}}^\dagger[X_N]\right]
 &=\mathcal U_{N,t}[L_{k,N}^\dagger]\,\mathcal U_{N,t}[X_N]\,\mathcal U_{N,t}[L_{k,N}]
 -\frac12\bigl\{\mathcal U_{N,t}[L_{k,N}^\dagger L_{k,N}],\mathcal U_{N,t}[X_N]\bigr\}\notag\\
 &=e^{-i\omega_{k,N}t}L_{k,N}^\dagger\,\mathcal U_{N,t}[X_N]\,e^{i\omega_{k,N}t}L_{k,N}
 -\frac12\bigl\{L_{k,N}^\dagger L_{k,N},\mathcal U_{N,t}[X_N]\bigr\}\notag\\
 &=\mathcal D_{L_{k,N}}^\dagger\!\left[\mathcal U_{N,t}[X_N]\right].
\end{align}
Substituting this into Eq.~\eqref{eq:sm:dissipator_sum} yields Eq.~\eqref{eq:sm:dissipator_time_translation_covariance}.\hfill\qed

By Lemma~\ref{lem:sm:covariance_mode_preservation} and $\mathcal U_{N,t}=e^{-t\mathcal H_N^\dagger}$, both $\mathcal H_N^\dagger$ and $\mathcal D_N^\dagger$ commute with time translations, and so does their sum $\mathcal L_N^\dagger$.
Hence, applying (c) of Proposition~\ref{prop:rta:modes} to $H=H_N$, we find that $\mathcal M_N^\dagger\in\{\mathcal H_N^\dagger,\mathcal D_N^\dagger,\mathcal L_N^\dagger\}$ satisfies
\begin{align}
 \mathcal M_N^\dagger[\mathfrak B_{N,\omega}]
 \subseteq\mathfrak B_{N,\omega}
 \qquad(\omega\in\mathcal{W}_N).
 \label{eq:sm:mode_invariance}
\end{align}
Moreover, with respect to the direct-sum decomposition~\eqref{eq:sm:mode_direct_sum} given by (a) of the proposition,
\begin{align}
 \mathcal M_N^\dagger
 =\bigoplus_{\omega\in\mathcal{W}_N}
 \left.\mathcal M_N^\dagger\right|_{\mathfrak B_{N,\omega}},
 \label{eq:sm:superoperator_mode_decomposition}
\end{align}
where $\left.\cdot\right|_{\mathfrak B_{N,\omega}}$ denotes the restriction to each mode space.
This covariance also implies that, for any $s\geq0$ and $t\in\bbR$,
\begin{align}
 e^{s\mathcal L_N}\circ\mathcal U_{N,t}
 =\mathcal U_{N,t}\circ e^{s\mathcal L_N},
 \label{eq:rta:semigroup_covariance}
\end{align}
so the time evolution $e^{s\mathcal L_N}$ at each time is a covariant operation, i.e., a free operation in the resource theory of asymmetry.
In this sense, condition (ii) of NRSC means that the time evolution itself supplies no time-translation asymmetry from outside.

\section{Basic properties of Lindbladians and the no-go theorem for Gibbs states at nonzero temperatures}
\label{sec:asymmetry_spectral_details}

In this section, we use the mode decomposition introduced in the previous section to prove basic properties of Lindbladians and the no-go theorem for Gibbs states at nonzero temperatures.
In Sec.~\ref{subsec:gksl_identity_and_spectral_bound}, we prove that the eigenvalues of Lindbladians have nonpositive real parts.
In Sec.~\ref{subsec:gibbs_dissipative_self_adjointness}, we introduce the Gibbs-weighted inner product and prove that the dissipative part is self-adjoint with respect to it.
Using these results, we construct an orthonormal eigenbasis (Sec.~\ref{subsec:sm:gibbs_eigenbasis}), expand the two-time correlation functions in this basis (Sec.~\ref{subsec:sm:gibbs_correlations}), and finally prove the no-go theorem from the resulting spectral representation (Sec.~\ref{subsec:sm:gibbs_nogo}).

\subsection{Bound on the real parts of eigenvalues}
\label{subsec:gksl_identity_and_spectral_bound}
\label{subsec:gksl_spectral_bound}

To show that the eigenvalues of a Lindbladian have nonpositive real parts, we first prove the following lemma.
Lemmas~\ref{lem:sm:gksl_product_identity} and \ref{lem:sm:gksl_spectral_bound} below hold for any finite-dimensional Lindbladian of the form of Eqs.~\eqref{eq:sm:gksl_decomposition}--\eqref{eq:sm:single_jump_dissipator}.

\begin{lemma}[Carré du champ identity]
\label{lem:sm:gksl_product_identity}
For any $X_N,Y_N\in\mathfrak B_N$,
\begin{align}
 \mathcal L_N^\dagger[X_N^\dagger Y_N]
 -\bigl(\mathcal L_N^\dagger[X_N]\bigr)^\dagger Y_N
 -X_N^\dagger\mathcal L_N^\dagger[Y_N]
 &=\sum_k\gamma_{k,N}
 [L_{k,N},X_N]^\dagger[L_{k,N},Y_N].
 \label{eq:sm:gksl_product_identity}
\end{align}
The same identity holds with every $\mathcal L_N^\dagger$ on the left-hand side replaced by $\mathcal D_N^\dagger$.
In particular, for $Y_N=X_N$,
\begin{align}
 \mathcal L_N^\dagger[X_N^\dagger X_N]
 -\bigl(\mathcal L_N^\dagger[X_N]\bigr)^\dagger X_N
 -X_N^\dagger\mathcal L_N^\dagger[X_N]
 &=\sum_k\gamma_{k,N}
 [L_{k,N},X_N]^\dagger[L_{k,N},X_N]\geq0.
 \label{eq:sm:carre_du_champ_positive}
\end{align}
\end{lemma}

\noindent\textit{Proof of Lemma~\ref{lem:sm:gksl_product_identity}.---}
Expanding the commutator and using $\mathcal H_N^\dagger[X_N^\dagger]=\bigl(\mathcal H_N^\dagger[X_N]\bigr)^\dagger$, we find that the Hamiltonian part satisfies
\begin{align}
 \mathcal H_N^\dagger[X_N^\dagger Y_N]
 &=i[H_N,X_N^\dagger Y_N]\notag\\
 &=i[H_N,X_N^\dagger]Y_N+iX_N^\dagger[H_N,Y_N]\notag\\
 &=\bigl(\mathcal H_N^\dagger[X_N]\bigr)^\dagger Y_N
 +X_N^\dagger\mathcal H_N^\dagger[Y_N].
 \label{eq:sm:hamiltonian_product_rule}
\end{align}
Hence, the Hamiltonian part cancels out on the left-hand side of Eq.~\eqref{eq:sm:gksl_product_identity}.

For the dissipative part, it suffices to compute each jump term separately.
Expanding the definition~\eqref{eq:sm:single_jump_dissipator}, we obtain
\begin{align}
 &\mathcal D_{L_{k,N}}^\dagger[X_N^\dagger Y_N]
 -\bigl(\mathcal D_{L_{k,N}}^\dagger[X_N]\bigr)^\dagger Y_N
 -X_N^\dagger\mathcal D_{L_{k,N}}^\dagger[Y_N]
 =L_{k,N}^\dagger X_N^\dagger Y_NL_{k,N}\notag\\
 &\qquad
 -\frac12L_{k,N}^\dagger L_{k,N}X_N^\dagger Y_N
 -\frac12X_N^\dagger Y_NL_{k,N}^\dagger L_{k,N}\notag\\
 &\qquad
 -L_{k,N}^\dagger X_N^\dagger L_{k,N}Y_N
 +\frac12X_N^\dagger L_{k,N}^\dagger L_{k,N}Y_N
 +\frac12L_{k,N}^\dagger L_{k,N}X_N^\dagger Y_N\notag\\
 &\qquad
 -X_N^\dagger L_{k,N}^\dagger Y_NL_{k,N}
 +\frac12X_N^\dagger L_{k,N}^\dagger L_{k,N}Y_N
 +\frac12X_N^\dagger Y_NL_{k,N}^\dagger L_{k,N}\notag\\
 &\quad=
 L_{k,N}^\dagger X_N^\dagger Y_NL_{k,N}
 -L_{k,N}^\dagger X_N^\dagger L_{k,N}Y_N
 -X_N^\dagger L_{k,N}^\dagger Y_NL_{k,N}
 +X_N^\dagger L_{k,N}^\dagger L_{k,N}Y_N\notag\\
 &\quad=
 (X_N^\dagger L_{k,N}^\dagger-L_{k,N}^\dagger X_N^\dagger)
 (L_{k,N}Y_N-Y_NL_{k,N})
 =[L_{k,N},X_N]^\dagger[L_{k,N},Y_N].
 \label{eq:sm:single_jump_product_calculation}
\end{align}
Multiplying by $\gamma_{k,N}$, summing over $k$, and using Eq.~\eqref{eq:sm:hamiltonian_product_rule}, we obtain Eq.~\eqref{eq:sm:gksl_product_identity}.
For $Y_N=X_N$, the operator $[L_{k,N},X_N]^\dagger[L_{k,N},X_N]$ is positive semidefinite for each $k$, and $\gamma_{k,N}\geq0$, so Eq.~\eqref{eq:sm:carre_du_champ_positive} follows.\hfill\qed

\begin{lemma}[Spectral bound]
\label{lem:sm:gksl_spectral_bound}
Any eigenvalue $z$ of a finite-dimensional Lindbladian $\mathcal L_N^\dagger$ satisfies
\begin{align}
 \operatorname{Re}z\leq0.
 \label{eq:sm:gksl_spectral_bound}
\end{align}
\end{lemma}

\noindent\textit{Proof of Lemma~\ref{lem:sm:gksl_spectral_bound}.---}
Let $z$ be any eigenvalue of $\mathcal L_N^\dagger$, and let $R_N$ be a corresponding eigenoperator, so that $\mathcal L_N^\dagger[R_N]=zR_N$.
To obtain the real part of $z$, we left-multiply this eigenvalue equation by $R_N^\dagger$ and add its adjoint, obtaining
\begin{align}
 R_N^\dagger\mathcal L_N^\dagger[R_N]
 +\bigl(\mathcal L_N^\dagger[R_N]\bigr)^\dagger R_N
 &=(z+z^*)R_N^\dagger R_N
 =2\operatorname{Re}z\,R_N^\dagger R_N.
 \label{eq:sm:eigenvalue_real_part}
\end{align}
Substituting $X_N=R_N$ into Eq.~\eqref{eq:sm:carre_du_champ_positive} and using Eq.~\eqref{eq:sm:eigenvalue_real_part} gives
\begin{align}
 \mathcal L_N^\dagger[R_N^\dagger R_N]
 -2\operatorname{Re}z\,R_N^\dagger R_N
 &=\sum_k\gamma_{k,N}[L_{k,N},R_N]^\dagger[L_{k,N},R_N]
 \geq0.
 \label{eq:sm:eigenoperator_product_identity}
\end{align}
Let $m_N>0$ be the largest eigenvalue of $R_N^\dagger R_N$, and let $|\psi_N\rangle$ be a corresponding normalized eigenvector.
Then $R_N^\dagger R_N|\psi_N\rangle=m_N|\psi_N\rangle$ and $R_N^\dagger R_N\leq m_N I$.
Taking the expectation value of the above inequality in $|\psi_N\rangle$ and using Eqs.~\eqref{eq:sm:gksl_decomposition}--\eqref{eq:sm:single_jump_dissipator}, we obtain
\begin{align}
 2m_N\operatorname{Re}z
 &\leq\langle\psi_N|\mathcal L_N^\dagger[R_N^\dagger R_N]|\psi_N\rangle\notag\\
 &=i\langle\psi_N|
 \bigl(H_NR_N^\dagger R_N-R_N^\dagger R_NH_N\bigr)
 |\psi_N\rangle\notag\\
 &\quad+\sum_k\gamma_{k,N}
 \langle\psi_N|\Bigl(
 L_{k,N}^\dagger R_N^\dagger R_NL_{k,N}
 -\frac12L_{k,N}^\dagger L_{k,N}R_N^\dagger R_N
 -\frac12R_N^\dagger R_NL_{k,N}^\dagger L_{k,N}
 \Bigr)|\psi_N\rangle\notag\\
 &=i(m_N-m_N)\langle\psi_N|H_N|\psi_N\rangle\notag\\
 &\quad+\sum_k\gamma_{k,N}\Bigl(
 \langle\psi_N|L_{k,N}^\dagger R_N^\dagger R_NL_{k,N}|\psi_N\rangle
 -\frac{m_N+m_N}{2}
 \langle\psi_N|L_{k,N}^\dagger L_{k,N}|\psi_N\rangle
 \Bigr)\notag\\
 &=\sum_k\gamma_{k,N}
 \langle\psi_N|L_{k,N}^\dagger
 (R_N^\dagger R_N-m_N I)L_{k,N}|\psi_N\rangle
 \leq0.
 \label{eq:sm:maximal_identity_expectation}
\end{align}
In the second equality, we used $R_N^\dagger R_N|\psi_N\rangle=m_N|\psi_N\rangle$ and its adjoint, $\langle\psi_N|R_N^\dagger R_N=m_N\langle\psi_N|$.
The last inequality follows from $R_N^\dagger R_N\leq m_N I$ and $\gamma_{k,N}\geq0$.
Since $m_N>0$, we conclude that $\operatorname{Re}z\leq0$.\hfill\qed

\subsection{Gibbs-weighted inner product and self-adjointness of the dissipative part}
\label{subsec:gibbs_dissipative_self_adjointness}

For $0\leq\beta<\infty$, we show that the dissipative part of a Lindbladian satisfying NRSC is self-adjoint on the whole operator space $\mathfrak B_N$ with respect to the Gibbs-weighted inner product.
We write the Gibbs state as $\pi_N:=e^{-\beta H_N}/Z_N$, where $Z_N:=\Tr[e^{-\beta H_N}]$ is the partition function.

\begin{lemma}[Gibbs-weighted inner product]
\label{lem:sm:gibbs_inner_product}
For $X_N,Y_N\in\mathfrak B_N$, define
\begin{align}
 \langle X_N,Y_N\rangle_{\pi_N}
 &:=\Tr[\pi_NX_N^\dagger Y_N].
 \label{eq:sm:gibbs_inner_product}
\end{align}
This is an inner product on $\mathfrak B_N$.
\end{lemma}

\noindent\textit{Proof of Lemma~\ref{lem:sm:gibbs_inner_product}.---}
By the linearity of the trace and the conjugate linearity of the adjoint, $\langle\cdot,\cdot\rangle_{\pi_N}$ is antilinear in the first argument and linear in the second.
Since $\pi_N=\pi_N^\dagger$, the cyclicity of the trace gives
\begin{align}
 \langle X_N,Y_N\rangle_{\pi_N}^*
 &=\Tr[\pi_NX_N^\dagger Y_N]^*
 =\Tr[Y_N^\dagger X_N\pi_N]
 =\langle Y_N,X_N\rangle_{\pi_N}.
 \label{eq:sm:gibbs_inner_product_symmetry}
\end{align}
Moreover, $\pi_N>0$ for $0\leq\beta<\infty$, and
\begin{align}
 \langle X_N,X_N\rangle_{\pi_N}
 =\|X_N\pi_N^{1/2}\|_2^2,
 \label{eq:sm:gibbs_inner_product_positive}
\end{align}
where $\|X_N\|_2^2:=\Tr[X_N^\dagger X_N]$.
Hence, if the right-hand side vanishes, then $X_N\pi_N^{1/2}=0$.
Since $\pi_N^{1/2}$ is invertible, this implies $X_N=0$.
Therefore, $\langle\cdot,\cdot\rangle_{\pi_N}$ is an inner product on $\mathfrak B_N$.\hfill\qed

\begin{lemma}[Gibbs-weighted self-adjointness]
\label{lem:sm:gibbs_dissipative_self_adjointness}
Let $\mathcal L_N^\dagger$ satisfy NRSC at nonzero temperature.
Then its dissipative part $\mathcal D_N^\dagger$ satisfies
\begin{align}
 \langle\mathcal D_N^\dagger[X_N],Y_N\rangle_{\pi_N}
 =\langle X_N,\mathcal D_N^\dagger[Y_N]\rangle_{\pi_N}
 \label{eq:sm:gibbs_dissipative_self_adjointness}
\end{align}
for any $X_N,Y_N\in\mathfrak B_N$.
Moreover, the Gibbs state is stationary under the Hamiltonian and dissipative parts separately:
\begin{align}
 \mathcal H_N[\pi_N]=\mathcal D_N[\pi_N]=\mathcal L_N[\pi_N]=0.
 \label{eq:sm:gibbs_stationarity}
\end{align}
\end{lemma}

\noindent\textit{Proof of Lemma~\ref{lem:sm:gibbs_dissipative_self_adjointness}.---}
Let $X_N,Y_N\in\mathfrak B_N$ be arbitrary.
By the definitions of the inner product and the dissipative part,
\begin{align}
 \langle\mathcal D_N^\dagger[X_N],Y_N\rangle_{\pi_N}
 &=\Tr\!\left[\pi_N\bigl(\mathcal D_N^\dagger[X_N]\bigr)^\dagger Y_N\right]\notag\\
 &=\sum_k\gamma_{k,N}
 \Tr\!\left[\pi_N(L_{k,N}^\dagger X_NL_{k,N})^\dagger Y_N\right]\notag\\
 &\quad-\frac12\sum_k\gamma_{k,N}
 \Tr\!\left[\pi_N(L_{k,N}^\dagger L_{k,N}X_N)^\dagger Y_N\right]
 -\frac12\sum_k\gamma_{k,N}
 \Tr\!\left[\pi_N(X_NL_{k,N}^\dagger L_{k,N})^\dagger Y_N\right].
 \label{eq:sm:weighted_dissipator_expansion}
\end{align}
We evaluate the three terms on the right-hand side separately.

For the first term, we use the relation between the jump operators and the Gibbs state.
Since $[L_{k,N},H_N]=\omega_{k,N}L_{k,N}$, we have $L_{k,N}=\sum_E P_{E,N}L_{k,N}P_{E+\omega_{k,N},N}$.
Hence,
\begin{align}
 L_{k,N}\pi_N
 &=\frac{1}{Z_N}\sum_E e^{-\beta(E+\omega_{k,N})}
 P_{E,N}L_{k,N}P_{E+\omega_{k,N},N}\notag\\
 &=e^{-\beta\omega_{k,N}}\pi_NL_{k,N}.
 \label{eq:sm:jump_gibbs_relation}
\end{align}
Using its adjoint, $\pi_NL_{k,N}^\dagger=e^{-\beta\omega_{k,N}}L_{k,N}^\dagger\pi_N$, and the cyclicity of the trace, we obtain
\begin{align}
 \Tr[\pi_N(L_{k,N}^\dagger X_NL_{k,N})^\dagger Y_N]
 &=\Tr[\pi_NL_{k,N}^\dagger X_N^\dagger L_{k,N}Y_N]\notag\\
 &=\Tr[X_N^\dagger L_{k,N}Y_N\pi_NL_{k,N}^\dagger]\notag\\
 &=e^{-\beta\omega_{k,N}}
 \Tr[X_N^\dagger L_{k,N}Y_NL_{k,N}^\dagger\pi_N]\notag\\
 &=e^{-\beta\omega_{k,N}}
 \Tr[\pi_NX_N^\dagger L_{k,N}Y_NL_{k,N}^\dagger].
 \label{eq:sm:weighted_jump_term}
\end{align}
The detailed-balance relation $\gamma_{k,N}e^{-\beta\omega_{k,N}}=\gamma_{-k,N}$ and the reverse-jump relation $L_{-k,N}=L_{k,N}^\dagger$ then give
\begin{align}
 \sum_k\gamma_{k,N}
 \Tr[\pi_N(L_{k,N}^\dagger X_NL_{k,N})^\dagger Y_N]
 &=\sum_k\gamma_{-k,N}
 \Tr[\pi_NX_N^\dagger L_{k,N}Y_NL_{k,N}^\dagger]\notag\\
 &=\sum_k\gamma_{-k,N}
 \Tr[\pi_NX_N^\dagger L_{-k,N}^\dagger Y_NL_{-k,N}]\notag\\
 &=\sum_k\gamma_{k,N}
 \Tr[\pi_NX_N^\dagger L_{k,N}^\dagger Y_NL_{k,N}].
 \label{eq:sm:weighted_jump_sum}
\end{align}
Here the last equality follows by relabeling $k\mapsto-k$.

For the second term, taking the adjoint gives
\begin{align}
 \Tr[\pi_N(L_{k,N}^\dagger L_{k,N}X_N)^\dagger Y_N]
 &=\Tr[\pi_NX_N^\dagger L_{k,N}^\dagger L_{k,N}Y_N].
 \label{eq:sm:weighted_left_loss_term}
\end{align}

For the third term, we use $[\pi_N,L_{k,N}^\dagger L_{k,N}]=0$, which follows from $[H_N,L_{k,N}^\dagger L_{k,N}]=0$, and the cyclicity of the trace to obtain
\begin{align}
 \Tr[\pi_N(X_NL_{k,N}^\dagger L_{k,N})^\dagger Y_N]
 &=\Tr[\pi_NL_{k,N}^\dagger L_{k,N}X_N^\dagger Y_N]\notag\\
 &=\Tr[L_{k,N}^\dagger L_{k,N}\pi_NX_N^\dagger Y_N]\notag\\
 &=\Tr[\pi_NX_N^\dagger Y_NL_{k,N}^\dagger L_{k,N}].
 \label{eq:sm:weighted_right_loss_term}
\end{align}

Substituting these three results into Eq.~\eqref{eq:sm:weighted_dissipator_expansion} gives
\begin{align}
 \langle\mathcal D_N^\dagger[X_N],Y_N\rangle_{\pi_N}
 &=\sum_k\gamma_{k,N}
 \Tr\!\left[\pi_NX_N^\dagger
 \left(L_{k,N}^\dagger Y_NL_{k,N}
 -\frac12\{L_{k,N}^\dagger L_{k,N},Y_N\}\right)\right]\notag\\
 &=\langle X_N,\mathcal D_N^\dagger[Y_N]\rangle_{\pi_N}.
 \label{eq:sm:weighted_dissipator_calculation}
\end{align}
Thus, $\mathcal D_N^\dagger$ is self-adjoint on $\mathfrak B_N$.

We next show that the Gibbs state is stationary under the Hamiltonian and dissipative parts separately.
Substituting $I$ into the definition of the dissipative part gives
\begin{align}
 \mathcal D_N^\dagger[I]
 &=\sum_k\gamma_{k,N}\left(
 L_{k,N}^\dagger L_{k,N}
 -\frac12L_{k,N}^\dagger L_{k,N}
 -\frac12L_{k,N}^\dagger L_{k,N}
 \right)=0.
\end{align}
For any $X_N\in\mathfrak B_N$, duality and the self-adjointness shown above give
\begin{align}
 \Tr[\mathcal D_N[\pi_N]X_N]
 &=\Tr[\pi_N\mathcal D_N^\dagger[X_N]]
 =\langle I,\mathcal D_N^\dagger[X_N]\rangle_{\pi_N}\notag\\
 &=\langle\mathcal D_N^\dagger[I],X_N\rangle_{\pi_N}=0.
 \label{eq:sm:gibbs_stationarity_calculation}
\end{align}
Since this holds for every $X_N$, we have $\mathcal D_N[\pi_N]=0$.
Since $\pi_N$ is a function of $H_N$, $\mathcal H_N[\pi_N]=-i[H_N,\pi_N]=0$, and hence $\mathcal L_N[\pi_N]=0$.\hfill\qed
\begin{corollary}
\label{cor:v9:gibbs_dissipative_identities}
Under the assumptions of Lemma~\ref{lem:sm:gibbs_dissipative_self_adjointness}, for any $X_N\in\mathfrak B_N$,
\begin{align}
 \mathcal D_N[\pi_NX_N]=\pi_N\mathcal D_N^\dagger[X_N].
 \label{eq:v4:gibbs_intertwining}
\end{align}
\end{corollary}

\noindent\textit{Proof of Corollary~\ref{cor:v9:gibbs_dissipative_identities}.---}
For any $Y_N\in\mathfrak B_N$, the duality and Lemma~\ref{lem:sm:gibbs_dissipative_self_adjointness} give
\begin{align}
 \Tr[\mathcal D_N[\pi_NX_N]Y_N]
 &=\Tr[\pi_NX_N\mathcal D_N^\dagger[Y_N]]
 =\langle X_N^\dagger,\mathcal D_N^\dagger[Y_N]\rangle_{\pi_N}\notag\\
 &=\langle\mathcal D_N^\dagger[X_N^\dagger],Y_N\rangle_{\pi_N}
 =\Tr[\pi_N\mathcal D_N^\dagger[X_N]Y_N].
\end{align}
In the last equality, we used $\mathcal D_N^\dagger[X_N^\dagger]=\bigl(\mathcal D_N^\dagger[X_N]\bigr)^\dagger$.
Since this equality holds for any $Y_N$, Eq.~\eqref{eq:v4:gibbs_intertwining} follows.\hfill\qed

\subsection{Construction of an orthonormal eigenbasis}
\label{subsec:sm:gibbs_eigenbasis}

Using the self-adjointness of the dissipative part shown in the previous subsection, we construct an orthonormal eigenbasis of the whole operator space $\mathfrak B_N$ that respects the mode decomposition.

\begin{lemma}[Orthonormal eigenbasis for Gibbs states at nonzero temperature]
\label{lem:sm:gibbs_eigenbasis}
We can choose a basis $\{R_{N,\omega,a}\}_{\omega,a}$ of $\mathfrak B_N$ that is orthonormal with respect to the Gibbs-weighted inner product and satisfies
\begin{align}
 R_{N,\omega,a}&\in\mathfrak B_{N,\omega},\notag\\
 \langle R_{N,\omega,a},R_{N,\omega',a'}\rangle_{\pi_N}
 &=\delta_{\omega,\omega'}\delta_{a,a'},\notag\\
 \mathcal D_N^\dagger[R_{N,\omega,a}]
 &=-\lambda_{N,\omega,a}R_{N,\omega,a},\notag\\
 \mathcal L_N^\dagger[R_{N,\omega,a}]
 &=(-i\omega-\lambda_{N,\omega,a})R_{N,\omega,a},
 \qquad\lambda_{N,\omega,a}\geq0.
 \label{eq:sm:gibbs_eigenoperators}
\end{align}
Here, $a$ labels the basis elements within each $\mathfrak B_{N,\omega}$.
\end{lemma}

\noindent\textit{Proof of Lemma~\ref{lem:sm:gibbs_eigenbasis}.---}
We first show that different modes are orthogonal.
For $X_{N,\omega}\in\mathfrak B_{N,\omega}$ and $Y_{N,\omega'}\in\mathfrak B_{N,\omega'}$, Proposition~\ref{prop:rta:modes}(b) gives
$[H_N,X_{N,\omega}^\dagger Y_{N,\omega'}]=(\omega-\omega')X_{N,\omega}^\dagger Y_{N,\omega'}$.
Together with $[\pi_N,H_N]=0$ and the cyclicity of the trace, this gives
\begin{align}
 (\omega-\omega')\langle X_{N,\omega},Y_{N,\omega'}\rangle_{\pi_N}
 &=\Tr[\pi_N[H_N,X_{N,\omega}^\dagger Y_{N,\omega'}]]\notag\\
 &=\Tr[[\pi_N,H_N]X_{N,\omega}^\dagger Y_{N,\omega'}]=0.
 \label{eq:sm:gibbs_mode_orthogonality}
\end{align}
Hence, $\langle X_{N,\omega},Y_{N,\omega'}\rangle_{\pi_N}=0$ for $\omega\neq\omega'$, so the mode decomposition~\eqref{eq:sm:mode_direct_sum} is orthogonal with respect to the Gibbs-weighted inner product.

Next, we diagonalize the dissipative part within each mode.
By Eq.~\eqref{eq:sm:mode_invariance},
\begin{align}
 X_{N,\omega}\in\mathfrak B_{N,\omega}
 \quad\Longrightarrow\quad
 \mathcal D_N^\dagger[X_{N,\omega}]\in\mathfrak B_{N,\omega},
 \label{eq:sm:gibbs_dissipator_mode_invariance}
\end{align}
so $\mathcal D_N^\dagger$ restricts to a linear operator on each $\mathfrak B_{N,\omega}$.
Restricting Eq.~\eqref{eq:sm:gibbs_dissipative_self_adjointness} to $\mathfrak B_{N,\omega}$ gives
\begin{align}
 \langle\mathcal D_N^\dagger[X_{N,\omega}],Y_{N,\omega}\rangle_{\pi_N}
 =\langle X_{N,\omega},\mathcal D_N^\dagger[Y_{N,\omega}]\rangle_{\pi_N}
 \qquad(X_{N,\omega},Y_{N,\omega}\in\mathfrak B_{N,\omega}).
 \label{eq:sm:gibbs_mode_self_adjointness}
\end{align}
Thus, this restriction is self-adjoint with respect to the Gibbs-weighted inner product.
By the finite-dimensional spectral theorem, it therefore admits an orthonormal basis $\{R_{N,\omega,a}\}_a$ of eigenoperators with real eigenvalues.
Equation~\eqref{eq:sm:gibbs_mode_orthogonality}, together with the direct-sum decomposition~\eqref{eq:sm:mode_direct_sum}, then implies that $\{R_{N,\omega,a}\}_{\omega,a}$ is an orthonormal basis of $\mathfrak B_N$.

Let $-\lambda_{N,\omega,a}$ denote the real eigenvalue of $\mathcal D_N^\dagger$ corresponding to $R_{N,\omega,a}$.
Since $R_{N,\omega,a}\in\mathfrak B_{N,\omega}$, we have
$\mathcal H_N^\dagger[R_{N,\omega,a}]=-i\omega R_{N,\omega,a}$, and hence
\begin{align}
 \mathcal L_N^\dagger[R_{N,\omega,a}]
 =\mathcal H_N^\dagger[R_{N,\omega,a}]
 +\mathcal D_N^\dagger[R_{N,\omega,a}]
 =(-i\omega-\lambda_{N,\omega,a})R_{N,\omega,a}.
\end{align}
Applying Lemma~\ref{lem:sm:gksl_spectral_bound} to this eigenvalue gives
$-\lambda_{N,\omega,a}\leq0$, i.e., $\lambda_{N,\omega,a}\geq0$.\hfill\qed

\subsection{Spectral representation of the two-time correlation functions}
\label{subsec:sm:gibbs_correlations}

We use the eigenbasis constructed in the preceding subsection to derive the spectral representation presented in the End Matter for the two-time correlation functions of Gibbs states at nonzero temperatures.
Let $\{A_N\}_N\in\mathfrak A_{\mathrm{loc}}$ be a family of order parameters.
Since $A_N$ and $e^{t\mathcal L_N^\dagger}[A_N]$ are Hermitian, the two-time correlation functions can be written as
\begin{align}
 G_{N,\pi_N}^{\mathcal L^\dagger}(t;A_N)
 &=\frac{1}{N^2}\Tr[\pi_Ne^{t\mathcal L_N^\dagger}[A_N]A_N]
 =\frac{1}{N^2}\langle e^{t\mathcal L_N^\dagger}[A_N],A_N\rangle_{\pi_N},
 \label{eq:sm:gibbs_open_inner_product}\\
 G_{N,\pi_N}^{\mathcal H^\dagger}(t;A_N)
 &=\frac{1}{N^2}\langle e^{t\mathcal H_N^\dagger}[A_N],A_N\rangle_{\pi_N}.
 \label{eq:sm:gibbs_closed_inner_product}
\end{align}
We expand $A_N$ in the basis of Lemma~\ref{lem:sm:gibbs_eigenbasis} as
\begin{align}
 A_N=\sum_{\omega,a}c_{N,\omega,a}R_{N,\omega,a},
 \qquad c_{N,\omega,a}:=\langle R_{N,\omega,a},A_N\rangle_{\pi_N}.
 \label{eq:sm:gibbs_observable_expansion}
\end{align}
By Eq.~\eqref{eq:sm:gibbs_eigenoperators}, each eigenoperator evolves as
\begin{align}
 e^{t\mathcal L_N^\dagger}[R_{N,\omega,a}]
 &=e^{(-i\omega-\lambda_{N,\omega,a})t}R_{N,\omega,a},\notag\\
 e^{t\mathcal H_N^\dagger}[R_{N,\omega,a}]
 &=e^{-i\omega t}R_{N,\omega,a}.
 \label{eq:sm:gibbs_eigenoperator_evolution}
\end{align}
Using the antilinearity of the inner product in the first argument and the orthonormality of the basis, we obtain
\begin{align}
 G_{N,\pi_N}^{\mathcal L^\dagger}(t;A_N)
 &=\frac{1}{N^2}\sum_{\omega,a}\sum_{\omega',a'}
 c_{N,\omega,a}^*c_{N,\omega',a'}e^{(i\omega-\lambda_{N,\omega,a})t}
 \langle R_{N,\omega,a},R_{N,\omega',a'}\rangle_{\pi_N}\notag\\
 &=\frac{1}{N^2}\sum_{\omega,a}|c_{N,\omega,a}|^2e^{(i\omega-\lambda_{N,\omega,a})t}
 =\sum_{\omega,a}v_{N,\omega,a}e^{(i\omega-\lambda_{N,\omega,a})t},
 \label{eq:sm:gibbs_open_spectral}\\
 G_{N,\pi_N}^{\mathcal H^\dagger}(t;A_N)
 &=\sum_{\omega,a}v_{N,\omega,a}e^{i\omega t},
 \label{eq:sm:gibbs_closed_spectral}
\end{align}
where $v_{N,\omega,a}:=|c_{N,\omega,a}|^2/N^2\geq0$.
In particular, at $t=0$,
$G_{N,\pi_N}^{\mathcal L^\dagger}(0;A_N)=G_{N,\pi_N}^{\mathcal H^\dagger}(0;A_N)$.

\subsection{Proof of the no-go theorem}
\label{subsec:sm:gibbs_nogo}

Suppose that the open-system two-time correlation function converges to a periodic function in the thermodynamic limit. We show that its difference from the two-time correlation function under unitary evolution generated by \(H_N\), evaluated for the same family of states, vanishes in this limit.

\begin{proposition}
\label{prop:sm:gibbs_correlation_comparison}
Let $\{\mathcal L_N\}_N$ be a family of open systems satisfying NRSC at inverse temperature $0\leq\beta<\infty$, with Hamiltonians $\{H_N\}_N$, and let $\{\pi_N\}_N$ be the family of Gibbs states at inverse temperature $\beta$.
If, for some family of order parameters $\{A_N\}_N\in\mathfrak A_{\mathrm{loc}}$, the thermodynamic limit of $G_{N,\pi_N}^{\mathcal L^\dagger}(t;A_N)$ exists for all $t\geq0$ and is a periodic function, then
\begin{align}
 \lim_{N\to\infty}
 \left|
 G_{N,\pi_N}^{\mathcal L^\dagger}(t;A_N)
 -
 G_{N,\pi_N}^{\mathcal H^\dagger}(t;A_N)
 \right|
 =0
 \qquad(\forall t\geq0).
 \label{eq:sm:gibbs_correlation_comparison}
\end{align}
\end{proposition}

\noindent\textit{Proof of Proposition~\ref{prop:sm:gibbs_correlation_comparison}.---}
Let $g(t)$ be the thermodynamic limit of the open-system two-time correlation function, and fix a period $\tau>0$ of $g$.
For any fixed positive integer $n$, we have $g(n\tau)=g(0)$, and hence
\begin{align}
 \lim_{N\to\infty}
 \operatorname{Re}\!\left[
 G_{N,\pi_N}^{\mathcal L^\dagger}(0;A_N)
 -
 G_{N,\pi_N}^{\mathcal L^\dagger}(n\tau;A_N)
 \right]
 =0.
 \label{eq:sm:gibbs_periodic_return}
\end{align}

On the other hand, Eq.~\eqref{eq:sm:gibbs_open_spectral}, together with $v_{N,\omega,a}\geq0$, $\lambda_{N,\omega,a}\geq0$, and $\cos(\omega n\tau)\leq1$, gives
\begin{align}
 0
 &\leq
 \sum_{\omega,a}v_{N,\omega,a}
 \left(1-e^{-\lambda_{N,\omega,a}n\tau}\right)
 \notag\\
 &\leq
 \sum_{\omega,a}v_{N,\omega,a}
 \left[
 1-e^{-\lambda_{N,\omega,a}n\tau}\cos(\omega n\tau)
 \right]
 \notag\\
 &=
 \operatorname{Re}\!\left[
 G_{N,\pi_N}^{\mathcal L^\dagger}(0;A_N)
 -
 G_{N,\pi_N}^{\mathcal L^\dagger}(n\tau;A_N)
 \right].
\end{align}
Since the last expression converges to zero as $N\to\infty$ by Eq.~\eqref{eq:sm:gibbs_periodic_return}, we have
\begin{align}
 \lim_{N\to\infty}
 \sum_{\omega,a}v_{N,\omega,a}
 \left(1-e^{-\lambda_{N,\omega,a}n\tau}\right)
 =0
 \label{eq:sm:gibbs_decay_vanishes}
\end{align}
for any fixed positive integer $n$.

Next, fix any $t\geq0$ and choose a positive integer $n$, independent of $N$, such that $t\leq n\tau$.
From Eqs.~\eqref{eq:sm:gibbs_open_spectral} and \eqref{eq:sm:gibbs_closed_spectral},
\begin{align}
 \left|
 G_{N,\pi_N}^{\mathcal L^\dagger}(t;A_N)
 -
 G_{N,\pi_N}^{\mathcal H^\dagger}(t;A_N)
 \right|
 &=
 \left|
 \sum_{\omega,a}v_{N,\omega,a}e^{i\omega t}
 \left(e^{-\lambda_{N,\omega,a}t}-1\right)
 \right|\notag\\
 &\leq
 \sum_{\omega,a}v_{N,\omega,a}
 \left(1-e^{-\lambda_{N,\omega,a}t}\right)\notag\\
 &\leq
 \sum_{\omega,a}v_{N,\omega,a}
 \left(1-e^{-\lambda_{N,\omega,a}n\tau}\right).
\end{align}
Here the last inequality uses $\lambda_{N,\omega,a}\geq0$ and $t\leq n\tau$.
By Eq.~\eqref{eq:sm:gibbs_decay_vanishes}, the right-hand side converges to zero as $N\to\infty$.
Therefore, Eq.~\eqref{eq:sm:gibbs_correlation_comparison} follows.\hfill\qed

\noindent\textit{Proof of Theorem~3 for $0\leq\beta<\infty$.---}
We prove the contrapositive.
Fix $0\leq\beta<\infty$, and let $\{\mathcal L_N\}_N$ be a family of open systems satisfying NRSC, with the Gibbs states $\{\pi_N\}_N$ of their Hamiltonians $H_N$.
By Lemma~\ref{lem:sm:gibbs_dissipative_self_adjointness}, these Gibbs states are stationary under both the open-system dynamics and the isolated-system dynamics generated by the same Hamiltonian $H_N$.

Suppose that the open system exhibits a time crystal in $\{\pi_N\}_N$.
By the definition of time crystals, there exist a family of order parameters $\{A_N\}_N\in\mathfrak A_{\mathrm{loc}}$, a nonconstant function $g_{\mathrm{op}}(t)$, and a period $\tau>0$ such that
\begin{align}
 \lim_{N\to\infty}
 G_{N,\pi_N}^{\mathcal L^\dagger}(t;A_N)
 &=g_{\mathrm{op}}(t)
 \qquad(\forall t\geq0),
 \label{eq:sm:gibbs_proof_open_limit}\\
 g_{\mathrm{op}}(t+\tau)
 &=g_{\mathrm{op}}(t)
 \qquad(\forall t\geq0).
 \label{eq:sm:gibbs_proof_periodicity}
\end{align}

For this family of order parameters, we consider the thermodynamic limit of the isolated-system two-time correlation function.
For any fixed $t\geq0$, the triangle inequality gives
\begin{align}
 \left|
 G_{N,\pi_N}^{\mathcal H^\dagger}(t;A_N)
 -g_{\mathrm{op}}(t)
 \right|
 &\leq
 \left|
 G_{N,\pi_N}^{\mathcal H^\dagger}(t;A_N)
 -
 G_{N,\pi_N}^{\mathcal L^\dagger}(t;A_N)
 \right|
 +
 \left|
 G_{N,\pi_N}^{\mathcal L^\dagger}(t;A_N)
 -g_{\mathrm{op}}(t)
 \right|.
\end{align}
The first term on the right-hand side converges to zero as $N\to\infty$ by Proposition~\ref{prop:sm:gibbs_correlation_comparison}, while the second does so by Eq.~\eqref{eq:sm:gibbs_proof_open_limit}.
Hence, the thermodynamic limit of the isolated-system two-time correlation function also exists and
\begin{align}
 \lim_{N\to\infty}
 G_{N,\pi_N}^{\mathcal H^\dagger}(t;A_N)
 =
 g_{\mathrm{op}}(t)
 \qquad(\forall t\geq0).
 \label{eq:sm:gibbs_proof_closed_limit}
\end{align}
Since $g_{\mathrm{op}}(t)$ is a nonconstant periodic function, the same family of order parameters $\{A_N\}_N$ shows that the isolated system also exhibits a time crystal in the same family of Gibbs states $\{\pi_N\}_N$.
Therefore, if the isolated system does not exhibit a time crystal in this family of Gibbs states, neither does the open system.\hfill\qed

\section{Operator spaces associated with stationary states and induced superoperators}
\label{sec:v4:stationary_structure}

In the previous section, we expressed the two-time correlation functions in terms of the Gibbs-weighted inner product at nonzero temperature and proved the no-go theorem from their spectral representation.
In this section, we prepare for extending this argument to arbitrary stationary states of open systems satisfying NRSC.

In Secs.~\ref{subsec:v4:finite_temperature} and \ref{subsec:v4:zero_temperature}, we show the properties of stationary states used to construct the inner product and the induced superoperators, at nonzero temperatures and at absolute zero, respectively.
The subsequent subsections treat nonzero temperatures and absolute zero simultaneously.
In Sec.~\ref{subsec:v4:weighted_space}, we define an inner product weighted by the stationary state on a restricted operator space. We use it to express the two-time correlation functions for the open-system dynamics and the unitary evolution generated by \(H_N\). In Sec.~\ref{subsec:v9:common_induced_operators}, we construct induced superoperators describing these two dynamics on this restricted space.

\subsection{Properties of stationary states of open systems satisfying NRSC at nonzero temperature}
\label{subsec:v4:finite_temperature}

For \(0\leq\beta<\infty\), we show that any stationary state \(\rho_{\mathrm{ss},N}\) of an open system satisfying NRSC is also stationary under the unitary evolution generated by \(H_N\).

\begin{lemma}\label{lem:finite_stationary_properties}
For any stationary state $\rho_{\mathrm{ss},N}$ satisfying $\mathcal L_N[\rho_{\mathrm{ss},N}]=0$, define
\begin{align}
 S_N:=\pi_N^{-1}\rho_{\mathrm{ss},N}.
 \label{eq:v4:finite_S_definition}
\end{align}
Then $S_N$ satisfies
\begin{align}
 S_N&\geq0,\notag\\
 [H_N,S_N]&=[L_{k,N},S_N]=[L_{k,N}^\dagger,S_N]=0.
 \label{eq:v4:finite_S_properties}
\end{align}
Moreover, let $Q_N$ be the orthogonal projection onto the support
$\operatorname{supp}\rho_{\mathrm{ss},N}:=(\ker\rho_{\mathrm{ss},N})^\perp$.
Then
\begin{align}
 \mathcal H_N[\rho_{\mathrm{ss},N}]&=\mathcal D_N[\rho_{\mathrm{ss},N}]=0,\notag\\
 [Q_N,H_N]&=[Q_N,L_{k,N}]=[Q_N,L_{k,N}^\dagger]=0.
 \label{eq:finite_support_commutation}
\end{align}
\end{lemma}

\noindent\textit{Proof of Lemma~\ref{lem:finite_stationary_properties}.---}
First, we show that \(\rho_{\mathrm{ss},N}\) is stationary under the unitary evolution generated by \(H_N\), i.e., \([H_N,\rho_{\mathrm{ss},N}]=0\).
Since $\rho_{\mathrm{ss},N}=\pi_NS_N$ and $[H_N,\pi_N]=0$, we have
$[H_N,\rho_{\mathrm{ss},N}]=\pi_N[H_N,S_N]$, so it suffices to prove $[H_N,S_N]=0$.
Using Eq.~\eqref{eq:v4:gibbs_intertwining}, the stationarity condition
$\mathcal L_N[\rho_{\mathrm{ss},N}]=0$ can be rewritten in terms of $S_N$ as
\begin{align}
 0
 &=\mathcal L_N[\rho_{\mathrm{ss},N}]\notag\\
 &=-i[H_N,\pi_NS_N]+\mathcal D_N[\pi_NS_N]\notag\\
 &=\pi_N\bigl(\mathcal D_N^\dagger[S_N]-i[H_N,S_N]\bigr).
 \label{eq:v4:generator_intertwining}
\end{align}
Since $\pi_N$ is invertible, we obtain
$\mathcal D_N^\dagger[S_N]=i[H_N,S_N]$.

Using the mode decomposition of Sec.~\ref{sec:sm:asymmetry_modes}, we write
\begin{align}
 S_N=\sum_{\omega\in\mathcal{W}_N}S_{N,\omega},
 \qquad S_{N,\omega}\in\mathfrak B_{N,\omega}.
\end{align}
It therefore suffices to show that $S_{N,\omega}=0$ for every $\omega\neq0$.
Substituting this decomposition into
$\mathcal D_N^\dagger[S_N]=i[H_N,S_N]$ gives
\begin{align}
 \sum_{\omega\in\mathcal{W}_N}
 \bigl(\mathcal D_N^\dagger[S_{N,\omega}]+i\omega S_{N,\omega}\bigr)=0.
\end{align}
Since $\mathcal D_N^\dagger[S_{N,\omega}]\in\mathfrak B_{N,\omega}$ by Eq.~\eqref{eq:sm:mode_invariance}, the direct-sum decomposition implies
\begin{align}
 \mathcal D_N^\dagger[S_{N,\omega}]
 =i[H_N,S_{N,\omega}]
 =-i\omega S_{N,\omega}.
\end{align}
Since $\mathcal D_N^\dagger$ is self-adjoint with respect to the Gibbs-weighted inner product,
\begin{align}
 -i\omega\langle S_{N,\omega},S_{N,\omega}\rangle_{\pi_N}
 &=\langle S_{N,\omega},\mathcal D_N^\dagger[S_{N,\omega}]\rangle_{\pi_N}\notag\\
 &=\langle\mathcal D_N^\dagger[S_{N,\omega}],S_{N,\omega}\rangle_{\pi_N}\notag\\
 &=i\omega\langle S_{N,\omega},S_{N,\omega}\rangle_{\pi_N}.
\end{align}
Since this inner product is positive definite, $S_{N,\omega}=0$ for every $\omega\neq0$.
Hence, $S_N=S_{N,0}$, implying
$[H_N,S_N]=0$ and $\mathcal D_N^\dagger[S_N]=0$.
Together with $\mathcal L_N[\rho_{\mathrm{ss},N}]=0$, this gives
\begin{align}
 \mathcal H_N[\rho_{\mathrm{ss},N}]
 &=-i\pi_N[H_N,S_N]=0,\notag\\
 \mathcal D_N[\rho_{\mathrm{ss},N}]
 &=\mathcal L_N[\rho_{\mathrm{ss},N}]
 -\mathcal H_N[\rho_{\mathrm{ss},N}]=0.
\end{align}

Next, we show that $S_N\geq0$ and
$[L_{k,N},S_N]=[L_{k,N}^\dagger,S_N]=0$.
Since $[H_N,S_N]=0$ implies $[\pi_N,S_N]=0$, we have
\begin{align}
 S_N
 =\pi_N^{-1/2}(\pi_NS_N)\pi_N^{-1/2}
 =\pi_N^{-1/2}\rho_{\mathrm{ss},N}\pi_N^{-1/2}\geq0.
\end{align}

Applying Eq.~\eqref{eq:sm:carre_du_champ_positive} with
$\mathcal L_N^\dagger$ replaced by $\mathcal D_N^\dagger$ to $X_N=S_N$, and using
$\mathcal D_N^\dagger[S_N]=0$ and $S_N=S_N^\dagger$, we obtain
\begin{align}
 \mathcal D_N^\dagger[S_N^2]
 =\sum_k\gamma_{k,N}[L_{k,N},S_N]^\dagger[L_{k,N},S_N].
\end{align}
Taking the expectation value in the Gibbs state and using
$\mathcal D_N[\pi_N]=0$ gives
\begin{align}
 0
 &=\Tr[\mathcal D_N[\pi_N]S_N^2]\notag\\
 &=\Tr[\pi_N\mathcal D_N^\dagger[S_N^2]]\notag\\
 &=\sum_k\gamma_{k,N}\|[L_{k,N},S_N]\pi_N^{1/2}\|_2^2.
 \label{eq:v16:stationary_commutator_sum}
\end{align}
Since every $\gamma_{k,N}>0$ and $\pi_N^{1/2}$ is invertible, each term must vanish, and hence
$[L_{k,N},S_N]=0$.
Taking the adjoint and using $S_N=S_N^\dagger$ also gives
$[L_{k,N}^\dagger,S_N]=0$.

Finally, we show that $Q_N$ commutes with $H_N$, $L_{k,N}$, and $L_{k,N}^\dagger$.
Since $\rho_{\mathrm{ss},N}=\pi_NS_N$ and $\pi_N$ is invertible,
$\ker\rho_{\mathrm{ss},N}=\ker S_N$.
Thus, using the spectral decomposition of $S_N$, we write
\begin{align}
 S_N=\sum_{s\geq0}sP_{N,s},
 \qquad Q_N=\sum_{s>0}P_{N,s}.
\end{align}
Since $[H_N,S_N]=0$,
\begin{align}
 0=P_{N,s}[H_N,S_N]P_{N,s'}
 =(s'-s)P_{N,s}H_NP_{N,s'},
\end{align}
so $P_{N,s}H_NP_{N,s'}=0$ whenever $s\neq s'$.
Using $\sum_{s'}P_{N,s'}=I$, we obtain
\begin{align}
 H_NP_{N,s}
 &=\sum_{s'}P_{N,s'}H_NP_{N,s}
 =P_{N,s}H_NP_{N,s},\notag\\
 P_{N,s}H_N
 &=\sum_{s'}P_{N,s}H_NP_{N,s'}
 =P_{N,s}H_NP_{N,s}.
\end{align}
Thus, $H_NP_{N,s}=P_{N,s}H_N$, and summing over $s>0$ gives
$[Q_N,H_N]=0$.
The same argument, using
$[L_{k,N},S_N]=[L_{k,N}^\dagger,S_N]=0$, gives
$[Q_N,L_{k,N}]=[Q_N,L_{k,N}^\dagger]=0$.\hfill\qed

The following lemma establishes an identity used later.

\begin{lemma}
\label{lem:v9:finite_weighted_action}
For any stationary state $\rho_{\mathrm{ss},N}$ of an open system satisfying NRSC at nonzero temperature and any $X_N\in\mathfrak B_N$,
\begin{align}
 \mathcal D_N[X_N\rho_{\mathrm{ss},N}]
 =\mathcal D_N^\dagger[X_N]\rho_{\mathrm{ss},N}.
 \label{eq:v9:finite_weighted_action}
\end{align}
\end{lemma}

\noindent\textit{Proof of Lemma~\ref{lem:v9:finite_weighted_action}.---}
Applying Eq.~\eqref{eq:v4:gibbs_intertwining} to $X_N^\dagger$ and taking the adjoint gives
\begin{align}
 \mathcal D_N[X_N\pi_N]
 &=\bigl(\mathcal D_N[\pi_NX_N^\dagger]\bigr)^\dagger
 =\bigl(\pi_N\mathcal D_N^\dagger[X_N^\dagger]\bigr)^\dagger
 =\mathcal D_N^\dagger[X_N]\pi_N.
 \label{eq:v9:gibbs_right_action}
\end{align}
Next, Eq.~\eqref{eq:v4:finite_S_properties} implies that $S_N$ commutes with each
$L_{k,N}$ and $L_{k,N}^\dagger$, so
\begin{align}
 \mathcal D_N^\dagger[X_NS_N]
 &=\sum_k\gamma_{k,N}\left(
 L_{k,N}^\dagger X_NS_NL_{k,N}
 -\frac12L_{k,N}^\dagger L_{k,N}X_NS_N
 -\frac12X_NS_NL_{k,N}^\dagger L_{k,N}
 \right)\notag\\
 &=\sum_k\gamma_{k,N}\left(
 L_{k,N}^\dagger X_NL_{k,N}
 -\frac12L_{k,N}^\dagger L_{k,N}X_N
 -\frac12X_NL_{k,N}^\dagger L_{k,N}
 \right)S_N\notag\\
 &=\mathcal D_N^\dagger[X_N]S_N.
 \label{eq:finite_dissipative_right_multiplication}
\end{align}
Since $[H_N,S_N]=0$ implies $[\pi_N,S_N]=0$, we have
$\rho_{\mathrm{ss},N}=\pi_NS_N=S_N\pi_N$.
Therefore,
\begin{align}
 \mathcal D_N[X_N\rho_{\mathrm{ss},N}]
 &=\mathcal D_N[X_NS_N\pi_N]
 =\mathcal D_N^\dagger[X_NS_N]\pi_N\notag\\
 &=\mathcal D_N^\dagger[X_N]S_N\pi_N
 =\mathcal D_N^\dagger[X_N]\rho_{\mathrm{ss},N}.
\end{align}
\hfill\qed

\subsection{Properties of stationary states of open systems satisfying NRSC at absolute zero}
\label{subsec:v4:zero_temperature}

In this subsection, we consider the absolute-zero case.
Condition (iii) of NRSC is understood as the $\beta\to\infty$ limit in which the rates converge to finite values.
By condition (ii) of NRSC, a jump operator $L_{k,N}$ with $\omega_{k,N}>0$ lowers the energy by $\omega_{k,N}$, whereas the reverse jump operator, $L_{-k,N}=L_{k,N}^\dagger$, raises it by the same amount.
At nonzero temperature, condition (iii) gives
\begin{align}
 \gamma_{-k,N}=e^{-\beta\omega_{k,N}}\gamma_{k,N},
\end{align}
so $\gamma_{-k,N}\to0$ as $\beta\to\infty$.
Thus, the terms corresponding to energy-raising jump operators vanish from the dissipative part in the absolute-zero limit.
We denote the jump operators with $\omega=0$ by $L_{0,j,N}$, where $j$ distinguishes them.\footnote{By condition (iii), jump operators with $\omega=0$ occur in pairs $L$ and $L^\dagger$ with equal rates. Since $\mathcal D_L^\dagger+\mathcal D_{L^\dagger}^\dagger=\mathcal D_X^\dagger+\mathcal D_Y^\dagger$ with $X:=(L+L^\dagger)/\sqrt2$ and $Y:=i(L-L^\dagger)/\sqrt2$, each pair can be replaced by Hermitian jump operators without changing the dissipative part.}
Then, at absolute zero, the dissipative part can be written as
\begin{align}
 \mathcal D_N^\dagger
 &=\sum_k\gamma_{k,N}\mathcal D_{L_{k,N}}^\dagger
 +\sum_j\gamma_{0,j,N}\mathcal D_{L_{0,j,N}}^\dagger,\notag\\
 [L_{k,N},H_N]&=\omega_{k,N}L_{k,N}\quad(\omega_{k,N}>0),\notag\\
 L_{0,j,N}&=L_{0,j,N}^\dagger,\qquad[L_{0,j,N},H_N]=0,
 \label{eq:zero_generator}
\end{align}
where $\gamma_{k,N},\gamma_{0,j,N}>0$ by the convention in Sec.~\ref{subsec:mode_decomposition_properties}.

\subsubsection{Stationarity of the Gibbs state at absolute zero}
\label{ssubsec:v8:zero_gibbs_stationarity}

For each finite $N$, let $E_{\min,N}$ be the lowest eigenvalue of $H_N$ and $P_{\min,N}$ the projection onto the corresponding eigenspace.
Then the absolute-zero Gibbs state is
\begin{align}
 \pi_N
 &:=\lim_{\beta\to\infty}\frac{e^{-\beta H_N}}{\Tr[e^{-\beta H_N}]}\notag\\
 &=\lim_{\beta\to\infty}
 \frac{P_{\min,N}+\sum_{E>E_{\min,N}}e^{-\beta(E-E_{\min,N})}P_{E,N}}
 {\Tr P_{\min,N}+\sum_{E>E_{\min,N}}e^{-\beta(E-E_{\min,N})}\Tr P_{E,N}}\notag\\
 &=\frac{P_{\min,N}}{\Tr P_{\min,N}}.
 \label{eq:zero_gibbs_state}
\end{align}

\begin{lemma}
\label{lem:zero_gibbs_stationarity}
The absolute-zero Gibbs state $\pi_N$ given by Eq.~\eqref{eq:zero_gibbs_state} satisfies
\begin{align}
 \mathcal H_N[\pi_N]=\mathcal D_N[\pi_N]=\mathcal L_N[\pi_N]=0,
 \label{eq:zero_gibbs_stationarity}
\end{align}
and is therefore a stationary state of the Lindblad master equation.
\end{lemma}

\noindent\textit{Proof of Lemma~\ref{lem:zero_gibbs_stationarity}.---}
We first show that $L_{k,N}P_{\min,N}=0$.
For any $|\psi\rangle$ satisfying $H_N|\psi\rangle=E_{\min,N}|\psi\rangle$,
\begin{align}
 H_NL_{k,N}|\psi\rangle
 &=(L_{k,N}H_N-\omega_{k,N}L_{k,N})|\psi\rangle
 =(E_{\min,N}-\omega_{k,N})L_{k,N}|\psi\rangle.
\end{align}
Since $\omega_{k,N}>0$ and $E_{\min,N}$ is the lowest eigenvalue, we have $L_{k,N}P_{\min,N}=0$.
Using this relation and its adjoint, we find that the dissipative term associated with $L_{k,N}$ becomes
\begin{align}
 \mathcal D_{L_{k,N}}[\pi_N]
 &=L_{k,N}\pi_NL_{k,N}^\dagger-\frac12\{L_{k,N}^\dagger L_{k,N},\pi_N\}\notag\\
 &=\frac{1}{\Tr P_{\min,N}}\left(
 L_{k,N}P_{\min,N}L_{k,N}^\dagger
 -\frac12L_{k,N}^\dagger L_{k,N}P_{\min,N}
 -\frac12P_{\min,N}L_{k,N}^\dagger L_{k,N}
 \right)\notag\\
 &=0.
\end{align}
Moreover, $[L_{0,j,N},H_N]=0$ implies $[L_{0,j,N},P_{\min,N}]=0$, so the dissipative term associated with $L_{0,j,N}$ also vanishes:
\begin{align}
 \mathcal D_{L_{0,j,N}}[\pi_N]
 &=L_{0,j,N}\pi_NL_{0,j,N}-\frac12\{L_{0,j,N}^2,\pi_N\}\notag\\
 &=\frac{1}{\Tr P_{\min,N}}\left(
 L_{0,j,N}P_{\min,N}L_{0,j,N}
 -\frac12L_{0,j,N}^2P_{\min,N}
 -\frac12P_{\min,N}L_{0,j,N}^2
 \right)\notag\\
 &=\frac{1}{\Tr P_{\min,N}}\left(
 L_{0,j,N}^2P_{\min,N}
 -\frac12L_{0,j,N}^2P_{\min,N}
 -\frac12L_{0,j,N}^2P_{\min,N}
 \right)=0.
\end{align}
Together with $[H_N,P_{\min,N}]=0$, this gives
\begin{align}
 \mathcal H_N[\pi_N]&=-i[H_N,\pi_N]=0,\notag\\
 \mathcal D_N[\pi_N]
 &=\sum_k\gamma_{k,N}\mathcal D_{L_{k,N}}[\pi_N]
 +\sum_j\gamma_{0,j,N}\mathcal D_{L_{0,j,N}}[\pi_N]=0,\notag\\
 \mathcal L_N[\pi_N]&=\mathcal H_N[\pi_N]+\mathcal D_N[\pi_N]=0.
 \label{eq:zero_gibbs_stationarity_calculation}
\end{align}
\hfill\qed

\subsubsection{Properties of stationary states}
\label{ssubsec:v8:zero_stationary_properties}

Next, we show that any stationary state of an open system satisfying NRSC at absolute zero is also stationary under the unitary evolution generated by \(H_N\).
At absolute zero, the Gibbs state is not necessarily of full rank, so $S_N=\pi_N^{-1}\rho_{\mathrm{ss},N}$ used in the proof of Lemma~\ref{lem:finite_stationary_properties} cannot be defined in general.

\begin{lemma}\label{lem:zero_stationary_properties}
For any stationary state $\rho_{\mathrm{ss},N}$ satisfying $\mathcal L_N[\rho_{\mathrm{ss},N}]=0$, let $Q_N$ be the orthogonal projection onto its support.
Then
\begin{align}
 L_{k,N}Q_N&=0,\notag\\
 [L_{0,j,N},\rho_{\mathrm{ss},N}]&=[H_N,\rho_{\mathrm{ss},N}]=0,\notag\\
 \mathcal H_N[\rho_{\mathrm{ss},N}]&=\mathcal D_N[\rho_{\mathrm{ss},N}]=0,\notag\\
 [Q_N,H_N]&=[Q_N,L_{0,j,N}]=0.
 \label{eq:zero_stationary_properties}
\end{align}
\end{lemma}

\noindent\textit{Proof of Lemma~\ref{lem:zero_stationary_properties}.---}
We first show that $\mathcal D_{L_{k,N}}[\rho_{\mathrm{ss},N}]=0$.
By $[L_{k,N},H_N]=\omega_{k,N}L_{k,N}$ and $[H_N,L_{k,N}^\dagger L_{k,N}]=0$,
\begin{align}
 \mathcal D_{L_{k,N}}^\dagger[H_N]
 &=L_{k,N}^\dagger H_NL_{k,N}
 -L_{k,N}^\dagger L_{k,N}H_N\notag\\
 &=L_{k,N}^\dagger[H_N,L_{k,N}]
 =-\omega_{k,N}L_{k,N}^\dagger L_{k,N}.
\end{align}
On the other hand, $[L_{0,j,N},H_N]=0$ implies $\mathcal D_{L_{0,j,N}}^\dagger[H_N]=0$, and $\mathcal H_N^\dagger[H_N]=0$.
Hence, $\mathcal L_N[\rho_{\mathrm{ss},N}]=0$ gives
\begin{align}
 0
 &=\Tr[H_N\mathcal L_N[\rho_{\mathrm{ss},N}]]
 =\Tr[\rho_{\mathrm{ss},N}\mathcal L_N^\dagger[H_N]]\notag\\
 &=-\sum_k\gamma_{k,N}\omega_{k,N}
 \Tr[\rho_{\mathrm{ss},N}L_{k,N}^\dagger L_{k,N}]
 =-\sum_k\gamma_{k,N}\omega_{k,N}
 \left\|L_{k,N}\rho_{\mathrm{ss},N}^{1/2}\right\|_2^2.
 \label{eq:zero_stationary_energy_balance}
\end{align}
Since all $\gamma_{k,N}\omega_{k,N}$ are positive, $L_{k,N}\rho_{\mathrm{ss},N}^{1/2}=0$.
Hence, $L_{k,N}\rho_{\mathrm{ss},N}=\rho_{\mathrm{ss},N}L_{k,N}^\dagger=0$, and
\begin{align}
 \mathcal D_{L_{k,N}}[\rho_{\mathrm{ss},N}]
 &=L_{k,N}\rho_{\mathrm{ss},N}L_{k,N}^\dagger
 -\frac12\{L_{k,N}^\dagger L_{k,N},\rho_{\mathrm{ss},N}\}=0.
 \label{eq:zero_lowering_stationarity}
\end{align}
Moreover, letting $P_{N,r}$ be the projection onto the eigenspace of $\rho_{\mathrm{ss},N}$ with positive eigenvalue $r$, we have
\begin{align}
 0=L_{k,N}\rho_{\mathrm{ss},N}^{1/2}P_{N,r}
 =\sqrt r\,L_{k,N}P_{N,r}.
\end{align}
Since $r>0$, we have $L_{k,N}P_{N,r}=0$, and since $Q_N=\sum_{r>0}P_{N,r}$, we obtain $L_{k,N}Q_N=0$.

Next, we show that $[L_{0,j,N},\rho_{\mathrm{ss},N}]=0$.
Since $\mathcal D_{L_{k,N}}[\rho_{\mathrm{ss},N}]=0$, the condition $\mathcal L_N[\rho_{\mathrm{ss},N}]=0$ becomes
\begin{align}
 -i[H_N,\rho_{\mathrm{ss},N}]
 +\sum_j\gamma_{0,j,N}\mathcal D_{L_{0,j,N}}[\rho_{\mathrm{ss},N}]
 =0.
\end{align}
We multiply this equation by $\rho_{\mathrm{ss},N}$ from the left and take the trace.
Using the cyclicity of the trace and $L_{0,j,N}=L_{0,j,N}^\dagger$, we obtain
\begin{align}
 0
 &=-i\Tr[\rho_{\mathrm{ss},N}[H_N,\rho_{\mathrm{ss},N}]]
 +\sum_j\gamma_{0,j,N}
 \Tr[\rho_{\mathrm{ss},N}\mathcal D_{L_{0,j,N}}[\rho_{\mathrm{ss},N}]]\notag\\
 &=\sum_j\gamma_{0,j,N}\left(
 \Tr[\rho_{\mathrm{ss},N}L_{0,j,N}\rho_{\mathrm{ss},N}L_{0,j,N}]
 -\Tr[\rho_{\mathrm{ss},N}^2L_{0,j,N}^2]
 \right)\notag\\
 &=-\frac12\sum_j\gamma_{0,j,N}\Tr\!\left[
 (\rho_{\mathrm{ss},N}L_{0,j,N}-L_{0,j,N}\rho_{\mathrm{ss},N})
 (L_{0,j,N}\rho_{\mathrm{ss},N}-\rho_{\mathrm{ss},N}L_{0,j,N})
 \right]\notag\\
 &=-\frac12\sum_j\gamma_{0,j,N}
 \|[L_{0,j,N},\rho_{\mathrm{ss},N}]\|_2^2.
 \label{eq:zero_purity_dissipation}
\end{align}
Since all $\gamma_{0,j,N}$ are positive, $[L_{0,j,N},\rho_{\mathrm{ss},N}]=0$.
This commutation relation gives
\begin{align}
 \mathcal D_{L_{0,j,N}}[\rho_{\mathrm{ss},N}]
 &=L_{0,j,N}\rho_{\mathrm{ss},N}L_{0,j,N}
 -\frac12L_{0,j,N}^2\rho_{\mathrm{ss},N}
 -\frac12\rho_{\mathrm{ss},N}L_{0,j,N}^2=0.
 \label{eq:zero_zero_frequency_stationarity}
\end{align}
Hence, summing the dissipative terms and using $\mathcal L_N[\rho_{\mathrm{ss},N}]=0$, we obtain
\begin{align}
 \mathcal D_N[\rho_{\mathrm{ss},N}]&=0,\notag\\
 \mathcal H_N[\rho_{\mathrm{ss},N}]
 &=\mathcal L_N[\rho_{\mathrm{ss},N}]
 -\mathcal D_N[\rho_{\mathrm{ss},N}]
 =0.
\end{align}
That is, $[H_N,\rho_{\mathrm{ss},N}]=0$.

Finally, since $[H_N,\rho_{\mathrm{ss},N}]=[L_{0,j,N},\rho_{\mathrm{ss},N}]=0$, the same argument as in the proof of Lemma~\ref{lem:finite_stationary_properties} gives $[H_N,P_{N,r}]=[L_{0,j,N},P_{N,r}]=0$.
Since $Q_N=\sum_{r>0}P_{N,r}$, we obtain
\begin{align}
 [Q_N,H_N]=[Q_N,L_{0,j,N}]=0.
\end{align}
\hfill\qed

Next, we show that the identity established at nonzero temperature in Lemma~\ref{lem:v9:finite_weighted_action} also holds for any stationary state at absolute zero.

\begin{lemma}
\label{lem:v9:zero_weighted_action}
For any stationary state $\rho_{\mathrm{ss},N}$ of an open system satisfying NRSC at absolute zero and any $X_N\in\mathfrak B_N$,
\begin{align}
 \mathcal D_N[X_N\rho_{\mathrm{ss},N}]
 =\mathcal D_N^\dagger[X_N]\rho_{\mathrm{ss},N}.
 \label{eq:v9:zero_weighted_action}
\end{align}
\end{lemma}

\noindent\textit{Proof of Lemma~\ref{lem:v9:zero_weighted_action}.---}
We use $L_{k,N}\rho_{\mathrm{ss},N}=\rho_{\mathrm{ss},N}L_{k,N}^\dagger=0$, established in the proof of Lemma~\ref{lem:zero_stationary_properties}.
For the dissipative term associated with $L_{k,N}$, evaluating both sides separately gives
\begin{align}
 \mathcal D_{L_{k,N}}[X_N\rho_{\mathrm{ss},N}]
 &=L_{k,N}X_N\rho_{\mathrm{ss},N}L_{k,N}^\dagger
 -\frac12L_{k,N}^\dagger L_{k,N}X_N\rho_{\mathrm{ss},N}
 -\frac12X_N\rho_{\mathrm{ss},N}L_{k,N}^\dagger L_{k,N}\notag\\
 &=-\frac12L_{k,N}^\dagger L_{k,N}X_N\rho_{\mathrm{ss},N},
 \label{eq:v9:zero_lowering_schrodinger_action}\\
 \mathcal D_{L_{k,N}}^\dagger[X_N]\rho_{\mathrm{ss},N}
 &=L_{k,N}^\dagger X_NL_{k,N}\rho_{\mathrm{ss},N}
 -\frac12L_{k,N}^\dagger L_{k,N}X_N\rho_{\mathrm{ss},N}
 -\frac12X_NL_{k,N}^\dagger L_{k,N}\rho_{\mathrm{ss},N}\notag\\
 &=-\frac12L_{k,N}^\dagger L_{k,N}X_N\rho_{\mathrm{ss},N},
 \label{eq:v9:zero_lowering_heisenberg_action}
\end{align}
and hence the two expressions coincide.
For the dissipative term associated with $L_{0,j,N}$, using $L_{0,j,N}=L_{0,j,N}^\dagger$ and $[L_{0,j,N},\rho_{\mathrm{ss},N}]=0$, we also obtain
\begin{align}
 \mathcal D_{L_{0,j,N}}[X_N\rho_{\mathrm{ss},N}]
 &=L_{0,j,N}X_N\rho_{\mathrm{ss},N}L_{0,j,N}
 -\frac12L_{0,j,N}^2X_N\rho_{\mathrm{ss},N}
 -\frac12X_N\rho_{\mathrm{ss},N}L_{0,j,N}^2\notag\\
 &=\left(
 L_{0,j,N}X_NL_{0,j,N}
 -\frac12L_{0,j,N}^2X_N
 -\frac12X_NL_{0,j,N}^2
 \right)\rho_{\mathrm{ss},N}\notag\\
 &=\mathcal D_{L_{0,j,N}}^\dagger[X_N]\rho_{\mathrm{ss},N}.
 \label{eq:v9:zero_frequency_weighted_action}
\end{align}
Multiplying the identities for each $k$ and $j$ by $\gamma_{k,N}$ and $\gamma_{0,j,N}$, respectively, and summing, we obtain Eq.~\eqref{eq:v9:zero_weighted_action}.\hfill\qed

\subsection{Inner product weighted by the stationary state and two-time correlation functions}
\label{subsec:v4:weighted_space}

We now treat nonzero temperatures and absolute zero within a unified framework.
Let $\rho_{\mathrm{ss},N}$ be any stationary state of a Lindblad master equation satisfying NRSC, and let $Q_N$ be the orthogonal projection onto its support.
To derive a spectral representation of the two-time correlation functions, we follow the approach used for Gibbs states at nonzero temperature in Sec.~\ref{subsec:sm:gibbs_correlations} and seek to express them in terms of an inner product weighted by the stationary state.
However, if $\rho_{\mathrm{ss},N}$ is not full rank, the sesquilinear form $\Tr[\rho_{\mathrm{ss},N}X_N^\dagger Y_N]$ is not positive definite on $\mathfrak B_N$ and therefore does not define an inner product on that space.\footnote{Indeed, if $Q_N\neq I$, then $\Tr[\rho_{\mathrm{ss},N}(I-Q_N)^\dagger(I-Q_N)]=0$ even though $I-Q_N\neq0$, so positive definiteness fails.}
We therefore restrict the operator space to
\begin{align}
 \mathfrak K_N:=\mathfrak B_NQ_N=\{X_NQ_N\mid X_N\in\mathfrak B_N\}.
 \label{eq:v4:weighted_space}
\end{align}
Each $\xi_N\in\mathfrak K_N$ satisfies $\xi_NQ_N=\xi_N$ and hence vanishes on $\ker\rho_{\mathrm{ss},N}$.
The following lemma shows that this form defines an inner product on $\mathfrak K_N$.

\begin{lemma}\label{lem:support_inner_product}
On $\mathfrak K_N$, define
\begin{align}
 \langle\xi_N,\zeta_N\rangle_{\rho_{\mathrm{ss},N}}
 :=\Tr[\rho_{\mathrm{ss},N}\xi_N^\dagger\zeta_N]
 \qquad(\xi_N,\zeta_N\in\mathfrak K_N).
 \label{eq:v4:stationary_inner_product}
\end{align}
This is an inner product on $\mathfrak K_N$.
\end{lemma}

\noindent\textit{Proof of Lemma~\ref{lem:support_inner_product}.---}
By the linearity of the trace and the antilinearity of the adjoint, $\langle\cdot,\cdot\rangle_{\rho_{\mathrm{ss},N}}$ is antilinear in the first argument and linear in the second.
Moreover, the Hermiticity of $\rho_{\mathrm{ss},N}$ and the cyclicity of the trace give
$\langle\xi_N,\zeta_N\rangle_{\rho_{\mathrm{ss},N}}^*=\langle\zeta_N,\xi_N\rangle_{\rho_{\mathrm{ss},N}}$.

Finally, we show positive definiteness.
For any $\xi_N\in\mathfrak K_N$,
\begin{align}
 \langle \xi_N,\xi_N\rangle_{\rho_{\mathrm{ss},N}}
 &=\Tr[\rho_{\mathrm{ss},N}\xi_N^\dagger\xi_N]
 =\Tr\!\left[
 \bigl(\xi_N\rho_{\mathrm{ss},N}^{1/2}\bigr)^\dagger
 \bigl(\xi_N\rho_{\mathrm{ss},N}^{1/2}\bigr)
 \right]
 \geq0.
\end{align}
If this quantity vanishes, then $\xi_N\rho_{\mathrm{ss},N}^{1/2}=0$.
Let $P_{N,r}$ be the projection onto the eigenspace of $\rho_{\mathrm{ss},N}$ with positive eigenvalue $r$.
Then
\begin{align}
 0=\xi_N\rho_{\mathrm{ss},N}^{1/2}P_{N,r}
 =\sqrt r\,\xi_NP_{N,r}.
\end{align}
Since $r>0$, we have $\xi_NP_{N,r}=0$.
Since $Q_N=\sum_{r>0}P_{N,r}$ and $\xi_N\in\mathfrak K_N$,
\begin{align}
 \xi_N=\xi_NQ_N=\sum_{r>0}\xi_NP_{N,r}=0.
\end{align}
Therefore, Eq.~\eqref{eq:v4:stationary_inner_product} defines an inner product on $\mathfrak K_N$.\hfill\qed

With this inner product, we express the two-time correlation functions for an order parameter $A_N$.
Using the Hermiticity of $A_N$ and $e^{t\mathcal L_N^\dagger}[A_N]$, $Q_N\rho_{\mathrm{ss},N}=\rho_{\mathrm{ss},N}Q_N=\rho_{\mathrm{ss},N}$, and the cyclicity of the trace, we have, for any $t\geq0$,
\begin{align}
 G_{N,\rho_{\mathrm{ss},N}}^{\mathcal L^\dagger}(t;A_N)
 &=\frac{1}{N^2}\Tr[\rho_{\mathrm{ss},N}e^{t\mathcal L_N^\dagger}[A_N]A_N]\notag\\
 &=\frac{1}{N^2}\Tr[\rho_{\mathrm{ss},N}Q_Ne^{t\mathcal L_N^\dagger}[A_N]A_NQ_N]\notag\\
 &=\frac{1}{N^2}\Tr[\rho_{\mathrm{ss},N}(e^{t\mathcal L_N^\dagger}[A_N]Q_N)^\dagger A_NQ_N]\notag\\
 &=\frac{1}{N^2}\langle e^{t\mathcal L_N^\dagger}[A_N]Q_N,A_NQ_N\rangle_{\rho_{\mathrm{ss},N}}.
 \label{eq:v6:open_projected_inner_product}
\end{align}
Similarly, \(e^{t\mathcal H_N^\dagger}[A_N]\) is Hermitian, so the correlation function for the unitary evolution generated by \(H_N\) satisfies:
\begin{align}
 G_{N,\rho_{\mathrm{ss},N}}^{\mathcal H^\dagger}(t;A_N)
 &=\frac{1}{N^2}\Tr[\rho_{\mathrm{ss},N}e^{t\mathcal H_N^\dagger}[A_N]A_N]\notag\\
 &=\frac{1}{N^2}\Tr[\rho_{\mathrm{ss},N}(e^{t\mathcal H_N^\dagger}[A_N]Q_N)^\dagger A_NQ_N]\notag\\
 &=\frac{1}{N^2}\langle e^{t\mathcal H_N^\dagger}[A_N]Q_N,A_NQ_N\rangle_{\rho_{\mathrm{ss},N}}.
 \label{eq:v6:closed_projected_inner_product}
\end{align}
Thus, the two-time correlation functions for the open-system dynamics and the unitary evolution generated by \(H_N\) can both be expressed as inner products on \(\mathfrak K_N\).

\subsection{Induced superoperators for the Hamiltonian and dissipative parts}
\label{subsec:v9:common_induced_operators}

Next, to derive the spectral representation of the correlation functions as in the Gibbs-state case, we express $e^{t\mathcal L_N^\dagger}[A_N]Q_N$ as the time evolution of $A_NQ_N$ under a superoperator on $\mathfrak K_N$.
We define the action on $X_NQ_N$ by applying $\mathcal H_N^\dagger$ or $\mathcal D_N^\dagger$ to $X_N$ and then multiplying the result by $Q_N$ from the right.
Specifically, we define the induced superoperators for the Hamiltonian part, the dissipative part, and the whole Lindbladian by
\begin{align}
 \widetilde{\mathcal H}_N^\dagger[X_NQ_N]
 &:=\mathcal H_N^\dagger[X_N]Q_N,\notag\\
 \widetilde{\mathcal D}_N^\dagger[X_NQ_N]
 &:=\mathcal D_N^\dagger[X_N]Q_N,\notag\\
 \widetilde{\mathcal{L}}_{N}^\dagger[X_NQ_N]
 &:=\mathcal L_N^\dagger[X_N]Q_N.
 \label{eq:v9:induced_definition}
\end{align}
The following lemma shows that these induced superoperators are well defined.

\begin{lemma}
\label{lem:v9:induced_well_defined}
The superoperators $\widetilde{\mathcal H}_N^\dagger$, $\widetilde{\mathcal D}_N^\dagger$, and $\widetilde{\mathcal L}_N^\dagger$ in Eq.~\eqref{eq:v9:induced_definition} are well defined as linear superoperators on $\mathfrak K_N$, independently of the choice of the representative $X_N$.
\end{lemma}

\noindent\textit{Proof of Lemma~\ref{lem:v9:induced_well_defined}.---}
Let $X_N,Y_N\in\mathfrak B_N$ satisfy $X_NQ_N=Y_NQ_N$.
By Lemmas~\ref{lem:finite_stationary_properties} and \ref{lem:zero_stationary_properties}, $[H_N,Q_N]=0$ at both nonzero temperature and absolute zero.
Hence, for the Hamiltonian part,
\begin{align}
 \mathcal H_N^\dagger[X_N]Q_N
 &=i[H_N,X_N]Q_N\notag\\
 &=i[H_N,X_NQ_N]\notag\\
 &=i[H_N,Y_NQ_N]\notag\\
 &=\mathcal H_N^\dagger[Y_N]Q_N.
 \label{eq:v9:hamiltonian_representative_independence}
\end{align}

For the dissipative part, $X_NQ_N=Y_NQ_N$ and $Q_N\rho_{\mathrm{ss},N}=\rho_{\mathrm{ss},N}$ imply $X_N\rho_{\mathrm{ss},N}=Y_N\rho_{\mathrm{ss},N}$.
Using Lemmas~\ref{lem:v9:finite_weighted_action} and \ref{lem:v9:zero_weighted_action}, we obtain
\begin{align}
 \mathcal D_N^\dagger[X_N]\rho_{\mathrm{ss},N}
 &=\mathcal D_N[X_N\rho_{\mathrm{ss},N}]\notag\\
 &=\mathcal D_N[Y_N\rho_{\mathrm{ss},N}]\notag\\
 &=\mathcal D_N^\dagger[Y_N]\rho_{\mathrm{ss},N}.
 \label{eq:v9:dissipative_weighted_representative_independence}
\end{align}
Let $P_{N,r}$ be the projection onto the eigenspace of $\rho_{\mathrm{ss},N}$ with positive eigenvalue $r$.
Multiplying the above equation by $P_{N,r}$ from the right gives
\begin{align}
 r\,\mathcal D_N^\dagger[X_N]P_{N,r}
 =r\,\mathcal D_N^\dagger[Y_N]P_{N,r}.
\end{align}
Dividing by $r>0$ and summing over $r$, using $Q_N=\sum_{r>0}P_{N,r}$, we obtain
\begin{align}
 \mathcal D_N^\dagger[X_N]Q_N
 =\mathcal D_N^\dagger[Y_N]Q_N.
 \label{eq:v9:dissipative_representative_independence}
\end{align}
Thus, neither the Hamiltonian nor the dissipative part depends on the choice of representative.
Moreover, since $\mathcal L_N^\dagger=\mathcal H_N^\dagger+\mathcal D_N^\dagger$, we also have $\mathcal L_N^\dagger[X_N]Q_N=\mathcal L_N^\dagger[Y_N]Q_N$.
Each right-hand side of Eq.~\eqref{eq:v9:induced_definition} belongs to $\mathfrak K_N$, and the linearity of $\mathcal H_N^\dagger$, $\mathcal D_N^\dagger$, and $\mathcal L_N^\dagger$ implies the linearity of the induced superoperators.\hfill\qed

By definition, $\widetilde{\mathcal L}_N^\dagger=\widetilde{\mathcal H}_N^\dagger+\widetilde{\mathcal D}_N^\dagger$.
Next, we show the correspondence between the time evolution generated by the induced superoperators and the original one.

\begin{lemma}[Time evolution generated by the induced superoperators]\label{lem:v4:induced_dynamics}
For any $X_N\in\mathfrak B_N$ and $t\geq0$,
\begin{align}
 e^{t\widetilde{\mathcal H}_N^\dagger}[X_NQ_N]
 &=e^{t\mathcal H_N^\dagger}[X_N]Q_N,\notag\\
 e^{t\widetilde{\mathcal L}_N^\dagger}[X_NQ_N]
 &=e^{t\mathcal L_N^\dagger}[X_N]Q_N.
 \label{eq:v4:induced_evolution}
\end{align}
\end{lemma}

\noindent\textit{Proof of Lemma~\ref{lem:v4:induced_dynamics}.---}
By the definition of the induced superoperators, induction gives, for any nonnegative integer $m$,
\begin{align}
 (\widetilde{\mathcal L}_N^\dagger)^m[X_NQ_N]
 =(\mathcal L_N^\dagger)^m[X_N]Q_N.
 \label{eq:v5:induced_lindbladian_powers}
\end{align}
Substituting this into the exponential series, we obtain
\begin{align}
 e^{t\widetilde{\mathcal L}_N^\dagger}[X_NQ_N]
 &=\sum_{m=0}^{\infty}\frac{t^m}{m!}(\widetilde{\mathcal L}_N^\dagger)^m[X_NQ_N]\notag\\
 &=\left(\sum_{m=0}^{\infty}\frac{t^m}{m!}(\mathcal L_N^\dagger)^m[X_N]\right)Q_N\notag\\
 &=e^{t\mathcal L_N^\dagger}[X_N]Q_N.
\end{align}
The same holds for the Hamiltonian part.\hfill\qed

\section{Spectral representation for arbitrary stationary states and proof of the no-go theorem}
\label{sec:v4:spectral_nogo}

Using the results of the previous section, we extend the proof of the no-go theorem for Gibbs states at nonzero temperature to arbitrary stationary states at arbitrary temperatures.

\subsection{Mode decomposition and construction of an orthonormal eigenbasis}
\label{subsec:v4:common_spectral}

\subsubsection{Mode decomposition of the operator space \texorpdfstring{$\mathfrak K_N$}{K\_N} and the induced superoperators}
\label{ssubsec:v8:support_modes}

In Sec.~\ref{sec:resource_theory_of_asymmetry}, we showed the mode decomposition of the whole operator space $\mathfrak B_N$.
Here, we consider the analogous decomposition of the operator space $\mathfrak K_N$ defined in Sec.~\ref{subsec:v4:weighted_space}.
We also show that different modes are orthogonal with respect to the inner product weighted by $\rho_{\mathrm{ss},N}$.
For each $\omega\in\mathcal{W}_N$, we define
\begin{align}
 \mathfrak K_{N,\omega}:=\mathfrak K_N\cap\mathfrak B_{N,\omega}.
 \label{eq:v4:stationary_mode_space}
\end{align}

\begin{lemma}[Mode decomposition of the operator space]
\label{lem:v9:stationary_mode_decomposition}
With respect to the inner product weighted by $\rho_{\mathrm{ss},N}$, the operator space $\mathfrak K_N$ decomposes into the orthogonal direct sum
\begin{align}
 \mathfrak K_N=\bigoplus_{\omega\in\mathcal{W}_N}\mathfrak K_{N,\omega}.
 \label{eq:v4:orthogonal_modes}
\end{align}
\end{lemma}

\noindent\textit{Proof of Lemma~\ref{lem:v9:stationary_mode_decomposition}.---}
By Lemmas~\ref{lem:finite_stationary_properties} and \ref{lem:zero_stationary_properties}, $[H_N,\rho_{\mathrm{ss},N}]=[H_N,Q_N]=0$ at both nonzero temperature and absolute zero.
Since $[H_N,Q_N]=0$, for any eigenvalue $E$ of $H_N$ and any corresponding eigenvector $|\psi_N\rangle$,
\begin{align}
 H_N|\psi_N\rangle=E|\psi_N\rangle
 \quad\Longrightarrow\quad
 H_NQ_N|\psi_N\rangle=Q_NH_N|\psi_N\rangle=EQ_N|\psi_N\rangle.
 \label{eq:sm:stationary_projection_eigenvector}
\end{align}
Hence, $Q_N$ also commutes with each spectral projection $P_{E,N}$.
Using this, the mode components of $\xi_N=X_NQ_N\in\mathfrak K_N$ are
\begin{align}
 \xi_{N,\omega}
 &:=\sum_EP_{E,N}\xi_NP_{E+\omega,N}
 =\sum_EP_{E,N}X_NQ_NP_{E+\omega,N}\notag\\
 &=\sum_EP_{E,N}X_NP_{E+\omega,N}Q_N
 =X_{N,\omega}Q_N,
 \label{eq:v4:support_mode}
\end{align}
which belong to $\mathfrak K_N$.
Moreover, since $[H_N,\xi_{N,\omega}]=-\omega\xi_{N,\omega}$ by (a) of Proposition~\ref{prop:rta:modes}, we have $\xi_{N,\omega}\in\mathfrak B_{N,\omega}$.
Hence, each $\xi_{N,\omega}$ belongs to $\mathfrak K_{N,\omega}$, and by (a) of the same proposition, $\xi_N$ is decomposed as
\begin{align}
 \xi_N=\sum_{\omega\in\mathcal{W}_N}\xi_{N,\omega}.
\end{align}
Since each $\mathfrak K_{N,\omega}$ is a subspace of $\mathfrak B_{N,\omega}$, the uniqueness of this decomposition also follows from (a) of the same proposition.
This gives the direct-sum decomposition~\eqref{eq:v4:orthogonal_modes}.

Next, we show that different modes are orthogonal with respect to the inner product weighted by $\rho_{\mathrm{ss},N}$.
For $\xi_{N,\omega}\in\mathfrak K_{N,\omega}$ and $\zeta_{N,\omega'}\in\mathfrak K_{N,\omega'}$, expanding the commutator gives
\begin{align}
 [H_N,\xi_{N,\omega}^\dagger\zeta_{N,\omega'}]
 &=[H_N,\xi_{N,\omega}^\dagger]\zeta_{N,\omega'}
 +\xi_{N,\omega}^\dagger[H_N,\zeta_{N,\omega'}]\notag\\
 &=(\omega-\omega')\xi_{N,\omega}^\dagger\zeta_{N,\omega'}.
\end{align}
This equation and the cyclicity of the trace yield
\begin{align}
 (\omega-\omega')\langle\xi_{N,\omega},\zeta_{N,\omega'}\rangle_{\rho_{\mathrm{ss},N}}
 &=\Tr[\rho_{\mathrm{ss},N}[H_N,\xi_{N,\omega}^\dagger\zeta_{N,\omega'}]]\notag\\
 &=\Tr[\rho_{\mathrm{ss},N}H_N\xi_{N,\omega}^\dagger\zeta_{N,\omega'}]
 -\Tr[\rho_{\mathrm{ss},N}\xi_{N,\omega}^\dagger\zeta_{N,\omega'}H_N]\notag\\
 &=\Tr[(\rho_{\mathrm{ss},N}H_N-H_N\rho_{\mathrm{ss},N})\xi_{N,\omega}^\dagger\zeta_{N,\omega'}]
 =0.
\end{align}
In the last equality, we used $[H_N,\rho_{\mathrm{ss},N}]=0$.
Hence, for $\omega\neq\omega'$,
\begin{align}
 \langle\xi_{N,\omega},\zeta_{N,\omega'}\rangle_{\rho_{\mathrm{ss},N}}=0,
 \label{eq:v4:mode_orthogonality}
\end{align}
and thus Eq.~\eqref{eq:v4:orthogonal_modes} is an orthogonal direct-sum decomposition with respect to this inner product.\hfill\qed

Lemma~\ref{lem:v9:stationary_mode_decomposition} gives the mode decomposition of the operator space $\mathfrak K_N$.
To treat the time evolution mode by mode using this decomposition, we show that the induced superoperators also decompose into direct sums over the modes.
We extend the results shown in Sec.~\ref{sec:resource_theory_of_asymmetry} for $\mathcal H_N^\dagger$, $\mathcal D_N^\dagger$, and $\mathcal L_N^\dagger$ on the full operator space $\mathfrak B_N$ to the induced superoperators on $\mathfrak K_N$.
\begin{lemma}[Mode decomposition of the induced superoperators]
\label{lem:v9:induced_mode_decomposition}
For any $\omega\in\mathcal{W}_N$,
\begin{align}
 \xi_{N,\omega}\in\mathfrak K_{N,\omega}
 \quad\Longrightarrow\quad
 \widetilde{\mathcal H}_N^\dagger[\xi_{N,\omega}],\;
 \widetilde{\mathcal D}_N^\dagger[\xi_{N,\omega}],\;
 \widetilde{\mathcal L}_N^\dagger[\xi_{N,\omega}]
 \in\mathfrak K_{N,\omega}.
 \label{eq:v9:stationary_mode_preservation}
\end{align}
Moreover, with respect to the orthogonal direct-sum decomposition~\eqref{eq:v4:orthogonal_modes},
\begin{align}
 \begin{gathered}[b]
 \widetilde{\mathcal L}_N^\dagger
 =\bigoplus_{\omega\in\mathcal{W}_N}
 \left.\widetilde{\mathcal L}_N^\dagger\right|_{\mathfrak K_{N,\omega}},\\[0.6em]
 \widetilde{\mathcal H}_N^\dagger
 =\bigoplus_{\omega\in\mathcal{W}_N}
 \left.\widetilde{\mathcal H}_N^\dagger\right|_{\mathfrak K_{N,\omega}},
 \qquad
 \widetilde{\mathcal D}_N^\dagger
 =\bigoplus_{\omega\in\mathcal{W}_N}
 \left.\widetilde{\mathcal D}_N^\dagger\right|_{\mathfrak K_{N,\omega}}.
 \end{gathered}
 \label{eq:v9:induced_mode_decomposition}
\end{align}
\end{lemma}

\noindent\textit{Proof of Lemma~\ref{lem:v9:induced_mode_decomposition}.---}
Let $\xi_{N,\omega}\in\mathfrak K_{N,\omega}$.
By Eq.~\eqref{eq:v4:support_mode}, we can write $\xi_{N,\omega}=X_{N,\omega}Q_N$ with some $X_{N,\omega}\in\mathfrak B_{N,\omega}$.
Equation~\eqref{eq:sm:mode_invariance} gives $\mathcal D_N^\dagger[X_{N,\omega}]\in\mathfrak B_{N,\omega}$, and $[H_N,Q_N]=0$ gives $Q_N\in\mathfrak B_{N,0}$.
Hence, the definition of the induced superoperators and (b) of Proposition~\ref{prop:rta:modes} give
\begin{align}
 \widetilde{\mathcal D}_N^\dagger[\xi_{N,\omega}]
 =\mathcal D_N^\dagger[X_{N,\omega}]Q_N
 \in\mathfrak B_{N,\omega}\cap\mathfrak K_N
 =\mathfrak K_{N,\omega}.
 \label{eq:sm:induced_dissipator_mode}
\end{align}
By the definition of the induced superoperators and $[H_N,Q_N]=0$, we have $\widetilde{\mathcal H}_N^\dagger[\xi_N]=i[H_N,\xi_N]$ for any $\xi_N\in\mathfrak K_N$.
Thus,
\begin{align}
 \widetilde{\mathcal H}_N^\dagger[\xi_{N,\omega}]
 =i[H_N,\xi_{N,\omega}]
 =-i\omega\xi_{N,\omega}.
 \label{eq:sm:induced_hamiltonian_mode}
\end{align}
Moreover, since $\widetilde{\mathcal L}_N^\dagger=\widetilde{\mathcal H}_N^\dagger+\widetilde{\mathcal D}_N^\dagger$, Eq.~\eqref{eq:v9:stationary_mode_preservation} holds.

Using the decomposition $\xi_N=\sum_{\omega\in\mathcal{W}_N}\xi_{N,\omega}$ given by Lemma~\ref{lem:v9:stationary_mode_decomposition}, we have
\begin{align}
 \widetilde{\mathcal D}_N^\dagger[\xi_N]
 &=\widetilde{\mathcal D}_N^\dagger
 \left[\sum_{\omega\in\mathcal{W}_N}\xi_{N,\omega}\right]
 =\sum_{\omega\in\mathcal{W}_N}
 \widetilde{\mathcal D}_N^\dagger[\xi_{N,\omega}].
\end{align}
Since each term belongs to the corresponding $\mathfrak K_{N,\omega}$, we obtain the direct-sum decomposition of $\widetilde{\mathcal D}_N^\dagger$.
Similarly, Eq.~\eqref{eq:v9:induced_mode_decomposition} holds for $\widetilde{\mathcal H}_N^\dagger$ and $\widetilde{\mathcal L}_N^\dagger$.\hfill\qed

\subsubsection{Self-adjointness and eigenvalues of the induced superoperator for the dissipative part}
\label{ssubsec:v9:induced_self_adjointness}

To choose an orthonormal basis of eigenoperators of $\widetilde{\mathcal D}_N^\dagger$ within each mode, we show that the induced dissipative part is self-adjoint with respect to the inner product weighted by $\rho_{\mathrm{ss},N}$.
\begin{lemma}
\label{lem:v9:induced_dissipative_symmetry}
For any $\xi_N,\zeta_N\in\mathfrak K_N$,
\begin{align}
 \langle\widetilde{\mathcal D}_N^\dagger[\xi_N],\zeta_N\rangle_{\rho_{\mathrm{ss},N}}
 &=\langle\xi_N,\widetilde{\mathcal D}_N^\dagger[\zeta_N]\rangle_{\rho_{\mathrm{ss},N}}.
 \label{eq:v9:induced_self_adjointness}
\end{align}
\end{lemma}

\noindent\textit{Proof of Lemma~\ref{lem:v9:induced_dissipative_symmetry}.---}
Let $\xi_N,\zeta_N\in\mathfrak K_N$.
Using the definitions of the induced superoperators and the inner product, $Q_N\rho_{\mathrm{ss},N}=\rho_{\mathrm{ss},N}Q_N=\rho_{\mathrm{ss},N}$, and duality, we obtain
\begin{align}
 \langle\widetilde{\mathcal D}_N^\dagger[\xi_N],\zeta_N\rangle_{\rho_{\mathrm{ss},N}}
 &=\Tr\!\left[\rho_{\mathrm{ss},N}
 \bigl(\mathcal D_N^\dagger[\xi_N]Q_N\bigr)^\dagger\zeta_N\right]
 =\Tr\!\left[\rho_{\mathrm{ss},N}Q_N
 \bigl(\mathcal D_N^\dagger[\xi_N]\bigr)^\dagger\zeta_N\right]\notag\\
 &=\Tr\!\left[\bigl(\mathcal D_N^\dagger[\xi_N]\bigr)^\dagger
 \zeta_N\rho_{\mathrm{ss},N}\right]
 =\Tr\!\left[\xi_N^\dagger
 \mathcal D_N[\zeta_N\rho_{\mathrm{ss},N}]\right]\notag\\
 &=\Tr\!\left[\xi_N^\dagger
 \mathcal D_N^\dagger[\zeta_N]\rho_{\mathrm{ss},N}\right]\notag\\
 &=\Tr\!\left[\rho_{\mathrm{ss},N}\xi_N^\dagger
 \mathcal D_N^\dagger[\zeta_N]Q_N\right]
 =\langle\xi_N,\widetilde{\mathcal D}_N^\dagger[\zeta_N]\rangle_{\rho_{\mathrm{ss},N}}.
 \label{eq:v9:induced_symmetry_calculation}
\end{align}
In the fourth equality, we used $\mathcal D_N^\dagger[\xi_N^\dagger]=\bigl(\mathcal D_N^\dagger[\xi_N]\bigr)^\dagger$ and duality.
In the fifth equality, we used Lemmas~\ref{lem:v9:finite_weighted_action} and \ref{lem:v9:zero_weighted_action}.
Hence, $\widetilde{\mathcal D}_N^\dagger$ is self-adjoint with respect to the inner product weighted by $\rho_{\mathrm{ss},N}$.\hfill\qed

Next, we show that the eigenvalues of the induced dissipative part are nonpositive.

\begin{lemma}
\label{lem:v16:induced_spectral_bound}
Every eigenvalue of $\widetilde{\mathcal D}_N^\dagger$ is real and nonpositive.
\end{lemma}

\noindent\textit{Proof of Lemma~\ref{lem:v16:induced_spectral_bound}.---}
By the self-adjointness in Lemma~\ref{lem:v9:induced_dissipative_symmetry}, every eigenvalue $z$ of $\widetilde{\mathcal D}_N^\dagger$ is real.
Taking an eigenoperator $\xi_N\in\mathfrak K_N$ with $\widetilde{\mathcal D}_N^\dagger[\xi_N]=z\xi_N$, we have
\begin{align}
 2z\langle\xi_N,\xi_N\rangle_{\rho_{\mathrm{ss},N}}
 &=2\operatorname{Re}\Tr\!\left[
 \rho_{\mathrm{ss},N}\xi_N^\dagger\mathcal D_N^\dagger[\xi_N]
 \right]\notag\\
 &\leq\Tr\!\left[
 \rho_{\mathrm{ss},N}\mathcal D_N^\dagger[\xi_N^\dagger\xi_N]
 \right]\notag\\
 &=\Tr\!\left[
 \mathcal D_N[\rho_{\mathrm{ss},N}]\xi_N^\dagger\xi_N
 \right]=0.
 \label{eq:v16:induced_eigenvalue_bound}
\end{align}
In the first equality, we used the definition of the induced superoperators and $Q_N\rho_{\mathrm{ss},N}=\rho_{\mathrm{ss},N}$.
In the inequality, we used Eq.~\eqref{eq:sm:carre_du_champ_positive}.
In the last equality, we used duality and $\mathcal D_N[\rho_{\mathrm{ss},N}]=0$, shown in Lemmas~\ref{lem:finite_stationary_properties} and \ref{lem:zero_stationary_properties}.
Since the inner product on $\mathfrak K_N$ is positive definite, $\langle\xi_N,\xi_N\rangle_{\rho_{\mathrm{ss},N}}>0$, and hence $z\leq0$.\hfill\qed

\subsubsection{Construction of an orthonormal eigenbasis}
\label{ssubsec:v8:orthonormal_eigenbasis}

\begin{lemma}\label{lem:v4:common_spectral}
We can choose a basis $\{R_{N,\omega,a}\}_{\omega,a}$ of $\mathfrak K_N$ that is orthonormal with respect to $\langle\cdot,\cdot\rangle_{\rho_{\mathrm{ss},N}}$ and satisfies
\begin{align}
 R_{N,\omega,a}&\in\mathfrak K_{N,\omega},\notag\\
 \langle R_{N,\omega,a},R_{N,\omega',a'}\rangle_{\rho_{\mathrm{ss},N}}
 &=\delta_{\omega,\omega'}\delta_{a,a'},\notag\\
 \widetilde{\mathcal D}_N^\dagger[R_{N,\omega,a}]
 &=-\lambda_{N,\omega,a}R_{N,\omega,a},\notag\\
 \widetilde{\mathcal L}_N^\dagger[R_{N,\omega,a}]
 &=(-i\omega-\lambda_{N,\omega,a})R_{N,\omega,a},
 \qquad\lambda_{N,\omega,a}\geq0.
 \label{eq:v4:common_eigenoperators}
\end{align}
Here, $a$ labels the basis elements within each $\mathfrak K_{N,\omega}$.
\end{lemma}

\noindent\textit{Proof of Lemma~\ref{lem:v4:common_spectral}.---}
By Lemma~\ref{lem:v9:induced_dissipative_symmetry}, $\widetilde{\mathcal D}_N^\dagger$ is self-adjoint with respect to the inner product weighted by $\rho_{\mathrm{ss},N}$.
Moreover, by Lemma~\ref{lem:v9:induced_mode_decomposition},
\begin{align}
 \xi_{N,\omega}\in\mathfrak K_{N,\omega}
 \quad\Longrightarrow\quad
 \widetilde{\mathcal D}_N^\dagger[\xi_{N,\omega}]
 \in\mathfrak K_{N,\omega},
 \label{eq:sm:induced_dissipator_mode_invariance}
\end{align}
so $\widetilde{\mathcal D}_N^\dagger$ restricts to a linear operator on each $\mathfrak K_{N,\omega}$.
Furthermore, applying Eq.~\eqref{eq:v9:induced_self_adjointness} to elements of $\mathfrak K_{N,\omega}$ gives
\begin{align}
 \langle\widetilde{\mathcal D}_N^\dagger[\xi_{N,\omega}],\zeta_{N,\omega}\rangle_{\rho_{\mathrm{ss},N}}
 =\langle\xi_{N,\omega},\widetilde{\mathcal D}_N^\dagger[\zeta_{N,\omega}]\rangle_{\rho_{\mathrm{ss},N}}
 \qquad(\xi_{N,\omega},\zeta_{N,\omega}\in\mathfrak K_{N,\omega}).
 \label{eq:sm:induced_mode_self_adjointness}
\end{align}
That is, this restriction is self-adjoint on the finite-dimensional inner product space $\mathfrak K_{N,\omega}$ equipped with the inner product weighted by $\rho_{\mathrm{ss},N}$.
Hence, by the finite-dimensional spectral theorem, all its eigenvalues are real, and it admits an orthonormal basis $\{R_{N,\omega,a}\}_a$ of $\mathfrak K_{N,\omega}$ consisting of eigenoperators of $\widetilde{\mathcal D}_N^\dagger$.
Writing the corresponding eigenvalues as $-\lambda_{N,\omega,a}$, we have $\lambda_{N,\omega,a}\geq0$ by Lemma~\ref{lem:v16:induced_spectral_bound}.
By Eq.~\eqref{eq:v4:mode_orthogonality}, basis elements in different modes are orthogonal, and by Eq.~\eqref{eq:v4:orthogonal_modes}, the bases of all modes together span $\mathfrak K_N$.
Thus, their union over all $\omega$ forms an orthonormal basis $\{R_{N,\omega,a}\}_{\omega,a}$ of $\mathfrak K_N$.
Moreover, since $R_{N,\omega,a}\in\mathfrak K_{N,\omega}\subset\mathfrak B_{N,\omega}$, we have
\begin{align}
 \widetilde{\mathcal H}_N^\dagger[R_{N,\omega,a}]
 =-i\omega R_{N,\omega,a}.
\end{align}
Therefore,
\begin{align}
 \widetilde{\mathcal L}_N^\dagger[R_{N,\omega,a}]
 &=
 \widetilde{\mathcal H}_N^\dagger[R_{N,\omega,a}]
 +\widetilde{\mathcal D}_N^\dagger[R_{N,\omega,a}]\notag\\
 &=(-i\omega-\lambda_{N,\omega,a})R_{N,\omega,a},
\end{align}
which proves the lemma.\hfill\qed

\subsection{Spectral representation of the two-time correlation functions}
\label{subsec:v4:common_correlations}

Let $\{A_N\}_N\in\mathfrak A_{\mathrm{loc}}$ be a family of order parameters, and define $\Xi_N:=A_NQ_N\in\mathfrak K_N$.
Lemma~\ref{lem:v4:induced_dynamics} with $X_N=A_N$ gives
\begin{align}
 e^{t\widetilde{\mathcal L}_N^\dagger}[\Xi_N]&=e^{t\mathcal L_N^\dagger}[A_N]Q_N,\notag\\
 e^{t\widetilde{\mathcal H}_N^\dagger}[\Xi_N]&=e^{t\mathcal H_N^\dagger}[A_N]Q_N.
 \label{eq:v4:projected_evolution}
\end{align}
Substituting Eq.~\eqref{eq:v4:projected_evolution} into Eqs.~\eqref{eq:v6:open_projected_inner_product} and \eqref{eq:v6:closed_projected_inner_product}, we obtain
\begin{align}
 G_{N,\rho_{\mathrm{ss},N}}^{\mathcal L^\dagger}(t;A_N)
 &=\frac{1}{N^2}\langle e^{t\widetilde{\mathcal L}_N^\dagger}[\Xi_N],\Xi_N\rangle_{\rho_{\mathrm{ss},N}},
 \label{eq:v4:open_inner_product}\\
 G_{N,\rho_{\mathrm{ss},N}}^{\mathcal H^\dagger}(t;A_N)
 &=\frac{1}{N^2}\langle e^{t\widetilde{\mathcal H}_N^\dagger}[\Xi_N],\Xi_N\rangle_{\rho_{\mathrm{ss},N}}.
 \label{eq:v4:closed_inner_product}
\end{align}

We expand $\Xi_N$ in the basis of Lemma~\ref{lem:v4:common_spectral} as
\begin{align}
 \Xi_N&=\sum_{\omega,a}c_{N,\omega,a}R_{N,\omega,a},\notag\\
 c_{N,\omega,a}&:=\langle R_{N,\omega,a},\Xi_N\rangle_{\rho_{\mathrm{ss},N}},
 \qquad v_{N,\omega,a}:=\frac{|c_{N,\omega,a}|^2}{N^2}\geq0.
 \label{eq:v4:coefficients}
\end{align}
From here on, it suffices to replace $\pi_N$ with $\rho_{\mathrm{ss},N}$, $A_N$ with $\Xi_N$, and $\mathcal L_N^\dagger$ and $\mathcal H_N^\dagger$ with their induced superoperators in the basis-expansion calculation of Sec.~\ref{subsec:sm:gibbs_correlations}.
Using the eigenvalues in Lemma~\ref{lem:v4:common_spectral}, we obtain
\begin{align}
 G_{N,\rho_{\mathrm{ss},N}}^{\mathcal L^\dagger}(t;A_N)
 &=\sum_{\omega,a}v_{N,\omega,a}e^{(i\omega-\lambda_{N,\omega,a})t},
 \label{eq:v4:open_spectral}\\
 G_{N,\rho_{\mathrm{ss},N}}^{\mathcal H^\dagger}(t;A_N)
 &=\sum_{\omega,a}v_{N,\omega,a}e^{i\omega t}.
 \label{eq:v4:closed_spectral}
\end{align}
In particular, at $t=0$,
\begin{align}
 G_{N,\rho_{\mathrm{ss},N}}^{\mathcal L^\dagger}(0;A_N)
 =G_{N,\rho_{\mathrm{ss},N}}^{\mathcal H^\dagger}(0;A_N)
\end{align}
holds for each finite $N$.
We have thus obtained, for any stationary state that is not necessarily full rank, a spectral representation of the same form as that for Gibbs states at nonzero temperature in Sec.~\ref{subsec:sm:gibbs_correlations}.

\subsection{Proof of the no-go theorem}
\label{subsec:v4:common_nogo}

Suppose that the open-system two-time correlation function converges to a periodic function in the thermodynamic limit. We show that its difference from the two-time correlation function under unitary evolution generated by \(H_N\), evaluated for the same family of states, vanishes in this limit.

\begin{proposition}
\label{prop:spectral_reduction_to_isolated}
Let $\{\mathcal L_N\}_N$ be a family of open systems satisfying NRSC at inverse temperature $0\leq\beta\leq\infty$, with Hamiltonians $\{H_N\}_N$, and let $\{\rho_{\mathrm{ss},N}\}_N$ be any family of stationary states of these open systems.
If, for some family of order parameters $\{A_N\}_N\in\mathfrak A_{\mathrm{loc}}$, the thermodynamic limit of $G_{N,\rho_{\mathrm{ss},N}}^{\mathcal L^\dagger}(t;A_N)$ exists for all $t\geq0$ and is a periodic function, then
\begin{align}
 \lim_{N\to\infty}
 \left|
 G_{N,\rho_{\mathrm{ss},N}}^{\mathcal L^\dagger}(t;A_N)
 -
 G_{N,\rho_{\mathrm{ss},N}}^{\mathcal H^\dagger}(t;A_N)
 \right|
 =0
 \qquad(\forall t\geq0).
 \label{eq:common_correlation_reduction}
\end{align}
\end{proposition}

\noindent\textit{Proof of Proposition~\ref{prop:spectral_reduction_to_isolated}.---}
Let $g(t)$ be the thermodynamic limit of the open-system two-time correlation function, and fix a period $\tau>0$ of $g$.
For any fixed positive integer $n$, we have $g(n\tau)=g(0)$, and hence
\begin{align}
 \lim_{N\to\infty}
 \operatorname{Re}\!\left[
 G_{N,\rho_{\mathrm{ss},N}}^{\mathcal L^\dagger}(0;A_N)
 -
 G_{N,\rho_{\mathrm{ss},N}}^{\mathcal L^\dagger}(n\tau;A_N)
 \right]
 =0.
 \label{eq:finite_stationary_periodic_return}
\end{align}

On the other hand, Eq.~\eqref{eq:v4:open_spectral}, together with $v_{N,\omega,a}\geq0$, $\lambda_{N,\omega,a}\geq0$, and $\cos(\omega n\tau)\leq1$, gives
\begin{align}
 0
 &\leq
 \sum_{\omega,a}v_{N,\omega,a}
 \left(1-e^{-\lambda_{N,\omega,a}n\tau}\right)
 \notag\\
 &\leq
 \sum_{\omega,a}v_{N,\omega,a}
 \left[
 1-e^{-\lambda_{N,\omega,a}n\tau}\cos(\omega n\tau)
 \right]
 \notag\\
 &=
 \operatorname{Re}\!\left[
 G_{N,\rho_{\mathrm{ss},N}}^{\mathcal L^\dagger}(0;A_N)
 -
 G_{N,\rho_{\mathrm{ss},N}}^{\mathcal L^\dagger}(n\tau;A_N)
 \right].
\end{align}
Since the last expression converges to zero as $N\to\infty$ by Eq.~\eqref{eq:finite_stationary_periodic_return}, we have
\begin{align}
 \lim_{N\to\infty}
 \sum_{\omega,a}v_{N,\omega,a}
 \left(1-e^{-\lambda_{N,\omega,a}n\tau}\right)
 =0
 \label{eq:finite_stationary_decay_vanishes}
\end{align}
for any fixed positive integer $n$.

Next, fix any $t\geq0$ and choose a positive integer $n$, independent of $N$, such that $t\leq n\tau$.
From Eqs.~\eqref{eq:v4:open_spectral} and \eqref{eq:v4:closed_spectral},
\begin{align}
 \left|
 G_{N,\rho_{\mathrm{ss},N}}^{\mathcal L^\dagger}(t;A_N)
 -
 G_{N,\rho_{\mathrm{ss},N}}^{\mathcal H^\dagger}(t;A_N)
 \right|
 &=
 \left|
 \sum_{\omega,a}v_{N,\omega,a}e^{i\omega t}
 \left(e^{-\lambda_{N,\omega,a}t}-1\right)
 \right|\notag\\
 &\leq
 \sum_{\omega,a}v_{N,\omega,a}
 \left(1-e^{-\lambda_{N,\omega,a}t}\right)\notag\\
 &\leq
 \sum_{\omega,a}v_{N,\omega,a}
 \left(1-e^{-\lambda_{N,\omega,a}n\tau}\right).
\end{align}
Here the last inequality uses $\lambda_{N,\omega,a}\geq0$ and $t\leq n\tau$.
By Eq.~\eqref{eq:finite_stationary_decay_vanishes}, the right-hand side converges to zero as $N\to\infty$.
Therefore, Eq.~\eqref{eq:common_correlation_reduction} follows.\hfill\qed

\noindent\textit{Proof of Theorem~4.---}
We prove the contrapositive.
Fix $0\leq\beta\leq\infty$, and let $\{\rho_{\mathrm{ss},N}\}_N$ be any family of stationary states of a family of open systems $\{\mathcal L_N\}_N$ satisfying NRSC.
By Lemmas~\ref{lem:finite_stationary_properties} and \ref{lem:zero_stationary_properties}, this family of states is also stationary under the isolated-system dynamics generated by the Hamiltonians $\{H_N\}_N$ of the open systems.

Suppose that the open system exhibits a time crystal in this family of stationary states.
By the definition of time crystals, there exist a family of order parameters $\{A_N\}_N\in\mathfrak A_{\mathrm{loc}}$, a nonconstant function $g_{\mathrm{op}}(t)$, and a period $\tau>0$ such that
\begin{align}
 \lim_{N\to\infty}
 G_{N,\rho_{\mathrm{ss},N}}^{\mathcal L^\dagger}(t;A_N)
 &=g_{\mathrm{op}}(t)
 \qquad(\forall t\geq0),
 \label{eq:finite_stationary_proof_open_limit}\\
 g_{\mathrm{op}}(t+\tau)
 &=g_{\mathrm{op}}(t)
 \qquad(\forall t\geq0).
 \label{eq:finite_stationary_proof_periodicity}
\end{align}

For this family of order parameters, we consider the thermodynamic limit of the isolated-system two-time correlation function.
For any fixed $t\geq0$, the triangle inequality gives
\begin{align}
 \left|
 G_{N,\rho_{\mathrm{ss},N}}^{\mathcal H^\dagger}(t;A_N)
 -g_{\mathrm{op}}(t)
 \right|
 &\leq
 \left|
 G_{N,\rho_{\mathrm{ss},N}}^{\mathcal H^\dagger}(t;A_N)
 -
 G_{N,\rho_{\mathrm{ss},N}}^{\mathcal L^\dagger}(t;A_N)
 \right|
 +
 \left|
 G_{N,\rho_{\mathrm{ss},N}}^{\mathcal L^\dagger}(t;A_N)
 -g_{\mathrm{op}}(t)
 \right|.
\end{align}
The first term on the right-hand side converges to zero as $N\to\infty$ by Proposition~\ref{prop:spectral_reduction_to_isolated}, while the second does so by Eq.~\eqref{eq:finite_stationary_proof_open_limit}.
Hence, the thermodynamic limit of the isolated-system two-time correlation function also exists and
\begin{align}
 \lim_{N\to\infty}
 G_{N,\rho_{\mathrm{ss},N}}^{\mathcal H^\dagger}(t;A_N)
 =
 g_{\mathrm{op}}(t)
 \qquad(\forall t\geq0).
 \label{eq:finite_stationary_proof_closed_limit}
\end{align}
Since $g_{\mathrm{op}}(t)$ is a nonconstant periodic function, the same family of order parameters $\{A_N\}_N$ shows that the isolated system also exhibits a time crystal in the same family of stationary states $\{\rho_{\mathrm{ss},N}\}_N$.
Therefore, if the isolated system does not exhibit a time crystal in this family of stationary states, neither does the open system.\hfill\qed

Finally, by Lemma~\ref{lem:zero_gibbs_stationarity}, the Gibbs state $\pi_N$ at $\beta=\infty$ is a stationary state of the open system.
Hence, setting $\rho_{\mathrm{ss},N}=\pi_N$ in the above result, we also obtain the case $\beta=\infty$ of Theorem~3 in the main text.

\section{Derivation of Lindbladians satisfying NRSC}
\label{sec:nrsc_lindbladian}

In this section, we derive Lindbladians satisfying NRSC in two ways: from the Born--Markov--secular approximation and from the continuous-time limit of a collision model.
We take $0<\beta<\infty$ throughout this section; as in the main text, $\beta=0,\infty$ are understood as the $\beta\to0,\infty$ limits.
As in the main text, for a system with Hamiltonian $H$, we quantify the thermodynamic resource of a state $\rho$ by its nonequilibrium free-energy difference from the Gibbs state $\pi_\beta:=e^{-\beta H}/\Tr[e^{-\beta H}]$~\cite{Esposito2011second,Brandao2013resource,Horodecki2013fundamental},
\begin{align}
 \Delta F:=F_\beta(\rho)-F_\beta(\pi_\beta)=\beta^{-1}D(\rho\|\pi_\beta)\geq0,
 \label{eq:nrsc:free_energy}
\end{align}
where $F_\beta(\rho):=\Tr[H\rho]-\beta^{-1}S(\rho)$ is the nonequilibrium free energy, and $S(\rho)$ and $D(\rho\|\sigma)$ are the von Neumann entropy and the quantum relative entropy, respectively; $\Delta F=0$ if and only if $\rho=\pi_\beta$.

\subsection{Born--Markov--secular approximation}
\label{subsec:nrsc_bms}

Applying the Born--Markov--secular approximation~\cite{Breuer2007theory,Spohn1978irreversible} to time-independent Hamiltonians and a Gibbs environment, we show that the resulting Lindbladian satisfies NRSC.
We fix the system size $N$ and write the isolated-system Hamiltonian as $H_S^{(0)}:=H_N^{(0)}$.

\subsubsection{Derivation of the Lindbladian}
The total Hamiltonian is
\begin{align}
 H_{\mathrm{tot}}
 &=H_S^{(0)}\otimes I_{\mathrm E}+I_S\otimes H_{\mathrm E}
 +\lambda\sum_{a=1}^{M}S_a\otimes B_a,
 \label{eq:nrsc:bms_bare_hamiltonian}
\end{align}
where $S_a=S_a^\dagger$ and $B_a=B_a^\dagger$ are Hermitian operators on the system and the environment, respectively, $M$ is the number of coupling terms, and $\lambda\in\mathbb R$ is the coupling strength.
All Hamiltonians and coupling operators are time independent.

The total system is initially in $\rho_S(0)\otimes\rho_{\mathrm E}$, where $\rho_{\mathrm E}=\pi_{\beta,\mathrm E}:=e^{-\beta H_{\mathrm E}}/\Tr[e^{-\beta H_{\mathrm E}}]$ is the Gibbs state of the environment at inverse temperature $\beta$.

We define the mean of $B_a$ and the centered operator by
\begin{align}
 \overline B_a&:=\Tr[\rho_{\mathrm E} B_a],\\
 \widetilde B_a&:=B_a-\overline B_a I_{\mathrm E}
 \label{eq:nrsc:bms_centered_bath}
\end{align}
and rewrite the interaction as
\begin{align}
 \lambda\sum_a S_a\otimes B_a
 &=\left(\lambda\sum_a\overline B_a S_a\right)\otimes I_{\mathrm E}
 +\lambda\sum_a S_a\otimes\widetilde B_a.
 \label{eq:nrsc:bms_mean_split}
\end{align}
With the environment-induced Hamiltonian term and the remaining interaction
\begin{align}
 K&:=\lambda\sum_a\overline B_a S_a,\\
 H_I&:=\sum_a S_a\otimes\widetilde B_a,
 \label{eq:nrsc:bms_mean_and_interaction}
\end{align}
and $H_S:=H_S^{(0)}+K$, the total Hamiltonian reads
\begin{align}
 H_{\mathrm{tot}}
 &=H_S\otimes I_{\mathrm E}+I_S\otimes H_{\mathrm E}+\lambda H_I.
 \label{eq:nrsc:bms_hamiltonian}
\end{align}
This $H_S$, including the environment-induced Hamiltonian term $K$, corresponds to $H_N$ in the main text up to the Lamb-shift correction discussed below.

In the interaction picture with respect to $H_S\otimes I_{\mathrm E}+I_S\otimes H_{\mathrm E}$, the operators become $S_a(t)=e^{iH_St}S_ae^{-iH_St}$, $\widetilde B_a(t)=e^{iH_{\mathrm E}t}\widetilde B_ae^{-iH_{\mathrm E}t}$, and $H_I(t)=\sum_a S_a(t)\otimes\widetilde B_a(t)$.
Let $\rho_{S\mathrm E,I}(t)$ be the total density operator and $\rho_I(t):=\Tr_{\mathrm E}[\rho_{S\mathrm E,I}(t)]$ the reduced state of the system.
Then, writing $\rho_S(t)$ for the system state in the Schr\"odinger picture, we have $\rho_I(t)=e^{iH_St}\rho_S(t)e^{-iH_St}$.
Integrating the von Neumann equation $\dot\rho_{S\mathrm E,I}(t)=-i\lambda[H_I(t),\rho_{S\mathrm E,I}(t)]$ once gives
\begin{align}
 \rho_{S\mathrm E,I}(t)&=\rho_S(0)\otimes\rho_{\mathrm E}
 -i\lambda\int_0^t du\,[H_I(u),\rho_{S\mathrm E,I}(u)].
 \label{eq:nrsc:bms_integrated_von_neumann}
\end{align}
Substituting Eq.~\eqref{eq:nrsc:bms_integrated_von_neumann} into the right-hand side of the von Neumann equation and tracing out the environment, we obtain
\begin{align}
 \frac{d\rho_I(t)}{dt}
 &=-i\lambda\Tr_{\mathrm E}\!\left[
 [H_I(t),\rho_S(0)\otimes\rho_{\mathrm E}]
 \right]\notag\\
 &\quad-\lambda^2\int_0^t du\,
 \Tr_{\mathrm E}\!\left[
 [H_I(t),[H_I(u),\rho_{S\mathrm E,I}(u)]]
 \right].
 \label{eq:nrsc:bms_after_substitution}
\end{align}
The definition of $\widetilde B_a$ and $[H_{\mathrm E},\rho_{\mathrm E}]=0$ give $\Tr[\rho_{\mathrm E}\widetilde B_a(t)]=\Tr[\rho_{\mathrm E}\widetilde B_a]=0$, and hence
\begin{align}
 \Tr_{\mathrm E}\!\left[[H_I(t),\rho_S(0)\otimes\rho_{\mathrm E}]\right]
 &=\sum_a[S_a(t),\rho_S(0)]\Tr[\rho_{\mathrm E} \widetilde B_a(t)]=0.
 \label{eq:nrsc:bms_first_order_zero}
\end{align}
Therefore, the first term on the right-hand side of Eq.~\eqref{eq:nrsc:bms_after_substitution} vanishes.
Changing variables to $s=t-u$ in the remaining second term, we obtain
\begin{align}
 \frac{d\rho_I(t)}{dt}
 &=-\lambda^2\int_0^t ds\,
 \Tr_{\mathrm E}\!\left[
 [H_I(t),[H_I(t-s),\rho_{S\mathrm E,I}(t-s)]]
 \right].
 \label{eq:nrsc:bms_exact_integral}
\end{align}
For weak coupling, the Born approximation replaces $\rho_{S\mathrm E,I}(t-s)$ on the right-hand side by $\rho_I(t-s)\otimes\rho_{\mathrm E}$, neglecting system--environment correlations and changes in the environment state.

Let $\tau_{\mathrm E}$ be the decay time of the environment correlations and $\tau_S$ the time scale on which $\rho_I(t)$ changes; we assume $\tau_{\mathrm E}\ll\tau_S$.
The Markov approximation then replaces $\rho_I(t-s)$ by $\rho_I(t)$, and for $t\gg\tau_{\mathrm E}$ we extend the integral to infinity, obtaining
\begin{align}
 \frac{d\rho_I(t)}{dt}
 &=-\lambda^2\int_0^\infty ds\,
 \Tr_{\mathrm E}\!\left[
 [H_I(t),[H_I(t-s),\rho_I(t)\otimes\rho_{\mathrm E}]]
 \right].
 \label{eq:nrsc:bms_markov_integral}
\end{align}

Writing $P_E:=P_{E,N}$ for the spectral projectors of $H_N=H_S$ in Eq.~\eqref{eq:sm:hamiltonian_resolution}, we define the frequency components of each $S_a$ by
\begin{align}
 S_a(\omega)&:=\sum_E P_E S_a P_{E+\omega}
 \qquad(\omega\in\mathcal{W}_N).
 \label{eq:nrsc:bms_frequency_components}
\end{align}
The relation $H_SP_E=EP_E$ gives $[S_a(\omega),H_S]=\omega S_a(\omega)$, and $S_a=S_a^\dagger$ gives $S_a(-\omega)=S_a(\omega)^\dagger$.
Thus, $S_a(t)=\sum_{\omega\in\mathcal{W}_N}e^{-i\omega t}S_a(\omega)$.

We expand the double commutator in Eq.~\eqref{eq:nrsc:bms_markov_integral} into four terms and trace out the environment.
The first term is
\begin{align}
 &\Tr_{\mathrm E}\!\left[H_I(t)H_I(t-s)(\rho_I(t)\otimes\rho_{\mathrm E})\right]\notag\\
 &\quad=\sum_{a,b}S_a(t)S_b(t-s)\rho_I(t)
 \Tr[\widetilde B_a(t)\widetilde B_b(t-s)\rho_{\mathrm E}].
 \label{eq:nrsc:bms_first_partial_trace}
\end{align}
To express the environment trace on the right-hand side, we define the two-time correlation functions $C_{ab}(s):=\Tr[\rho_{\mathrm E}\widetilde B_a(s)\widetilde B_b]$.
Since $[H_{\mathrm E},\rho_{\mathrm E}]=0$ and the trace is cyclic, we have
\begin{align*}
 \Tr[\widetilde B_a(t)\widetilde B_b(t-s)\rho_{\mathrm E}]&=C_{ab}(s),\\
 \Tr[\widetilde B_a(t)\rho_{\mathrm E}\widetilde B_b(t-s)]&=C_{ba}(-s).
\end{align*}
Moreover, $\widetilde B_a^\dagger=\widetilde B_a$ and $[H_{\mathrm E},\rho_{\mathrm E}]=0$ give $C_{ba}(-s)=C_{ab}(s)^*$.
Computing the remaining three terms similarly, we obtain the time evolution of $\rho_I(t)$
\begin{align}
 \frac{d\rho_I(t)}{dt}
 &=-\lambda^2\sum_{a,b}\int_0^\infty ds\,\Bigl(
 C_{ab}(s)[S_a(t),S_b(t-s)\rho_I(t)]\notag\\
 &\qquad\qquad+C_{ba}(-s)[\rho_I(t)S_b(t-s),S_a(t)]
 \Bigr).
 \label{eq:nrsc:bms_markov_equation}
\end{align}

We reverse the order of the two commutators in Eq.~\eqref{eq:nrsc:bms_markov_equation} to absorb the overall minus sign and substitute the frequency decomposition.
In the term containing $C_{ab}(s)$, we expand
\begin{align}
 S_a(t)&=\sum_{\omega'\in\mathcal{W}_N}
 e^{i\omega't}S_a(\omega')^\dagger,\\
 S_b(t-s)&=\sum_{\omega\in\mathcal{W}_N}
 e^{-i\omega(t-s)}S_b(\omega).
 \label{eq:nrsc:bms_frequency_assignment}
\end{align}
Since the phase $e^{i(\omega'-\omega)t}$ and the commutator are independent of $s$, they can be taken outside the integral.
Integrating the remaining factor $e^{i\omega s}C_{ab}(s)$, we define the coefficients
\begin{align}
 \Gamma_{ab}(\omega)
 &:=\lambda^2\int_0^\infty e^{i\omega s}C_{ab}(s)\,ds.
 \label{eq:nrsc:bms_one_sided_rates}
\end{align}
In the term containing $C_{ba}(-s)$, we interchange $a$ and $b$ and expand $S_b(t)$ and $S_a(t-s)$ in $S_b(\omega)$ and $S_a(\omega')^\dagger$, respectively.
Since $C_{ab}(-s)=C_{ba}(s)^*$, the $s$ integral in this term gives
\begin{align}
 \lambda^2\int_0^\infty e^{-i\omega's}C_{ab}(-s)\,ds
 &=\Gamma_{ba}(\omega')^*.
 \label{eq:nrsc:bms_second_frequency_integral}
\end{align}
Therefore, combining the two terms gives
\begin{align}
 \frac{d\rho_I(t)}{dt}
 &=\sum_{\omega,\omega'\in\mathcal{W}_N}\sum_{a,b}
 e^{i(\omega'-\omega)t}\Bigl\{
 \Gamma_{ab}(\omega)
 [S_b(\omega)\rho_I(t),S_a(\omega')^\dagger]\notag\\
 &\qquad\qquad+\Gamma_{ba}(\omega')^*
 [S_b(\omega),\rho_I(t)S_a(\omega')^\dagger]
 \Bigr\}.
 \label{eq:nrsc:bms_frequency_equation}
\end{align}

The terms with $\omega\neq\omega'$ in Eq.~\eqref{eq:nrsc:bms_frequency_equation} carry the phase $e^{i(\omega'-\omega)t}$.
If this phase oscillates much faster than the system relaxes, the secular approximation keeps only the terms with $\omega=\omega'$, giving
\begin{align}
 \frac{d\rho_I(t)}{dt}
 &=\sum_{\omega\in\mathcal{W}_N}\sum_{a,b}\Bigl[
 \Gamma_{ab}(\omega)
 \bigl(S_b(\omega)\rho_I(t)S_a(\omega)^\dagger
       -S_a(\omega)^\dagger S_b(\omega)\rho_I(t)\bigr)\notag\\
 &\qquad\qquad+\Gamma_{ba}(\omega)^*
 \bigl(S_b(\omega)\rho_I(t)S_a(\omega)^\dagger
       -\rho_I(t)S_a(\omega)^\dagger S_b(\omega)\bigr)
 \Bigr].
 \label{eq:nrsc:bms_secular_equation}
\end{align}

We now define the Hermitian matrices
\begin{align}
 \gamma_{ab}(\omega)
 &:=\Gamma_{ab}(\omega)+\Gamma_{ba}(\omega)^*,
 \label{eq:nrsc:bms_rate_matrix}\\
 \sigma_{ab}(\omega)
 &:=\frac{\Gamma_{ab}(\omega)-\Gamma_{ba}(\omega)^*}{2i}.
 \label{eq:nrsc:bms_shift_matrix}
\end{align}
In terms of $\gamma$ and $\sigma$, Eq.~\eqref{eq:nrsc:bms_secular_equation} reads $d\rho_I(t)/dt=-i[H_{\mathrm{LS}},\rho_I(t)]+\mathcal D[\rho_I(t)]$.
Here, we define the Lamb-shift term and the dissipative term by
\begin{align}
 H_{\mathrm{LS}}
 &:=\sum_{\omega\in\mathcal{W}_N}\sum_{a,b}
 \sigma_{ab}(\omega)S_a(\omega)^\dagger S_b(\omega),
 \label{eq:nrsc:bms_lamb_shift}\\
 \mathcal D[\rho]
 &:=\sum_{\omega\in\mathcal{W}_N}\sum_{a,b}\gamma_{ab}(\omega)
 \left(
 S_b(\omega)\rho S_a(\omega)^\dagger
 -\frac12\{S_a(\omega)^\dagger S_b(\omega),\rho\}
 \right).
 \label{eq:nrsc:bms_matrix_dissipator}
\end{align}
Since $C_{ba}(s)^*=C_{ab}(-s)$, the substitution $s\to-s$ turns $\Gamma_{ba}(\omega)^*$ into an integral over $s\leq0$, so the dissipation coefficients are the two-sided Fourier transform
\begin{align}
 \gamma_{ab}(\omega)
 &=\lambda^2\int_{-\infty}^{\infty}e^{i\omega s}C_{ab}(s)\,ds.
 \label{eq:nrsc:bms_full_fourier}
\end{align}

We show that $\gamma(\omega)$ is positive semidefinite by using the energy eigenbasis of the environment.
Since $[H_{\mathrm E},\rho_{\mathrm E}]=0$, we can diagonalize $H_{\mathrm E}$ and $\rho_{\mathrm E}$ simultaneously.
We write $H_{\mathrm E}\ket{m}=\epsilon_m\ket{m}$ and $\rho_{\mathrm E}=\sum_m p_m\ket{m}\bra{m}$ with $p_m\geq0$, and set $(\widetilde B_a)_{mn}:=\bra{m}\widetilde B_a\ket{n}$.
Expanding the correlation function in this energy basis and evaluating the Fourier integral in Eq.~\eqref{eq:nrsc:bms_full_fourier}, we find that, in the sense of distributions, the dissipation coefficients are
\begin{align}
 \gamma_{ab}(\omega)
 &=2\pi\lambda^2\sum_{m,n}p_m
 (\widetilde B_a)_{mn}(\widetilde B_b)_{nm}
 \delta(\omega+\epsilon_m-\epsilon_n),
 \label{eq:nrsc:bms_spectral_rates}
\end{align}
where $\delta$ is the Dirac delta function.
Using $\widetilde B_a^\dagger=\widetilde B_a$, we obtain, for any $z=(z_1,\ldots,z_M)\in\mathbb C^M$,
\begin{align}
\sum_{a,b}z_a^*\gamma_{ab}(\omega)z_b
&=
2\pi\lambda^2\sum_{m,n}p_m
\left|\sum_b z_b(\widetilde B_b)_{nm}\right|^2
\delta(\omega+\epsilon_m-\epsilon_n).
 \label{eq:nrsc:bms_rate_positivity}
\end{align}
The right-hand side is nonnegative in the sense of distributions, and hence $\gamma(\omega)\geq0$.

Next, we use that $\rho_{\mathrm E}$ is a Gibbs state, $p_m\propto e^{-\beta\epsilon_m}$.
Interchanging $m$ and $n$ in $\gamma_{ab}(-\omega)$ and using $p_n=e^{-\beta\omega}p_m$ on the support of $\delta(\omega+\epsilon_m-\epsilon_n)$, we obtain
\begin{align}
 \gamma_{ab}(-\omega)&=e^{-\beta\omega}\gamma_{ba}(\omega).
 \label{eq:nrsc:bms_kms}
\end{align}

Since $\gamma(\omega)\geq0$, we diagonalize it with nonnegative eigenvalues $\gamma_{\omega r}$ and orthonormal eigenvectors $u_r(\omega)$ as
\begin{align}
\gamma_{ab}(\omega)
&=\sum_{r=1}^{M}\gamma_{\omega r}
u_{ar}(\omega)u_{br}(\omega)^*.
 \label{eq:nrsc:bms_rate_diagonalization}
\end{align}
We define the jump operators
\begin{align}
L_{\omega r}&:=\sum_a u_{ar}(\omega)^*S_a(\omega).
 \label{eq:nrsc:bms_jump_operators}
\end{align}
By Eq.~\eqref{eq:nrsc:bms_frequency_components}, they satisfy $[L_{\omega r},H_S]=\omega L_{\omega r}$.

From Eq.~\eqref{eq:nrsc:bms_kms}, $\gamma(-\omega)=e^{-\beta\omega}\gamma(\omega)^{\mathsf T}$, so for $\omega>0$ we can choose $u_r(-\omega)=u_r(\omega)^*$.
This choice and $S_a(-\omega)=S_a(\omega)^\dagger$ give
\begin{align}
L_{-\omega r}&=L_{\omega r}^\dagger,\qquad
\gamma_{-\omega r}=e^{-\beta\omega}\gamma_{\omega r}.
 \label{eq:nrsc:bms_reverse_channels}
\end{align}

At zero frequency, $\gamma(0)$ is real symmetric, so we choose $u_r(0)$ to be real.
Since $S_a(0)^\dagger=S_a(0)$, the operators $L_{0r}$ are also Hermitian.

Writing $k=(\omega,r)$, $-k=(-\omega,r)$, and $(L_k,\gamma_k,\omega_k):=(L_{\omega r},\gamma_{\omega r},\omega)$, we substitute this eigendecomposition into Eq.~\eqref{eq:nrsc:bms_matrix_dissipator} and obtain
\begin{align}
\mathcal D[\rho]
&=\sum_k\gamma_k
\left(
L_k\rho L_k^\dagger
-\frac12\{L_k^\dagger L_k,\rho\}
\right).
 \label{eq:nrsc:bms_diagonal_dissipator}
\end{align}
In the Schr\"odinger picture, the reduced dynamics thus reads
\begin{align}
\frac{d\rho_S(t)}{dt}
&=-i[H_S+H_{\mathrm{LS}},\rho_S(t)]+\mathcal D[\rho_S(t)].
 \label{eq:nrsc:bms_nrsc_lindbladian}
\end{align}

The Lamb-shift Hamiltonian $H_{\mathrm{LS}}$ is either neglected or renormalized into the system Hamiltonian~\cite{Winczewski2021renormalization}, and the resulting Lindbladian satisfies NRSC in both cases.
If it is neglected, $H_S$, $L_k$, and $\gamma_k$ are time independent, satisfying condition (i), and the commutation relation and the reverse channels above give conditions (ii) and (iii).
If it is renormalized, the coupling operators are frequency-decomposed with the corrected Hamiltonian and the rates are defined at its Bohr frequencies, so that conditions (ii) and (iii) hold with respect to this Hamiltonian and its Bohr frequencies.

\subsubsection{Relation to external resource supply}\label{subsubsec:nrsc_bms_resource}
We now identify which assumptions give each condition of NRSC.
Besides the Born--Markov--secular approximation and the treatment of $H_{\mathrm{LS}}$ above, the derivation relies on three assumptions:
\begin{enumerate}
 \item[(B1)] Time independence: $H_S^{(0)}$, $H_{\mathrm E}$, $S_a$, and $B_a$ are time independent.
 \item[(B2)] Stationary environment: $[H_{\mathrm E},\rho_{\mathrm E}]=0$.
 \item[(B3)] Gibbs environment: $\rho_{\mathrm E}=\pi_{\beta,\mathrm E}$.
\end{enumerate}
In the derivation above, conditions (i) and (ii) follow from (B1) and (B2), and condition (iii) follows from (B3), which implies (B2).

Within the Born--Markov--secular approximation, NRSC therefore follows from (B1)--(B3) alone, and violating it requires violating at least one of them.
Each such violation requires an external supply of thermodynamic resources.
Violating (B1) requires time dependence in one of $H_S^{(0)}$, $H_{\mathrm E}$, $S_a$, and $B_a$, and hence an external clock~\cite{Goold2016role,Woods2019autonomous} asymmetric under time translations~\cite{Marvian2019nobroadcasting,Kwon2018clock}: the state $\rho_{\mathrm C}$ of the clock does not commute with its Hamiltonian $H_{\mathrm C}$, $[\rho_{\mathrm C},H_{\mathrm C}]\neq0$.
Violating (B2) means that the environment state is itself asymmetric under time translations, $[\rho_{\mathrm E},H_{\mathrm E}]\neq0$.
Violating (B3) while keeping (B2) requires a nonequilibrium environment, $\rho_{\mathrm E}\neq\pi_{\beta,\mathrm E}$.
In all three cases, the state $\rho$ of the clock or the environment differs from the Gibbs state $\pi_\beta$ of its Hamiltonian $H$: in the first two cases because $[\pi_\beta,H]=0$, and in the third by assumption.
Preparing such a state therefore requires $\Delta F=\beta^{-1}D(\rho\|\pi_\beta)>0$ in Eq.~\eqref{eq:nrsc:free_energy}.

\subsection{Collision model}
\label{subsec:nrsc_collision}

Next, we derive a GKSL equation from a collision model~\cite{Scarani2002thermalizing,Ciccarello2022quantum} and show that the resulting Lindbladian satisfies NRSC under conditions (C1)--(C4) below.
Therefore, if the resulting Lindbladian violates NRSC, at least one of these conditions fails.

In this subsection, we fix $N$ and write the isolated-system Hamiltonian as $H_S^{(0)}:=H_N^{(0)}$.
We assume that the system and the ancillas are finite dimensional.

\subsubsection{Model and assumptions}
In a collision model, the environment consists of identical ancillas $E_1,E_2,\dots$, with which the system $S$ interacts sequentially, each for a short time.
Each ancilla has the Hamiltonian $H_{\mathrm E}=\sum_i\epsilon_i\dm{i}$, where $\{\ket{i}\}$ is an orthonormal eigenbasis of $H_{\mathrm E}$.
We assume that, before its interaction, each ancilla is in the state $\rho_{\mathrm E}$ and uncorrelated with the system.
We set $H_0:=H_S^{(0)}\otimes I_{\mathrm E}+I_S\otimes H_{\mathrm E}$, where $I_S$ and $I_{\mathrm E}$ are the identity operators on the system and the ancilla, respectively.
In each interaction of duration $\delta t$, the system and one ancilla evolve under
\begin{align}
 H_{\mathrm{tot}}^{(\delta t)}=H_0+V_{\delta t},
 \qquad V_{\delta t}:=\frac{1}{\sqrt{\delta t}}H_I+G,
 \label{eq:nrsc:collision_hamiltonian}
\end{align}
after which the ancilla is discarded.
Here, $H_I=H_I^\dagger$ and $G=G^\dagger$ are interaction Hamiltonians.
To avoid divergence as $\delta t\to0$, we assume
\begin{align}
 \Tr_{\mathrm E}\!\left[H_I(I_S\otimes\rho_{\mathrm E})\right]=0.
 \label{eq:nrsc:zero_mean}
\end{align}
We also define
\begin{align}
 K:=\Tr_{\mathrm E}\!\left[G(I_S\otimes\rho_{\mathrm E})\right],
 \qquad H_S:=H_S^{(0)}+K=H_N.
 \label{eq:nrsc:mean_field}
\end{align}
With $U:=e^{-iH_{\mathrm{tot}}^{(\delta t)}\delta t}$, a single interaction changes the state of the system as
\begin{align}
 \rho_S(t+\delta t)
=\Tr_{\mathrm E}\!\left[U\bigl(\rho_S(t)\otimes\rho_{\mathrm E}\bigr)U^\dagger\right].
 \label{eq:nrsc:collision_map}
\end{align}

We impose four assumptions on this model.
\begin{enumerate}
\item[(C1)] Time independence: $H_S^{(0)}$, $H_{\mathrm E}$, $H_I$, $G$, and $\rho_{\mathrm E}$ are independent of the time $t$ of each interaction.
 \item[(C2)] Energy conservation: $[H_I,H_0]=[G,H_0]=0$.
 \item[(C3)] Equilibrium environment: $\rho_{\mathrm E}$ is the Gibbs state at inverse temperature $\beta$,
 \begin{align}
  \rho_{\mathrm E}=\pi_{\beta,\mathrm E}:=\frac{e^{-\beta H_{\mathrm E}}}{\Tr[e^{-\beta H_{\mathrm E}}]}=\sum_ip_i\dm{i},
  \label{eq:nrsc:ancilla_gibbs}
 \end{align}
 where $p_i:=e^{-\beta\epsilon_i}/\sum_je^{-\beta\epsilon_j}$.
 \item[(C4)] Zero average power of ancilla replacement: For an input state $\rho$ of the system, we set
 \begin{align}
  \rho'_{S\mathrm E}:=U(\rho\otimes\rho_{\mathrm E})U^\dagger,
  \qquad \Phi_{\delta t}[\rho]:=\Tr_{\mathrm E}[\rho'_{S\mathrm E}].
  \label{eq:nrsc:replacement_output}
 \end{align}
 Defining the average work for replacing the used ancilla with a fresh one~\cite{Ciccarello2022quantum} as
 \begin{align}
  W_{\mathrm{rep}}^{(\delta t)}(\rho)
  :=\Tr\!\left[V_{\delta t}(\Phi_{\delta t}[\rho]\otimes\rho_{\mathrm E})\right]
  -\Tr\!\left[V_{\delta t}\rho'_{S\mathrm E}\right],
  \label{eq:nrsc:replacement_work}
 \end{align}
 we have
 \begin{align}
  P_{\mathrm{rep}}(\rho)
  :=\lim_{\delta t\to0}\frac{W_{\mathrm{rep}}^{(\delta t)}(\rho)}{\delta t}=0
  \qquad\text{for all }\rho.
  \label{eq:nrsc:zero_replacement_power}
 \end{align}
\end{enumerate}

\subsubsection{Derivation of the Lindbladian}
To take the continuous-time limit, we compute the one-step state change to first order in $\delta t$.
Since the exponent of $U$ contains a term proportional to $\sqrt{\delta t}$, we expand $U$ to second order in $\sqrt{\delta t}$,
\begin{align}
 U&=\exp\!\left[-i\left((H_0+G)\delta t+H_I\sqrt{\delta t}\right)\right]\notag\\
 &=I-i\sqrt{\delta t}\,H_I-i\delta t\,(H_0+G)-\frac{\delta t}{2}H_I^2+O\!\left((\delta t)^{3/2}\right),\notag\\
 U^\dagger&=I+i\sqrt{\delta t}\,H_I+i\delta t\,(H_0+G)-\frac{\delta t}{2}H_I^2+O\!\left((\delta t)^{3/2}\right),
 \label{eq:nrsc:U_expansion}
\end{align}
where $I$ is the identity on the system and the ancilla.
For $\rho_{S\mathrm E}:=\rho_S(t)\otimes\rho_{\mathrm E}$, this gives
\begin{align}
 U\rho_{S\mathrm E}U^\dagger
 &=\rho_{S\mathrm E}
 -i\sqrt{\delta t}\,[H_I,\rho_{S\mathrm E}]
 -i\delta t\,[H_0+G,\rho_{S\mathrm E}]\notag\\
 &\quad
 +\delta t\left(
 H_I\rho_{S\mathrm E}H_I-\frac12\{H_I^2,\rho_{S\mathrm E}\}
 \right)
 +O\!\left((\delta t)^{3/2}\right).
 \label{eq:nrsc:one_step_expansion}
\end{align}

We now trace out the ancilla term by term.
By Eq.~\eqref{eq:nrsc:zero_mean}, the term proportional to $\sqrt{\delta t}$ vanishes,
\begin{align}
 \Tr_{\mathrm E}\!\left[[H_I,\rho_{S\mathrm E}]\right]
 =\left[\Tr_{\mathrm E}\!\left[H_I(I_S\otimes\rho_{\mathrm E})\right],\rho_S(t)\right]
 =0,
 \label{eq:nrsc:first_order_vanishes}
\end{align}
and by $H_0=H_S^{(0)}\otimes I_{\mathrm E}+I_S\otimes H_{\mathrm E}$ and Eq.~\eqref{eq:nrsc:mean_field}, $\Tr_{\mathrm E}[[H_0+G,\rho_{S\mathrm E}]]=[H_S^{(0)}+K,\rho_S(t)]=[H_S,\rho_S(t)]$.
For the remaining terms, we expand the interaction in the eigenbasis of the ancilla,
\begin{align}
 H_I=\sum_{i,j}L_{ij}\otimes\ketbra{i}{j},
 \label{eq:nrsc:jump_definition}
\end{align}
where $L_{ij}:=(I_S\otimes\bra{i})H_I(I_S\otimes\ket{j})$ acts on the system when the ancilla makes a transition from $\ket{j}$ to $\ket{i}$, and satisfies $L_{ij}^\dagger=L_{ji}$ because $H_I=H_I^\dagger$.
Moreover, under (C2), $L_{ij}$ changes the energy of $H_S^{(0)}$ by a definite amount.
Indeed, substituting the expansion in Eq.~\eqref{eq:nrsc:jump_definition} into $[H_0,H_I]=0$ gives
\begin{align}
 0=[H_0,H_I]
 =\sum_{i,j}\left([H_S^{(0)},L_{ij}]+(\epsilon_i-\epsilon_j)L_{ij}\right)\otimes\ketbra{i}{j}.
\end{align}
Taking the $\ketbra{i}{j}$ component and setting $\omega_{ij}:=\epsilon_i-\epsilon_j$, we obtain
\begin{align}
 [L_{ij},H_S^{(0)}]=\omega_{ij}L_{ij}.
 \label{eq:nrsc:bare_jump_covariance}
\end{align}
Thus, $L_{ij}$ removes from the system the energy $\omega_{ij}$ that the ancilla gains in the transition $\ket{j}\to\ket{i}$.
By (C3), $\bra{j}\rho_{\mathrm E}\ket{k}=p_j\delta_{jk}$.
Using this, we obtain
\begin{align}
 \Tr_{\mathrm E}\!\left[H_I\rho_{S\mathrm E}H_I\right]
 &=\sum_{i,j,k,l}L_{ij}\rho_S(t)L_{kl}\Tr\!\left[\ketbra{i}{j}\rho_{\mathrm E}\ketbra{k}{l}\right]
 =\sum_{i,j}p_jL_{ij}\rho_S(t)L_{ij}^\dagger,
 \label{eq:nrsc:jump_term}\\
 \Tr_{\mathrm E}\!\left[H_I^2\rho_{S\mathrm E}\right]
 &=\sum_{i,j,l}L_{ij}L_{jl}\rho_S(t)\Tr\!\left[\ketbra{i}{l}\rho_{\mathrm E}\right]
 =\sum_{i,j}p_jL_{ij}^\dagger L_{ij}\rho_S(t),
 \label{eq:nrsc:anticommutator_term_left}\\
 \Tr_{\mathrm E}\!\left[\rho_{S\mathrm E}H_I^2\right]
 &=\sum_{i,j,l}\rho_S(t)L_{ij}L_{jl}\Tr\!\left[\rho_{\mathrm E}\ketbra{i}{l}\right]
 =\sum_{i,j}p_j\rho_S(t)L_{ij}^\dagger L_{ij}.
 \label{eq:nrsc:anticommutator_term_right}
\end{align}
We used $L_{ij}^\dagger=L_{ji}$ and, in the last two equations, interchanged $i$ and $j$.
Substituting these into the partial trace of Eq.~\eqref{eq:nrsc:one_step_expansion}, subtracting $\rho_S(t)$ from both sides, dividing by $\delta t$, and taking $\delta t\to0$, we obtain the GKSL equation
\begin{align}
 \frac{d\rho_S(t)}{dt}
 =-i[H_S,\rho_S(t)]
 +\sum_{i,j}\gamma_{ij}\left(
 L_{ij}\rho_S(t)L_{ij}^\dagger
 -\frac12\{L_{ij}^\dagger L_{ij},\rho_S(t)\}
 \right)
 \label{eq:nrsc:collision_gksl}
\end{align}
with $\gamma_{ij}:=p_j$.
We denote its right-hand side by $\mathcal L[\rho_S(t)]$ and the dissipative part by $\mathcal D[\rho_S(t)]$.

Assumptions (C2)--(C4) imply $[L_{ij},K]=0$, as shown at the end of this subsection, so that
\begin{align}
[L_{ij},H_S]
&=[L_{ij},H_S^{(0)}]+[L_{ij},K]\notag\\
&=\omega_{ij}L_{ij}.
 \label{eq:nrsc:jump_covariance}
\end{align}

We now check NRSC for Eq.~\eqref{eq:nrsc:collision_gksl}.
We label the channels by $k=(i,j)$ with $L_k:=L_{ij}$, $\gamma_k:=\gamma_{ij}$, and $\omega_k:=\omega_{ij}$, so that the reverse channel is $-k=(j,i)$.
Since $L_k$ is fixed by $H_I$ and the eigenbasis of $H_{\mathrm E}$, and $\gamma_k$ by $\rho_{\mathrm E}$, assumption (C1) makes $H_S$ and $\{\sqrt{\gamma_k}L_k\}_k$ time independent, giving condition (i).
Equation~\eqref{eq:nrsc:jump_covariance} is condition (ii), $[L_k,H_S]=\omega_kL_k$.
Finally, $L_{-k}=L_{ji}=L_{ij}^\dagger=L_k^\dagger$ and, by (C3), $\gamma_k=p_j=e^{\beta\omega_{ij}}p_i=e^{\beta\omega_k}\gamma_{-k}$, which is condition (iii) because $\omega_k$ is the frequency of $L_k$ by Eq.~\eqref{eq:nrsc:jump_covariance}.
Hence, Eq.~\eqref{eq:nrsc:collision_gksl} satisfies NRSC.

It remains to derive $[L_{ij},K]=0$ from (C2)--(C4).
By (C2) and $[H_{\mathrm E},\rho_{\mathrm E}]=0$, we have
\begin{align}
 [K,H_S^{(0)}]
 &=\Tr_{\mathrm E}\!\left[[G,H_S^{(0)}\otimes I_{\mathrm E}](I_S\otimes\rho_{\mathrm E})\right]\notag\\
 &=-\Tr_{\mathrm E}\!\left[[G,I_S\otimes H_{\mathrm E}](I_S\otimes\rho_{\mathrm E})\right]=0.
 \label{eq:nrsc:mean_field_commutes}
\end{align}
Since (C2) also gives $[V_{\delta t},H_{\mathrm{tot}}^{(\delta t)}]=0$ and hence $[U,V_{\delta t}]=0$, Eq.~\eqref{eq:nrsc:zero_mean} yields
\begin{align}
 \Tr\!\left[V_{\delta t}\rho'_{S\mathrm E}\right]
 &=\Tr\!\left[V_{\delta t}(\rho\otimes\rho_{\mathrm E})\right]
 =\Tr[K\rho],\notag\\
 \Tr\!\left[V_{\delta t}(\Phi_{\delta t}[\rho]\otimes\rho_{\mathrm E})\right]
 &=\Tr\!\left[K\Phi_{\delta t}[\rho]\right],
 \label{eq:nrsc:replacement_energy}
\end{align}
so the replacement work~\eqref{eq:nrsc:replacement_work} becomes
\begin{align}
 W_{\mathrm{rep}}^{(\delta t)}(\rho)
 =\Tr\!\left[K\bigl(\Phi_{\delta t}[\rho]-\rho\bigr)\right].
 \label{eq:nrsc:replacement_work_mean_field}
\end{align}
Because Eq.~\eqref{eq:nrsc:mean_field_commutes} gives $[K,H_S]=0$, assumption (C4) then reads
\begin{align}
 P_{\mathrm{rep}}(\rho)
 =\Tr\!\left[K\mathcal L[\rho]\right]
 =\Tr\!\left[\rho\mathcal D^\dagger[K]\right]=0
 \qquad\text{for all }\rho,
 \label{eq:nrsc:replacement_power_mean_field}
\end{align}
which implies
\begin{align}
 \mathcal D^\dagger[K]=0.
 \label{eq:nrsc:mean_field_fixed}
\end{align}

Next, let $\sigma_0:=e^{-\beta H_S^{(0)}}/\Tr[e^{-\beta H_S^{(0)}}]$.
By (C2) and (C3), $[U,\sigma_0\otimes\rho_{\mathrm E}]=0$, and hence $\Phi_{\delta t}[\sigma_0]=\sigma_0$ for all $\delta t$, which gives $\mathcal L[\sigma_0]=0$.
Moreover, Eq.~\eqref{eq:nrsc:mean_field_commutes} gives $[H_S,\sigma_0]=0$, and thus $\mathcal D[\sigma_0]=0$.
Finite dimensionality and $0<\beta<\infty$ give $\sigma_0>0$.
Applying Eq.~\eqref{eq:sm:gksl_product_identity} to $K=K^\dagger$, we obtain
\begin{align}
 \mathcal D^\dagger[K^2]
 -\mathcal D^\dagger[K]K-K\mathcal D^\dagger[K]
 =\sum_{i,j}p_j[L_{ij},K]^\dagger[L_{ij},K].
 \label{eq:nrsc:mean_field_product_identity}
\end{align}
Taking its expectation value in $\sigma_0$ and using Eq.~\eqref{eq:nrsc:mean_field_fixed} and $\mathcal D[\sigma_0]=0$, we obtain
\begin{align}
 0=\Tr\!\left[\sigma_0\mathcal D^\dagger[K^2]\right]
 =\sum_{i,j}p_j\Tr\!\left[\sigma_0[L_{ij},K]^\dagger[L_{ij},K]\right].
 \label{eq:nrsc:mean_field_positive_sum}
\end{align}
Since each term is nonnegative and $p_j>0$ and $\sigma_0>0$, we conclude $[L_{ij},K]=0$ for all $i$ and $j$.

\subsubsection{Relation to external resource supply}\label{subsubsec:nrsc_collision_resource}
In the derivation above, condition (i) follows from (C1), condition (ii) from (C2)--(C4), and condition (iii) from (C3).
Within the collision model, NRSC therefore follows from (C1)--(C4), and violating it requires violating at least one of them.
Each such violation requires an external supply of thermodynamic resources.
Violating (C1) requires time dependence in one of $H_S^{(0)}$, $H_{\mathrm E}$, $H_I$, $G$, and $\rho_{\mathrm E}$, and hence an external clock~\cite{Goold2016role,Woods2019autonomous} asymmetric under time translations~\cite{Marvian2019nobroadcasting,Kwon2018clock}: the state $\rho_{\mathrm C}$ of the clock does not commute with its Hamiltonian $H_{\mathrm C}$, $[\rho_{\mathrm C},H_{\mathrm C}]\neq0$.
If (C2) is violated, $[H_I,H_0]\neq0$ or $[G,H_0]\neq0$, and Eq.~\eqref{eq:nrsc:U_expansion} gives
\begin{align}
 [U,H_0]=-i\sqrt{\delta t}\,[H_I,H_0]-i\delta t\,[G,H_0]-\frac{\delta t}{2}[H_I^2,H_0]+O\!\left((\delta t)^{3/2}\right).
 \label{eq:nrsc:U_H0_commutator}
\end{align}
If $[H_I,H_0]\neq0$, the term of order $\sqrt{\delta t}$ is nonzero; otherwise, $[H_I^2,H_0]=0$, and the term of order $\delta t$ is $-i\delta t\,[G,H_0]\neq0$.
In either case, $[U,H_0]\neq0$ for sufficiently small $\delta t$.
Suppose that this $U$ is implemented for arbitrary states of the system and ancilla by adding an external system $X$ with Hamiltonian $H_X$ and a unitary that conserves the total energy $H_0+H_X$.
Then the state $\rho_X$ of $X$ must be asymmetric under time translations~\cite{TSS,TSS2,TTK}, $[\rho_X,H_X]\neq0$.
Violating (C3) requires a nonequilibrium environment, $\rho_{\mathrm E}\neq\pi_{\beta,\mathrm E}$.
If (C4) is violated, $P_{\mathrm{rep}}(\rho)\neq0$ for at least one input state $\rho$: replacing the used ancilla with a fresh one changes the mean interaction energy by $W_{\mathrm{rep}}^{(\delta t)}(\rho)$ in Eq.~\eqref{eq:nrsc:replacement_work}, so that work is exchanged with an external work source at the nonzero average rate $P_{\mathrm{rep}}(\rho)$~\cite{Ciccarello2022quantum}.
If (C2) holds, Eq.~\eqref{eq:nrsc:replacement_work_mean_field} shows that this work equals the change $\Tr[K(\Phi_{\delta t}[\rho]-\rho)]$ in the mean value of $K$ during one interaction.

In the first three cases, the state $\rho$ of the clock, the external system, or the environment differs from the Gibbs state $\pi_\beta$ of its Hamiltonian $H$: for the clock and the external system because $[\pi_\beta,H]=0$, and for the environment by assumption.
Preparing such a state therefore requires $\Delta F=\beta^{-1}D(\rho\|\pi_\beta)>0$ in Eq.~\eqref{eq:nrsc:free_energy}.
In the case of (C4), the resource is the work exchanged with the work source at the rate $P_{\mathrm{rep}}(\rho)$.
\bibliographystyle{apsrev4-1}
\bibliography{GO_and_NOGO_TC/go_and_nogo_ref}